\documentclass{article}
\usepackage[a4paper,margin=1in]{geometry}

\usepackage[utf8]{inputenc}
\usepackage[english]{babel}

\usepackage{booktabs, tabularx}
\newcolumntype{Y}{>{\raggedright\arraybackslash}X}
\newcolumntype{M}[1]{>{\centering\arraybackslash}m{#1}}
\usepackage[table, dvipsnames]{xcolor}
\usepackage{multirow}
\usepackage{nicematrix}
\renewcommand{\arraystretch}{1.2}

\usepackage{pdflscape} % For rotating a page in the PDF output
\usepackage{adjustbox} % For precise alignment

\usepackage{xr} % cross-referencing across files

\usepackage[title]{appendix}
\usepackage{authblk}
\usepackage[space]{grffile}
\usepackage{latexsym}
\usepackage{textcomp}
\usepackage{longtable}
\usepackage{amsfonts, amsmath, amssymb}
\usepackage{nicefrac}
\usepackage[round]{natbib}
\let\cite\citep
\usepackage{url}
\usepackage{siunitx}
\usepackage{bm}
\usepackage{enumitem}
\usepackage{algorithm}
\usepackage{algpseudocode}
\usepackage{gensymb}
\usepackage{graphicx}
\usepackage{tikz}
\usetikzlibrary{shapes, arrows, positioning, calc}
\usepackage{placeins}
\usepackage{lscape}
\usepackage{rotating}
\usepackage{subcaption}
\usepackage{hyperref}
\hypersetup{colorlinks=false, pdfborder={0 0 0}}
\usepackage{etoolbox}

\usepackage{lineno}
\usepackage{mycolors}

\newcommand{\CR}[1]{{\color{red}{#1}}}

\newcommand{\CGR}[1]{{\color{lightgray}{#1}}}
\definecolor{softyellow}{HTML}{FFF9DB}

\def\drawline#1#2{\raise 2.5pt\vbox{\hrule width #1pt height #2pt}}
\def\spacce#1{\hskip #1pt}
\def\solid{\drawline{24}{.7}}
\def\sol{\drawline{12}{.7}\nobreak}
\def\bdash{\hbox{\spacce{1}\drawline{4}{.7}\spacce{1}}}

\def\dashed{\bdash\bdash\bdash\bdash}

\def\trian{\raise 1.25pt\hbox{$\scriptscriptstyle\triangle$}\nobreak\ }
\def\trian{\raise 1.25pt\hbox{$\scriptscriptstyle\triangle$}\nobreak\ }

\DeclareRobustCommand{\legendsquare}[1]{%
  \textcolor{#1}{\rule{1.5ex}{1.5ex}}%
}
\newcommand{\loss}{\mathcal{L}}
\newcommand*{\V}[1]{\bm{#1}}% bold vectors

\title{Hybrid physics-data-driven modeling for sea ice thermodynamics and transfer learning}
\author[1]{G. De Cillis}
\author[1,5]{A. Carrassi}
\author[2]{J. Brajard}
\author[2]{L. Bertino}
\author[3]{M. Broccoli}
\author[3]{D. Iovino}
\author[4]{T.S. Finn}
\author[4]{M. Bocquet}

\affil[1]{Department of Physics and Astronomy ‘‘Augusto Righi’’, Bologna, 40126, Italy}
\affil[2]{Nansen Environmental and Remote Sensing Center, Jahnebakken 3, Bergen, N-5007, Norway}
\affil[3]{CMCC Foundation - Euro-Mediterranean Center on Climate Change, Bologna, Italy}
\affil[4]{CEREA, ENPC, EDF R\&D, Institut Polytechnique de Paris,
6-8 avenue Blaise Pascal, Cit\'{e} Descartes, 
Marne-la-Vall\'{e}e, 77455, France}
\affil[5]{Department of Meteorology, University of Reading, Reading, UK}

\begin{document}
    \maketitle
    \begin{abstract}
        This study explores a physics–data driven hybrid approach for sea-ice column physics models, in which a machine learning (ML) component acts as a state-dependent parameterization of forecast errors. We examine how perturbations in snow thermodynamics and sea-ice radiative properties affect forecast errors, and train dedicated neural networks (NNs) for each model configuration.
        The performance of the hybrid models is evaluated for long lead-time forecasts and compared against a benchmark system based on climatological forecast-error estimates. The \mbox{NN-based} hybrids prove to be stable, robust to initial condition and atmospheric forcing errors, and consistently outperform their climatology-based counterpart.
        To derive guiding principles for efficiently handling possible physical model updates, we perform transfer learning experiments to test whether pretrained NNs optimized for one model configuration can be successfully adapted to another. Results indicate that direct evaluation of pretrained networks on the target task provides useful insights into their adaptability, recommending transfer learning whenever performance exceeds a trivial baseline.
        Finally, a feature-importance analysis shows that atmospheric forcing inputs have negligible influence on NN predictive skill, while ice-layer enthalpies play a key role in achieving satisfactory performance.
        
    \end{abstract}

    %%%%%%%%%%%%%%%%%%%%%%%%%%%%%%%%%%%%%%%%%%%%%%%%%%%%%%%%%%%%%%%%%%%%%%%%%
    \section{Introduction}
    %%%%%%%%%%%%%%%%%%%%%%%%%%%%%%%%%%%%%%%%%%%%%%%%%%%%%%%%%%%%%%%%%%%%%%%%%
    Over the past 15 years, machine learning (ML) has experienced exceptional
    growth, driven by remarkable progress in computational power and the pivotal
    introduction of GPU-accelerated training for deep neural networks as demonstrated with the breakthrough in image classification by \citet{krizhevsky2012imagenet}
    with AlexNet. Since then, ML applications have permeated a wide range of
    scientific and technological domains, among which numerical weather and climate
    modeling has emerged as a prominent example, enabled by the extension of computer vision techniques to the geosciences. \\
    Recent advancements have enabled the development of global, fully data-driven,
    ML-based weather prediction (MLWP) systems~\cite{pathak2022fourcastnet,
    lam2023learning-graphcast, bi2023accurate-pangu, chen2025operational-fengwu, bodnar2025foundation}.
    Trained primarily on ERA5 reanalysis data~\cite{hersbach2020era5}, these models
    have shown comparable and sometimes superior skills with respect to ECMWF’s deterministic
    operational forecasting system, IFS-HRES, on short-to-medium range forecasts~\cite{rasp2024weatherbench}.
    The main advantage of these models is their exceptional computational speed and efficiency, compared to classical physics-based models.
    However, they could remain limited by their lack of physical consistency and realism~\cite{Bonavita2024},
    which are crucial for meteorology in complex terrain, extreme event, coupled phenomena, as well as for seasonal and climate predictions. Additionally, the majority of them rely
    on high resolution physics-based models and data assimilation (DA) for their
    training and initialization. \\
    Another pragmatic approach, more conservative, is represented by hybrid modeling that combines ML with traditional physics-based models, in an attempt to extract the best of both while mitigating their individual weaknesses. This combination is often achieved
    through the replacement of physics-based or empirical parameterizations with
    data-driven ML components, with the aim of improving accuracy or reducing computational
    cost \cite{chevallier1998neural, ogorman2018, rasp2018deep, Bolton2019}.
    More recently, \citet{kochkov2024neural} introduced NeuralGCM, an end-to-end differentiable hybrid general atmospheric circulation model that combines a dynamical core with learned physics and achieves competitive performance for both weather forecasting and climate simulation.
    In the context of sea ice modeling, \citet{driscoll2024parameter} trained Neural Networks (NNs) to emulate an advanced physics-based melt pond parameterization within the ice column model Icepack~\cite{elizabeth_hunke_2025_16422921}, while \citet{Horvat2022} focused on wave-induced ice floe fracturing through a neural-network emulation of a computationally expensive super-parameterization.\\
    Despite continuous progress, numerical weather and climate model predictions inevitably
    remain affected by systematic errors, due to discretization errors,
    imperfect subgrid-scale parameterizations or inaccurate boundary conditions, initial conditions and forcings. 
    Therefore, instead of focusing on specific process parameterizations, another research branch on hybrid modeling targeted the model
    error, developing ML-based model error parameterizations that complement physically derived models during execution \cite{Watson2019}.
    Similar to MLWP models, these hybrid models for model error correction leverage DA that, along with the increasing volume and
    quality of observational data, plays a fundamental role in geoscience, by
    providing optimal estimates (\emph{analyses}) of the state of the earth system~\cite{Carrassi2018}.
    In particular, \emph{analysis increments} (differences
    between analysis and forecast states) are exploited to infer state-dependent systematic model
    errors~\cite{carrassi2011treatment}.
    Key contributions along these lines include the work of \citet{brajard2021combining} and \citet{Farchi2021},
    who trained NNs as additive integrated correction term between two
    forecast times, i.e. in the \emph{resolvent} of geophysical models,
    learning from analysis increments, with particular focus on the role of
    observation noise and sparseness. \citet{Farchi2021comparison} further investigated
    the potential of tendency correction and introduced an \emph{online
    learning} strategy, in which both the model states and the model parameter
    are estimated at the same time as soon as new observations become available.
    Similar approaches to hybrid modeling have also been applied to more realistic
    scenarios, such as the work of \citet{Watt-Meyer2021}, who trained a ML model
    to emulate nudging tendency terms within a weather forecast system; \citet{farchi2025} who learned a neural network correction to the Integrated Forecasting System of the ECMWF both offline and online; \citet{du2025reducing}, who developed ML-based corrections for ocean mixed-layer temperature biases in a global climate model; and \citet{chapman2025improving}, who improved multidecadal atmospheric predictions through online bias correction trained on nudging increments.
    This type of hybrid modeling has also been explored for sea ice.
    For example, \citet{Finn2023} employed convolutional U-nets to parametrize model error due to unresolved subgrid scale of sea-ice dynamics within an unstructured finite element model with a Maxwell elasto-brittle rheology.
    More recently, \citet{Gregory2024} successfully trained
    Convolutional Neural Networks (CNNs) to correct systematic errors of a
    global coupled ocean-sea ice model, demonstrating improvements achieved by iterating DA and ML steps~\cite{bocquet2020_fds}.
    Building on this work, \citet{gregory2026advancing} developed a hybrid global coupled ice–atmosphere–ocean climate model, highlighting the importance of fully coupled training for generalization in free forecasts.
    \citet{He2025} applied ML for online error correction within an earth system model for seasonal arctic sea ice prediction and compared it with an offline approach where ML post-processes and calibrates model predictions after the simulation.

    In the present study, we build upon previous works to develop state-dependent, ML-based resolvent corrections for the sea-ice column model, Icepack. Model errors are introduced by perturbing parameters related to snow thermodynamics and radiative properties, while synthetic observations are generated using a reference configuration. Neural networks are trained to predict systematic forecast errors, conditioned on the forecast model state. The online execution of the hybrid models, therefore, combines consecutive forecast steps with NN-based corrections, in a similar fashion as  sequential DA algorithms. This hybrid framework enables a detailed assessment of how specific parametric model errors influence forecasts and the corresponding ML bias corrections.
    Insights into the relative importance of predictors for estimating model error, and indications for a shared minimal model, are obtained through recursive feature elimination analyses applied across all perturbed configurations.\\
    An important, yet often overlooked, challenge in the operational deployment of
    model-error-correcting hybrid systems is the impact of model updates (i.e., when a new model enters into play, with e.g. new parameterizations, numerical schemes, etc.), which can
    cause abrupt shifts in error statistics. Unlike gradual changes induced by
    climate variability, which can be addressed with online learning strategies,
    model version updates require dedicated measures. The most straightforward approach is
    to generate a reanalysis with the updated model over a sufficiently long period
    and retrain the ML component. However, this is computationally demanding,
    and transfer learning strategies offer a promising alternative to reduce the
    amount of new data required for network adaptation.
    In this study, we conduct detailed novel transfer learning experiments, focusing on investigating the extent to which an ML model optimized for bias correction in one physical model can then be successfully adapted to another through fine-tuning. The effectiveness of this approach is evaluated as a function of the available new training dataset size and we propose a criterion to determine when fine-tuning is more advantageous than retraining from scratch. This may guide strategic choice on whether and when a new complete retraining may be avoided while keeping the model performance satisfactory.

    This article is structured as follows. Section~\ref{sec:icepack_intro} introduces the numerical physics-based model, including reference and perturbed configurations. Section~\ref{sec:hybrid_framework} illustrates the hybrid framework, covering training data generation, NNs training, and online execution of hybrid models. Neural networks validation and
    hybrid models testing are presented in Section~\ref{sec:NN-hyb_test}. Section~\ref{sec:transfer_learning} presents transfer learning experiments, and a feature importance study is provided in Section~\ref{sec:feat_imp}.
    Finally, summary and conclusions are given in Section~\ref{sec:conclusions}.

    %%%%%%%%%%%%%%%%%%%%%%%%%%%%%%%%%%%%%%%%%%%%%%%%%%%%%%%%%%%%%%%%%%%%%%%%%
    \section{The Icepack column physics model} \label{sec:icepack_intro}
    %%%%%%%%%%%%%%%%%%%%%%%%%%%%%%%%%%%%%%%%%%%%%%%%%%%%%%%%%%%%%%%%%%%%%%%%%

    %------------------------------------------------------------------------
    \subsection{Model description}
    %------------------------------------------------------------------------
    %
    Icepack is a submodule of the Los Alamos Sea Ice Model (CICE)~\cite{elizabeth_hunke_2024_11223920},
    designed to simulate all vertical processes within sea ice, such as thermodynamics
    and mechanical redistribution (ridging), and related changes in ice area and
    thickness. In addition to the column physics code, the Icepack module includes a
    driver code, which enables standalone testing of the physics code on individual,
    independent grid points. The present work is based on pointwise sea ice simulations
    using the Icepack standalone driver.
    To represent subgrid-scale thickness variability, Icepack uses the Ice Thickness
    Distribution (ITD) parameterization~\cite{thorndike1975thickness}, whose
    transport equation is solved using the remapping method of \citet{Lipscomb2001}.
    In this work, $C=5$ thickness categories are used, each discretized into
    seven vertical ice layers and a single snow layer. These categories represent bins of the underlying ITD. For each category, both ice concentration (area fraction) and ice volume are defined, with the set of concentration values providing a discrete approximation of the ITD.
    The model is driven by atmospheric forcing consisting of downward radiative fluxes,
    air temperature, relative humidity, total precipitations and wind speed from
    ERA5 reanalysis data~\cite{hersbach2020era5}.
    The coupled ocean is modeled using a mixed-layer model with deep ocean heat flux
    set to zero and constant salinity. All simulations in this study employ the
    level-ice melt pond parameterization~\cite{hunke2013level}, the mushy layer thermodynamics~\cite{turner2013two},
    and a 3-band Delta-Eddington radiative transfer scheme.
    
    %------------------------------------------------------------------------
    \subsection{Reference and perturbed configurations}
\begin{figure}
        \centering
        \includegraphics[width=0.5\linewidth]{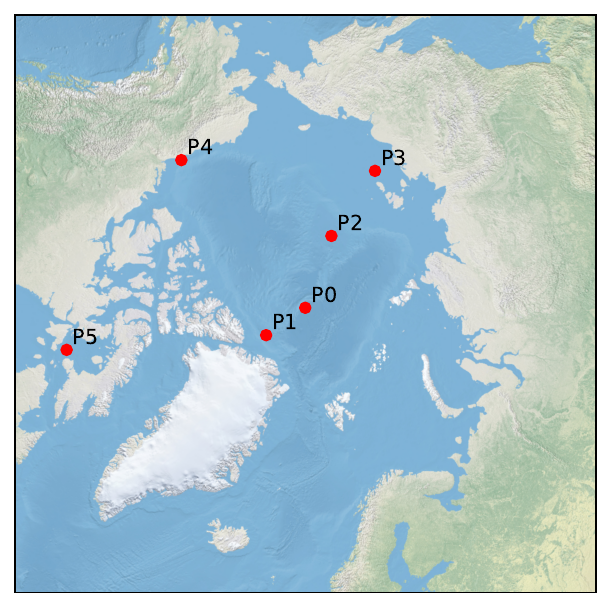}
        \caption{Locations of Icepack simulations, selected from the ERA5
        grid downsampled to 1\degree  resolution, at latitudes between 65\degree
        N and 90\degree N.}
        \label{fig:locations}
    \end{figure}

\begin{figure}
        \centering
        \begin{subfigure}
            {0.49\textwidth}
            \centering
            \captionsetup{position=above,justification=raggedright, skip=0pt, margin=25pt}
            \caption{P0: 90\degree N}
            \label{fig:long_forecasts_PC2a}
            \includegraphics[scale=0.55, trim={0 0.95cm 0 0.2cm}, clip]{
                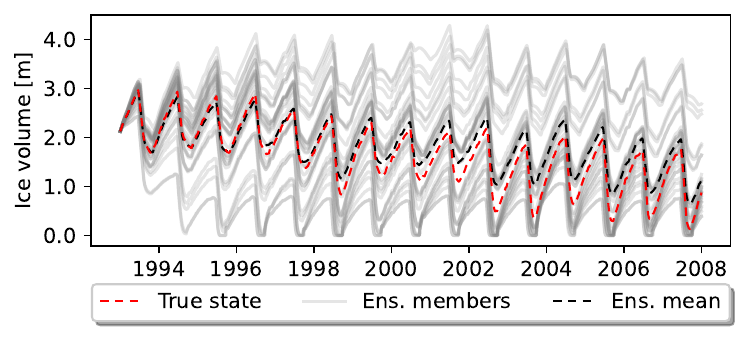
            }
        \end{subfigure}
        \begin{subfigure}
            {0.49\textwidth}
            \centering
            \captionsetup{position=above,justification=raggedright, skip=0pt, margin=25pt}
            \caption{P3: 74\degree N 153\degree E}
            \label{fig:long_forecasts_PC2b}
            \includegraphics[scale=0.55, trim={0 0.95cm 0 0.2cm}, clip]{
                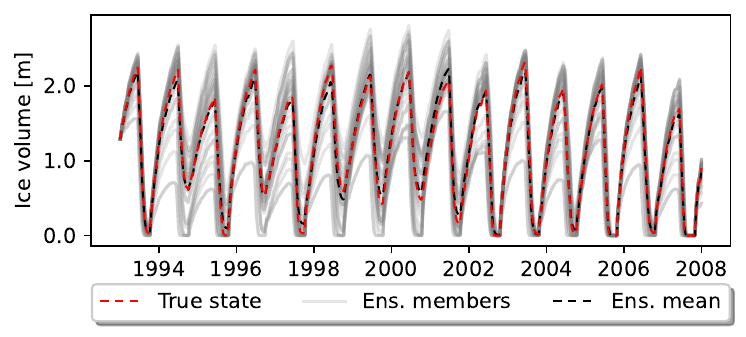
            }
        \end{subfigure}\\
        \vspace{0.1cm}
        \begin{subfigure}
            {0.49\textwidth}
            \centering
            \captionsetup{position=above,justification=raggedright, skip=0pt, margin=25pt}
            \caption{P1: 85\degree N 55\degree W}
            \label{fig:long_forecasts_PC2c}
            \includegraphics[scale=0.55, trim={0 0.95cm 0 0.2cm}, clip]{
                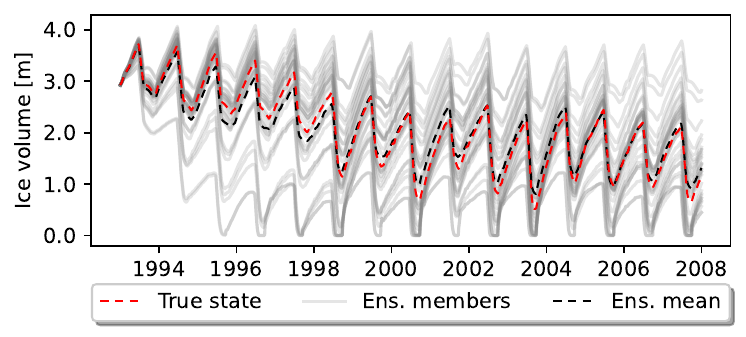
            }
        \end{subfigure}
        \begin{subfigure}
            {0.49\textwidth}
            \centering
            \captionsetup{position=above,justification=raggedright, skip=0pt, margin=25pt}
            \caption{P4: 70\degree N 140\degree W}
            \label{fig:long_forecasts_PC2d}
            \includegraphics[scale=0.55, trim={0 0.95cm 0 0.2cm}, clip]{
                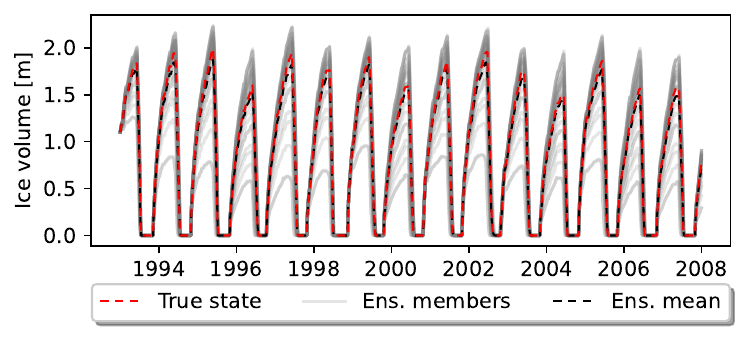
            }
        \end{subfigure}\\
        \vspace{0.1cm}
        \begin{subfigure}
            {0.49\textwidth}
            \centering
            \captionsetup{position=above,justification=raggedright, skip=0pt, margin=25pt}
            \caption{P2: 82\degree N 160\degree E}
            \label{fig:long_forecasts_PC2e}
            \includegraphics[scale=0.55, trim={0 0.95cm 0 0.2cm}, clip]{
                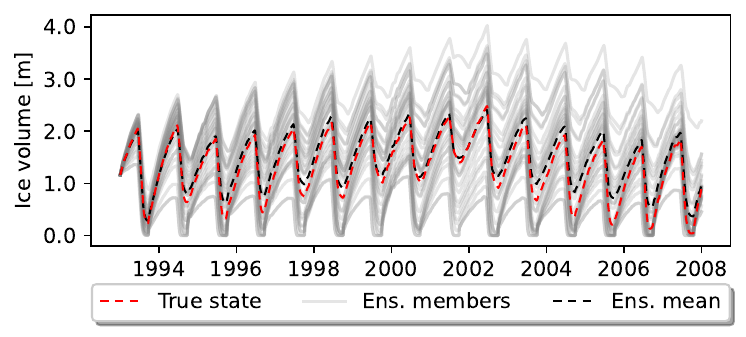
            }
        \end{subfigure}
        \begin{subfigure}
            {0.49\textwidth}
            \centering
            \captionsetup{position=above,justification=raggedright, skip=0pt, margin=25pt}
            \caption{P5: 65\degree N 80\degree W}
            \label{fig:long_forecasts_PC2f}
            \includegraphics[scale=0.55, trim={0 0.95cm 0 0.2cm}, clip]{
                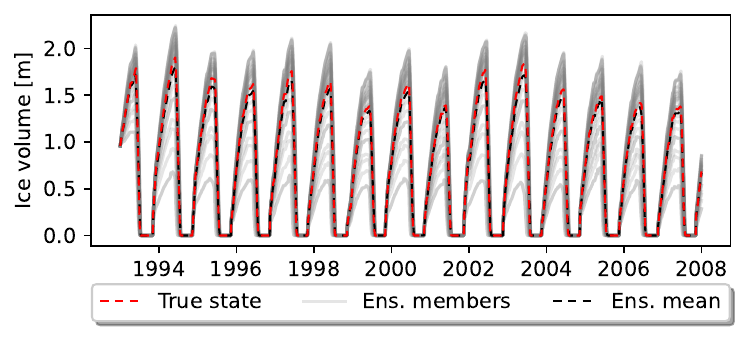
            }
        \end{subfigure}\\
        \vspace{0.1cm}
        \caption{Ice volume time-series over a 15-year period at the 6 selected
        locations: red dashed lines (\CR{\dashed}) indicate the truth; solid grey
        lines (\CGR{\solid}) show forecasts from individual models with perturbed parameters; black dashed lines (\dashed) denote the
        forecast ensemble means.}
        \label{fig:long_forecasts_PC2}
    \end{figure}

    %------------------------------------------------------------------------
    %
    We introduce model error by perturbing the
    reference model configuration. Specifically, we focus on errors arising
    from the misspecification of two key model parameters: (1) the thermal snow conductivity (\texttt{ksno}), which controls the heat flux through the snow layer and thus its insulating effect, and (2) the maximum melting
    snow grain size (\texttt{rsnw\_mlt}), a parameter of the delta-Eddington
    radiation scheme that influences snow albedo estimates. This choice was guided
    by the findings of~\citet{urrego2016uncertainty}, who demonstrated that
    model predictions are particularly sensitive to these two parameters. Additionally,
    the authors provided uncertainty estimates for model parameters using
    statistical distributions, as detailed in Table~\ref{tab:param_stats}. The
    reference configuration, used to produce trajectories assumed as true,
    utilizes the default values of these parameters~(cf.~Tab.~\ref{tab:param_stats}).
    \begin{table}[ht]
        \footnotesize
        \caption{Parameter descriptions and statistical distributions~\cite{urrego2016uncertainty}}
        \label{tab:param_stats}
        \begin{tabularx}
            {\textwidth}{c Y c c c c c} \textbf{Symbol} & \textbf{Description} &
            \textbf{Default value} & \textbf{Distribution} & \textbf{Min value} &
            \textbf{Max value} & \textbf{Mode} \\
            \midrule \texttt{ksno} & Thermal snow conductivity (W m\textsuperscript{-1}K\textsuperscript{-1})
            & 0.3 & Uniform & 0.03 & 0.65 & \\
            \texttt{rsnw\_mlt} & Max. melting snow grain size (\textmu m) & 1500
            & Triangular & 250 & 3000 & 1500 \\
            \bottomrule
        \end{tabularx}
    \end{table}
    To explore the perturbed parameters subspace, an ensemble of $M=30$ independent model
    configurations is generated by sampling each parameter from its respective
    distribution. Each model parameter configuration is uniquely identified by its
    member ID (\texttt{m\_01, m\_02, $\dots$, m\_30}), which is used hereafter. The amplitude of the perturbations
    and the degree of similarity across configurations are examined in Section~\ref{sec:param_err} of the Supporting Information.
    Simulations are run in $L=6$ locations (Figure~\ref{fig:locations}), selected as representative of the diverse sea-ice conditions observed across the Arctic.
    The reference configuration is firstly run for 30 years, applying cyclically
    the first year (1993) atmospheric forcings (Figure~\ref{fig:long_runs}).
    This allows us to obtain a stable sea ice annual cycle, which provides initial
    conditions for the 15-year reference runs (Figure~\ref{fig:long_forecasts}).
    Prediction errors over long lead times due to parameter perturbations are
    evaluated by running the 30 perturbed models, initialized with true states,
    for 15 years. Resulting ice volume forecasts are presented in Figure~\ref{fig:long_forecasts_PC2}.
    It can be noted that at higher latitudes (left panels), ice volume errors
    pile up over years, reaching values up to 2 meters. On the other hand, at lower latitudes (right panels), ice volume errors do not accumulate over years as
    the sea ice is completely melted every summer. Nonetheless, short-term errors in these regions cannot be neglected, especially in 2D configurations where advection may propagate them to other areas. The effect of climate change driven by atmospheric
    forcing is also evident in Figures~\ref{fig:long_forecasts_PC2a} and
    \ref{fig:long_forecasts_PC2c}, which show a gradual decline in ice volume over
    the years.
    As we shall see, this non-stationarity challenges the extrapolation
    capabilities of the ML algorithm for
    forecast error correction.
    %

    %%%%%%%%%%%%%%%%%%%%%%%%%%%%%%%%%%%%%%%%%%%%%%%%%%%%%%%%%%%%%%%%%%%%%%%%%
    \section{Hybrid framework} \label{sec:hybrid_framework}
    %%%%%%%%%%%%%%%%%%%%%%%%%%%%%%%%%%%%%%%%%%%%%%%%%%%%%%%%%%%%%%%%%%%%%%%%%

    The hybrid model developed in this work combines the physics-based Icepack model with an additive ML component that acts as a correction to the model resolvent. In the framework considered here, the ML component is trained on forecast errors computed with respect to reference true states at a prescribed lead time, and is then used to sequentially correct the model state. This corresponds to an idealized case of a more general approach in which DA analysis increments are used as training targets. The rationale for such an approach is that, when the analyses are accurate enough to serve as proxies for the unknown true states (e.g., negligible analysis error), the corresponding analysis increments largely reflect forecast errors arising from model deficiencies~\cite{carrassi2011treatment}.
    Under these assumptions, analysis increments provide suitable training targets for ML models that estimate state-dependent model error conditioned on forecast states~\cite{bocquet2020_fds,brajard2021combining, Farchi2021, Farchi2021comparison, bocquet2023surrogate,Finn2023, farchi2025}.
    The resulting hybrid approach is, therefore, reminiscent of a sequential DA cycle, with the ML corrections in place of analysis increments, as detailed in Section~\ref{ssec:hybrid_model}.
    A related, but distinct, approach is proposed by \citet{peng2024hybrid} who use deep learning within a DA framework for parameter-error estimation of a chaotic toy model, also in the presence of initial-condition errors.
    
    In this study, training data are generated in an idealized setting with perfect, complete observations: under this assumption, DA is bypassed, and forecast errors relative to the reference true state are used directly as training targets, thereby eliminating the effect of initial-condition error.
    This intentional simplification allows us to focus on ML design and performance and to study in detail the strength and limitation of the proposed approach without the confounding effect derived from the presence of initial condition error and its interplay with the model error. The latter issue is addressed in the online hybrid model experiments~(Section~\ref{sec:online_testing}).

    %------------------------------------------------------------------------
    \subsection{Training data generation} \label{sssec:dataset_gen}
    %------------------------------------------------------------------------
    %
    Given the assumption of perfect and complete observations, for each location and model configuration, incorrect forecast trajectories
    $\V{x}$ are produced by initializing models from "true" reference full states
    $\V{x}^{\textrm{t}}$ at $S^{\textrm{tr}}\!=522$ dates, $\{t_{s}\}$, spanning the period from 1 January 1994 to 31 December 2003, with a weekly stride $\tau_{w} = 7$ days:
    \begin{subequations}
        \begin{gather}
            \V{x}_{t_s + \tau}= \mathcal{M}_{t_s : t_s + \tau}\left( \V{x}_{t_s}^{\textrm{t}}, \V{F}_{t_s : t_s + \tau}\right), \\
            t_{s} = t_{0} + (s-1) \tau_{w}, \quad s=1, \dots, S^{\textrm{tr}}
        \end{gather}
    \end{subequations}
    where $\mathcal{M}_{t_s : t_s + \tau}$ denotes the resolvent of the physics-based model from time
    $t_{s}$ to $t_{s} + \tau$, and $\V{F}_{t_s : t_s + \tau}$ the corresponding atmospheric forcings over the forecast window.
    
    Thus, $\V{x}_{t_s+\tau}$ represents the forecast state at lead time $\tau$, initialized from the true state at time $t_s$. The corresponding forecast error $\V{e}_{s, \tau}$ at lead time $\tau$ is defined as the difference between the forecast state and the true state at the same time:
    \(
        \V{e}_{s, \tau} = \V{x}_{t_s + \tau}- \V{x}^{\textrm{t}}_{t_s + \tau}.
    \)
    
    Each simulation is run up to a maximum lead time
    $\tau = 180$ days, with outputs stored every 5 days. 
    \\
    The analysis of forecast error statistics as a function of lead time and initialization date~(Section~\ref{sec:lead-sdata_errors}, Supporting Information) led to selecting a fixed training lead time of $\tau_{\textrm{tr}}=60$ days, which provides a suitable balance between forecast degradation and signal strength.
    In more realistic settings, the forecast is expected to degrade faster, leading to a shorter training lead time. In general, a suitable balance should exist for any reasonable forecast model.
    Machine learning datasets are then constructed by
    pairing features with targets at the selected lead time $\tau_{\textrm{tr}}$. The targets are the forecast
    errors,
    \(
        \V{e}_{s} = \V{x}_{t_s + \tau_{\textrm{tr}}}- \V{x}^{\textrm{t}}_{t_s + \tau_{\textrm{tr}}},
    \)
    while the feature vector,
    \(
        \V{z}_{s} = \mathcal{G}\!\left(\V{x}_{t_s+\tau_{\textrm{tr}}}, \; \V{F}_{t_s+\tau_{\textrm{tr}}}\right
        ),
    \)
    comprises model state variables, atmospheric
    forcings and diagnostics at time $t_{s} + \tau_{\textrm{tr}}$,
    where $\mathcal{G}$ selects and/or derives the predictors used as ML inputs.
    The resulting supervised learning dataset, combining data from all locations,
    for a single perturbed configuration is
    \(
        \mathcal{D}= \big\{ \left( \V{z}_{i}, \V{e}_{i} \right) \big\}_{i=1}^{N},
    \)
    and counts $N=L\times S^{\textrm{tr}}=3132$ instances.
    %

    %------------------------------------------------------------------------
    \subsection{Machine learning regression task} \label{ssec:machlearn_model}
    %------------------------------------------------------------------------
    %
    The objective of the ML model is to approximate the mapping between input features $\V{z}$ and the forecast error $\V{e}$ through a NN, $g(\V{\theta}
    ,\V{z})$, where $\V{\theta}$ denotes the network trainable parameters. In the following, we denote by $\hat{\V{\theta}}$ the trained parameters obtained after optimization. In this study, we focus on predicting the component-wise forecast errors in ice concentration, $\V{a}=(a_1, \dots, a_C)$, and ice volume, $\V{v}=(v_1, \dots, v_C)$. Two independent NNs, for ice concentration and ice volume, are implemented to prevent cross-task interference and ensure stable learning. Each network is trained by minimizing the following cost function,
    \begin{equation}
        \loss(\V{\theta}^{\varphi}) = 
        \overline{\left( g(\V{\theta}^{\varphi}, \V{z}_{i}) - \V{e}^{\varphi}_{i} \right)^{2}}, 
        \qquad \varphi \in \{\V{a}, \V{v}\}
    \end{equation}
    where $\overline{(\cdot)}$ denotes the average over the dataset instances and the five ice categories, and $\V{e}^{\varphi}_{i}$ represents the forecast error, with
    $\varphi$ indicating either ice concentration or ice
    volume. The NN functional form $g$ (i.e., its architecture) is identical for both ice concentration and volume; the distinction arises solely from the respective trained weights $\V{\theta}^{\V{a}}$ and $\V{\theta}^{\V{v}}$. The inputs and targets are normalized by per-variable means and standard deviations, estimated from the training dataset. Table~\ref{tab:feat_targ}
    summarizes the target and features variables used. \\
    At inference time, the combined NN error prediction is given by
    \begin{equation}
        g \! \left(\hat{\V{\theta}},\, \V{z}\right) = \left[ g\! \left(\hat{\V{\theta}}^{\V{a}},\, \V{z}\right), g \!\left(\hat{\V{\theta}}^{\V{v}},\, \V{z}\right)\right]
    \end{equation}
    where $g\! \left(\hat{\V{\theta}}^{\V{a}},\, \V{z}\right)$ and $g \!\left(\hat{\V{\theta}}^{\V{v}},\, \V{z}\right)$ denote the trained networks’ predictions for ice concentration and ice volume, respectively.

    \begin{table}[ht]
    \centering
    \normalsize
    \caption{Features and targets of NN model. Variables with dimension 5 are defined over the ITD categories used by default in Icepack.}
    \label{tab:feat_targ}
    \begin{NiceTabularX}{0.7\textwidth}{>{\columncolor{black!10}}c|cYc}[code-before = \rowcolors{1}{black!10}{}]
    \textbf{} & \textbf{Symbol} & \textbf{Description} & \textbf{Dimension} \\
    \midrule
    
    \Block{17-1}{\textbf{Features}} & \texttt{frain}   & Precipitation              & 1 \\
                                   & \texttt{fswabs}  & Absorbed short wave        & 1 \\
                                   & \texttt{flw}     & Longwave radiation         & 1 \\
                                   & \texttt{flwout}  & Outgoing longwave          & 1 \\
                                   & \texttt{fsnow}   & Snowfall                   & 1 \\
                                   & \texttt{Tair}    & 2m air temperature         & 1 \\
                                   & \texttt{aicen}   & Ice concentrations         & 5 \\
                                   & \texttt{vicen}   & Ice volumes                & 5 \\
                                   & \texttt{vsnon}   & Snow volumes               & 5 \\
                                   & \texttt{Tsfcn}   & Ice surface temperature    & 5 \\
                                   & \texttt{qice001} & Ice enthalpy - layer 1     & 5 \\
                                   & \texttt{qice002} & Ice enthalpy - layer 2     & 5 \\
                                   & \texttt{qice003} & Ice enthalpy - layer 3     & 5 \\
                                   & \texttt{qice004} & Ice enthalpy - layer 4     & 5 \\
                                   & \texttt{qice005} & Ice enthalpy - layer 5     & 5 \\
                                   & \texttt{qice006} & Ice enthalpy - layer 6     & 5 \\
                                   & \texttt{qice007} & Ice enthalpy - layer 7     & 5 \\
    \midrule
    
    \Block{1-1}{\textbf{Targets}} & \texttt{err\_a} (\texttt{err\_v})
    & Ice concentration (volume) errors & 5 \\
    \bottomrule
    \end{NiceTabularX}
    
    \end{table}

    %------------------------------------------------------------------------
    \paragraph{Neural network architecture}
    %------------------------------------------------------------------------
    %
    As a column model, Icepack is horizontally pointwise, involving only the
    vertical and categorical dimensions in addition to time. Based on this structure, Multilayer Perceptrons (MLPs) are trained, using LeakyReLU activation
    functions in the hidden layers and linear activations in the output layer. Depth and width of the hidden layers, as well as the learning rate and batch size
    are selected through hyperparameter optimization using Asynchronous Successive Halving (ASHA)~\cite{li2020asha}.
    The search space includes MLP architectures having from one to three hidden layers, with hidden-layer widths between $10$ and $60$ neurons, learning rates sampled on a logarithmic scale in the range $10^{-4}$ to $10^{-1}$, and batch sizes drawn from a small set of standard values.
    The resulting configuration consists of two hidden layers with 20 neurons each, trained
    using Adam optimization~\cite{kingma2017adammethodstochasticoptimization}, with
    learning rate $\gamma = 8 \times 10^{-4}$, weight decay $\lambda = 10^{-4}$
    and batch size $N_{b} = 32$.
    %

    %------------------------------------------------------------------------
    \subsection{Icepack–NN hybrid model} \label{ssec:hybrid_model}
\begin{figure}
        \centering
        \includegraphics[width=\linewidth]{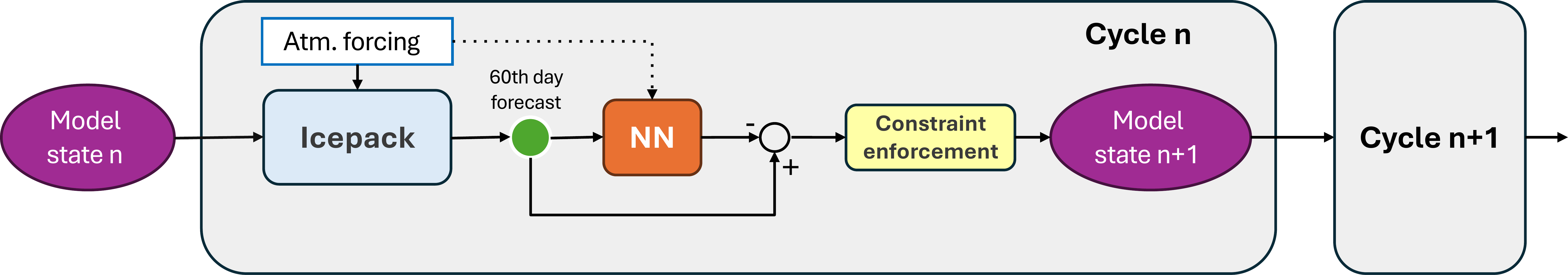}
        \caption{Schematic of the hybrid model loop.}
        \label{fig:hybrid_loop}
    \end{figure}

    %------------------------------------------------------------------------
    %
    The hybrid model we develop is designed to iteratively correct forecast
    errors during model execution. Among all the Icepack state variables, we predict
    and correct errors for the most relevant: the ice concentration
    and the ice volume for the five ice thickness categories. Figure~\ref{fig:hybrid_loop}
    illustrates a single hybrid model iteration. It consists of a forecast step,
    \begin{subequations}
        \begin{gather}
            \V{x}_{t_{k+1}}= \mathcal{M}_{t_k:t_{k+1}}\!\left(\V{x}^{\textrm{hyb}}_{t_k}, \V{F}_{t_k:t_{k+1}}\right), \\
            t_{k+1} - t_k = \tau_{\textrm{tr}},
        \end{gather}
    \end{subequations}
    where, starting from initial conditions $\V{x}^{\textrm{hyb}}_{t_k}$, the model is integrated
    forward under atmospheric forcings $\V{F}_{t_k:t_{k+1}}$, to produce the uncorrected forecast state $\V
    {x}_{t_{k+1}}$. The second step involves the forecast error correction:
    \begin{equation}
        \tilde{\V{x}}^{\textrm{hyb}}_{t_{k+1}}= \V{x}_{t_{k+1}}- g \!\left(\hat{\V{\theta}},\, \V{z}
        _{t_{k+1}}\right),
    \end{equation}
    where NN error predictions $g \! \left(\hat{\V{\theta}},\, \V{z}_{t_{k+1}}
    \right)$ are subtracted
    from the forecast state. This produces an intermediate hybrid model state
    $\tilde{\V{x}}^{\textrm{hyb}}_{t_{k+1}}$, that is adjusted in a final step:
    \begin{equation}
        \V{x}^{\textrm{hyb}}_{t_{k+1}}= \Pi^{\textrm{con}}\!\left(\tilde{\V{x}}^{\textrm{hyb}}_{t_{k+1}}\right),
    \end{equation}
    where a post-processing, $\Pi^{\textrm{con}}$, is applied to obtain the final hybrid model
    state $\V{x}^{\textrm{hyb}}_{t_{k+1}}$. The post-processing step, corresponding to the
    box \textit{Constraint enforcement} in Figure~\ref{fig:hybrid_loop}, is necessary
    to enforce specific constraints, such as non-negativity of ice concentration
    and volume, as well as the requirement that ice concentrations must not exceed one.
    All adjustments introduced to ensure physical and model consistency are
    detailed in paragraph below, \emph{Treatment of bounded variables}. The model state obtained after post processing
    represents the initial condition for the subsequent iteration.
    %

    %------------------------------------------------------------------------
    \paragraph{Treatment of bounded variables}
    %------------------------------------------------------------------------
    %
    Ice volumes and concentrations, their total values as well as those associated with each thickness
    category, are bounded quantities:
    \(
        v_{c} \in \left[0, + \infty\right), \; a_{c} \in \left[0, 1 \right], \; \sum_c
        a_{c} \in \left[0, 1 \right].
    \)
    Recall that $c=1,\dots,5$ indicates the ice category. When applying ML-based corrections, these bounds are
    not guaranteed to be respected. Violations typically occur when both the forecast concentration or volume and the true error are close to zero; in these cases, minor inaccuracies in the NN correction can produce slightly negative values. Although these violations are generally of negligible magnitude,
    it is essential for the hybrid model that all prognostic variables remain
    within their physical bounds to ensure successful numerical integration. The motivation
    for enforcing bounds differs here from what motivates their implementation in, e.g. ML surrogates for the sea ice~\cite{finn2024generative,durand2025}, whereby the autoregressive use of the NNs obliges their adoption for stabilizing the time evolution. In our case, NN inputs are always
    provided by the physical model, so the possible instability of the NN is not a primary concern.
    Instead, our objective is to guarantee physical model consistency and suitable initial conditions for the restart:
    without proper enforcement of bounds, Icepack may fail to restart or diverge
    rapidly. 
    Physical consistency (ensuring agreement between volume and concentration, their simultaneous vanishing, and adherence to thickness bounds) can be achieved either by enforcing constraints during training or by applying corrections after the NN outputs. While the latter approach may introduce systematic biases, the former would require a bivariate NN and constrained optimization strategies, adding substantial complexity for limited benefit. For these reasons, and since constrained optimization is not the focus here, constraints are enforced as a post-processing step on the corrected states. The procedure is detailed in the pseudo-code format in Algorithm~\ref{alg:postproc}, where $h_{c}^{\min}$ and $h_{c}^{\max}$ denote the prescribed minimum and maximum
    ice thickness for category $c$, $\mathbf{1}_{\{\cdot\}}$ denotes the indicator function, and a tolerance of
    $\epsilon = 10^{-2}\mathrm{m}$ is introduced to prevent assigning thickness
    values exactly at the bounds.
    \begin{algorithm}[h!]
    \centering
    \caption{Post-processing of category variables $a_c$ and $v_c$}
    \label{alg:postproc}
    \begin{algorithmic}[1]
    
    \For{$c = 1, \dots, C$} \Comment{Enforce non-negativity and ensure consistency between $\V{a}$ and $\V{v}$}
        \State $a_c \gets \max(0,\, a_c)$
        \State $v_c \gets \max(0,\, v_c)$
        \State $a_c \gets a_c \cdot \mathbf{1}_{\{v_c > 0\}}$
        \State $v_c \gets v_c \cdot \mathbf{1}_{\{a_c > 0\}}$
    \EndFor
    
    \State
    \If{$\sum_c a_c > 1$} \label{code:renorm_a} \Comment{Renormalize concentrations}
        \State $\V{a} \gets \V{a} / \sum_c a_c$
    \EndIf
    
    \State
    \For{$c = 1, \dots, C$} \label{code:thick_bounds} \Comment{Enforce ice thickness bounds}
        \State $h_c \gets v_c / a_c$
        \If{$h_c \geq h_c^{\max}$}
            \State $v_c \gets a_c \cdot (h_c^{\max} - \epsilon)$
        \ElsIf{$h_c < h_c^{\min}$}
            \State $a_c \gets v_c / (h_c^{\min} + \epsilon)$
        \EndIf
    \EndFor
    
    \end{algorithmic}
    \end{algorithm}
    Note that the enforcement of thickness bounds preserves the validity of the total concentration constraint.
    Finally, additional consistency adjustments are applied to related state variables. Specifically, 
    snow volume and enthalpy are set to zero whenever ice is entirely removed
    from a given category, while in previously ice-free categories that gain ice,
    the salinity and enthalpy of the new ice layers are initialized using
    Icepack’s default values.
    %

    %------------------------------------------------------------------------
    \section{Neural Network and Hybrid Model Evaluation} \label{sec:NN-hyb_test}
    %------------------------------------------------------------------------
    %
    The development of the NN-based hybrid Icepack models, hereafter referred to as \mbox{Icepack-NN},
    involves an initial analysis of the data, followed by NNs
    training, offline testing, and ultimately, hybrid model validation and performance assessment.
    Performance is evaluated relative to both the uncorrected
    Icepack model and a benchmark hybrid configuration, termed \mbox{Icepack-wclim}, which
    replaces NN predictions with location-independent weekly climatological corrections. Separate NN-based and climatological correction schemes are trained for each Icepack model configuration.
    The error climatologies are derived from the same training data as the NNs and averaged across locations, ensuring a fair comparison to the location-independent NN predictions.
    Note that the term \emph{weekly} refers to the
    averaging time window; the averaged data are still forecast errors at
    a lead time of 60 days. For the preliminary data analysis and performance
    evaluation purposes, we consider the root mean square error $\text{RMSE}^{m}$ and
    bias $\text{BIAS}^{m}$ for each model configuration $m$, as well as the model ensemble
    RMSE, $\text{RMSE}^{\textrm{ens}}$, all computed as function of lead time $\tau$:
    \begin{align}
        \text{RMSE}^{m}(\tau)   & = \sqrt{\frac{1}{N}\sum_{c, l, s}\left( \varphi_{c}^{m(s,l)}(\tau) - \varphi_{c}^{\textrm{t}(l)}(t_s + \tau)\right)^{2}}, \label{eq:rmse_met} \\
        \text{BIAS}^{m}(\tau)   & = \frac{1}{N}\sum_{c, l, s}\left( \varphi_{c}^{m(s,l)}(\tau) - \varphi_{c}^{\textrm{t}(l)}(t_s + \tau)\right), \label{eq:bias_met}             \\
        \text{RMSE}^{\textrm{ens}}(\tau) & = \sqrt{\frac{1}{M}\sum_{m} \text{RMSE}^{m}(\tau)^{2}}, \label{eqn:ens_rmse}
    \end{align}
    where $\varphi_{c}^{m(s,l)}(\tau)$ represents the forecast ice
    volume or concentration, relative to model configuration $m$, ice category $c$, location $l$ and initialized at start date $t_s$;
    $\varphi_{c}^{\textrm{t}(l)}(t_s + \tau)$ denotes the
    corresponding true value. Summation indices $c$, $l$, and $s$ span the $C$ ice categories, $L$ locations, and $S$ simulations (each initialized at a different start date), yielding $N$ instances, whereas in Eq.~\eqref{eqn:ens_rmse} the index $m$ ranges over the $M$ model configurations.
    In addition, RMSE metrics are evaluated under alternative averaging
    strategies by restricting the dataset to specific subsets (e.g., forecast month,
    start month, or geographical locations). To quantify the RMSE reduction achieved by
    the corrected Icepack models (\mbox{Icepack-NN} or \mbox{Icepack-wclim}), across the model configurations, relative to the uncorrected baseline, the ensemble-averaged normalized RMSE, $\langle \text{nRMSE} \rangle$, is considered,
    \begin{subequations}
        \label{eqn:nrmse_leadtime}
        \begin{gather}
            \langle \text{nRMSE} \rangle = \frac{\sum_{m} \text{RMSE}^{\textrm{u},m}\cdot \text{RMSE}^{\textrm{hyb},m}}{\sum_{m}
            \left[\text{RMSE}^{\textrm{u},m}\right]^{2}}= \sum_{m} w^{m} \cdot \text{nRMSE}^{m}, \\
            w^{m} = \frac{\left[\text{RMSE}^{\textrm{u},m}\right]^{2}}{\sum_{m} \left[\text{RMSE}^{\textrm{u},m}\right]^{2}}
            , \\
            \text{nRMSE}^{m}= \frac{\text{RMSE}^{\textrm{hyb},m}}{\text{RMSE}^{\textrm{u},m}},
        \end{gather}
    \end{subequations}
    where the superscripts $\textrm{u}$ and $\textrm{hyb}$ denote, respectively, the uncorrected
    Icepack and a corrected Icepack model (either \mbox{Icepack-NN} or \mbox{Icepack-wclim}). For readability,
    the explicit dependence on the lead time $\tau$ has been omitted.
    %

    %------------------------------------------------------------------------
    \subsection{Preliminary data analysis} \label{sec:data_analysis}
    %------------------------------------------------------------------------
    %
    %
    % \begin{landscape}
    %     \begin{figure}
        \begin{sidewaysfigure}
            \centering
            \begin{minipage}[t]{0.94\linewidth}
                \includegraphics[width=\linewidth]{
                    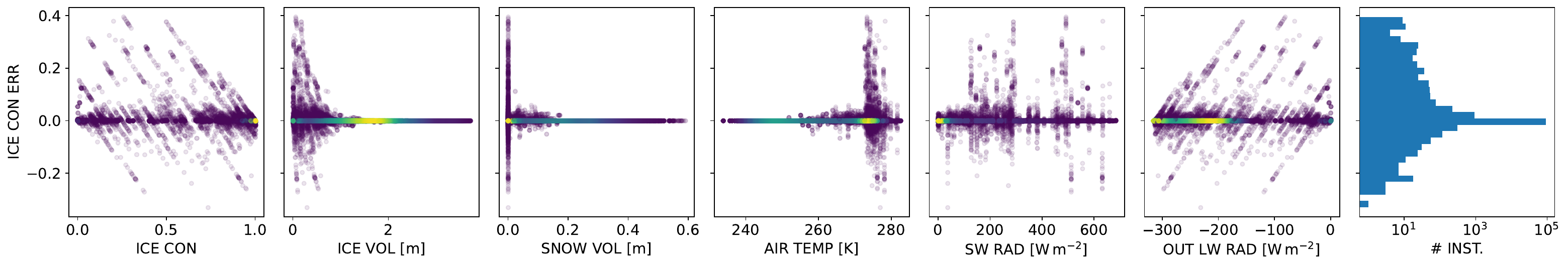
                }
            \end{minipage}
            \hfill
            \begin{minipage}[t]{0.05\textwidth} % Adjust width for the text box
                \raisebox{2cm}{(a)}
            \end{minipage}\\
            \begin{minipage}[t]{0.94\linewidth}
                \includegraphics[width=\linewidth]{
                    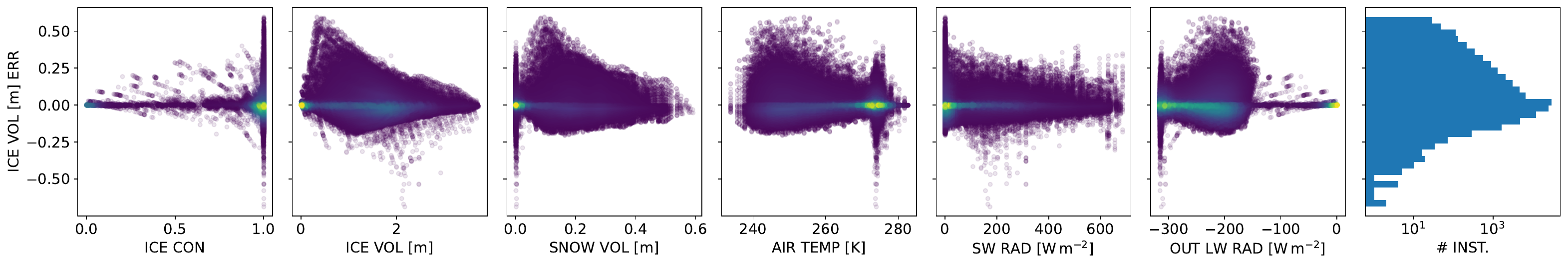
                }
            \end{minipage}
            \hfill
            \begin{minipage}[t]{0.05\textwidth} % Adjust width for the text box
                \raisebox{2cm}{(b)}
            \end{minipage}\\
            \begin{minipage}[t]{0.94\linewidth}
                \includegraphics[width=\linewidth]{
                    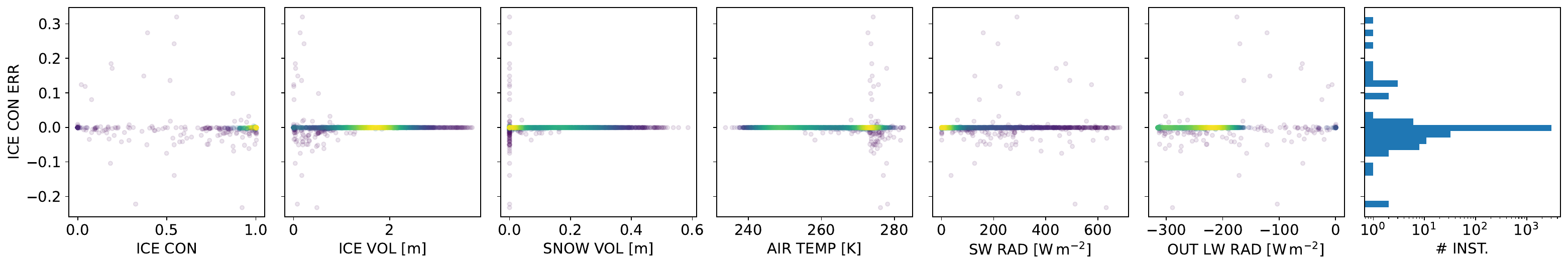
                }
            \end{minipage}
            \hfill
            \begin{minipage}[t]{0.05\textwidth} % Adjust width for the text box
                \raisebox{2cm}{(c)}
            \end{minipage}\\
            \begin{minipage}[t]{0.94\linewidth}
                \includegraphics[width=\linewidth]{
                    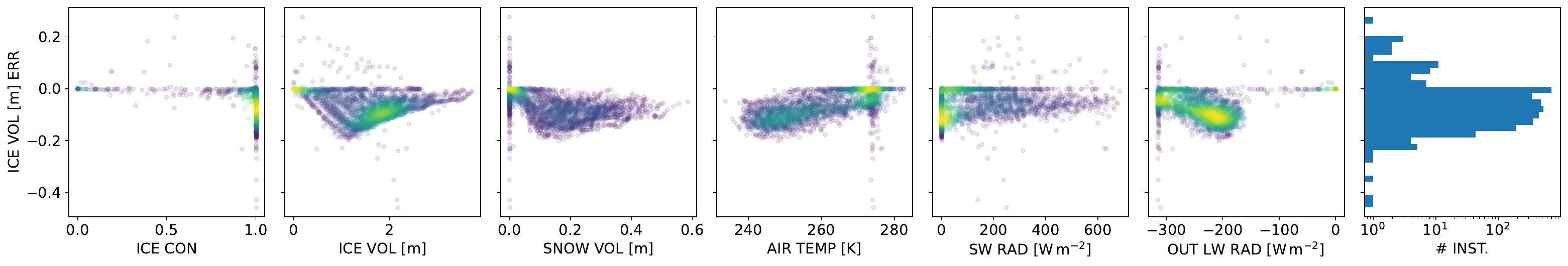
                }
            \end{minipage}
            \hfill
            \begin{minipage}[t]{0.05\textwidth} % Adjust width for the text box
                \raisebox{2cm}{(d)}
            \end{minipage}\\
            \caption{Scatter plots of total ice concentration and volume errors
            at lead time $\tau = 60$ days, as functions of selected forecast state
            variables and atmospheric forcings.
            Point colors represent density estimated via Gaussian KDE.
            Rows (a) and (b) show data
            aggregated across all model configurations, whereas panels (c) and (d)
            show results from model $\mathrm{m}\_05$ only. }
            \label{fig:scatter_inout_PC2}
            \end{sidewaysfigure}
\begin{figure}[h]
        \centering
        \begin{subfigure}
            {0.49\textwidth}            \captionsetup{position=above,justification=raggedright, skip=-1pt, margin=15pt}
            \caption{Ice volume}
            \includegraphics[scale=0.57, trim={0cm 0.3cm 0 0.0cm}, clip]{
                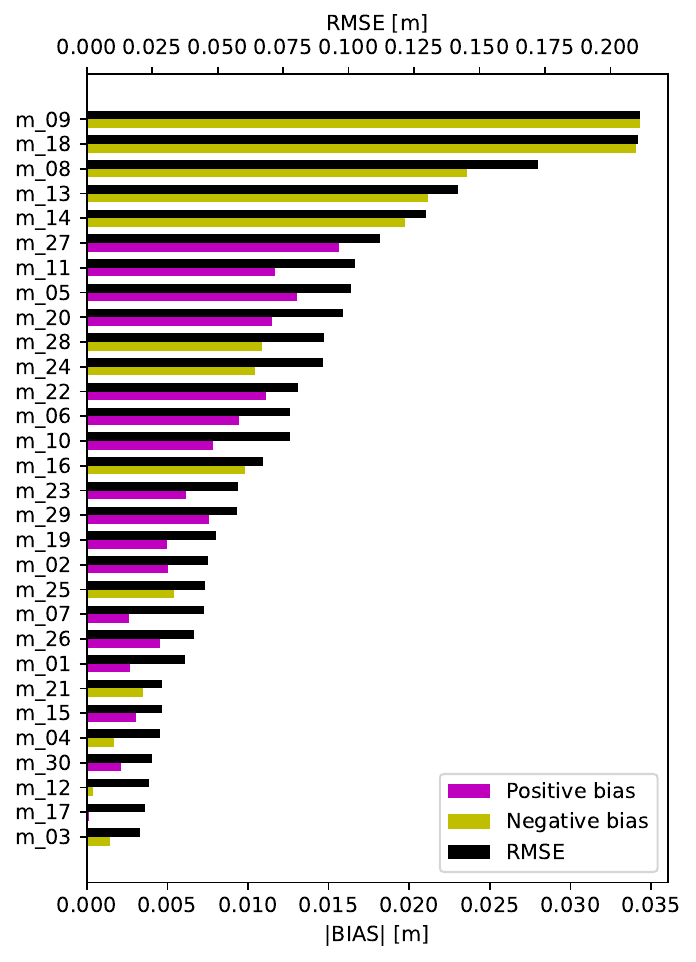
            }
        \end{subfigure}
        \begin{subfigure}
            {0.49\textwidth}            \captionsetup{position=above,justification=raggedright, skip=-1pt, margin=15pt}
            \caption{Ice concentration}
            \includegraphics[scale=0.57, trim={0cm 0.3cm 0 0.0cm}, clip]{
                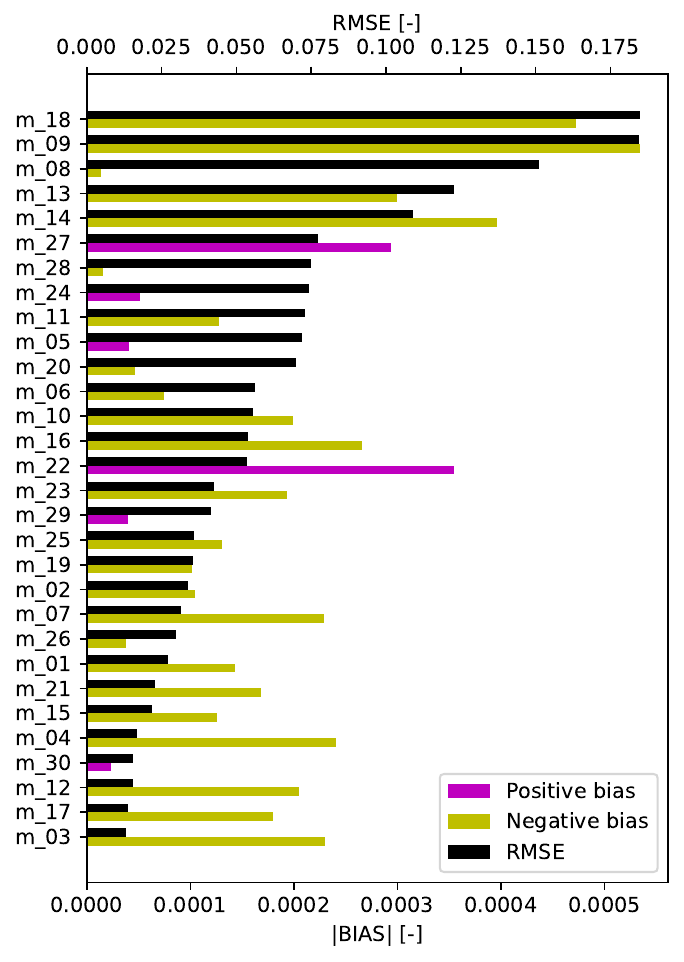
            }
        \end{subfigure}
        \caption{Ice volume (a) and concentration (b) RMSE and BIAS at a lead
        time of 60 days for all model configurations sorted by the RMSE. }
        \label{fig:rmse-bias_lt60data}
    \end{figure}

    %     \end{figure}
    % \end{landscape}
    %
    Figure~\ref{fig:scatter_inout_PC2} displays marginal scatter plots extracted from the underlying multivariate distributions relating targets and features, aimed at illustrating the nature of their relationships (e.g. linear, nonlinear, and of which sort). In particular, it shows scatter plots of the errors in ice concentration
    (rows a and c) and ice volume (rows b and d), cumulated over ice categories, against
    a selection of forecast model variables and atmospheric forcings. In rows (a) and
    (b), all data are put together, combining locations, model configurations, and start dates, whereas rows (c) and (d) show the data from a single, arbitrarily chosen, yet representative, configuration (\texttt{m\_05}).
    When data from all perturbed models are aggregated, features and targets appear largely uncorrelated, suggesting that training a neural network to correct forecast errors irrespective of the specific model configuration is not feasible.
    In contrast, when data associated to a single model configuration are selected, nonlinear
    relationships between features and ice volume errors become apparent~(cf. Figure~\ref{fig:scatter_inout_PC2}d), indicating that configuration-specific training is more appropriate and consistent with the practical setting in which training data are typically available for a single forecast model.
    This behavior is not observed for total ice concentration errors~(Figure~\ref{fig:scatter_inout_PC2}c), which are zero in most instances.
    This apparent lack of variability in total concentration errors stems from the bounded nature of ice concentration: during winters, when the true total concentration reaches 1, the forecast typically does so too, either earlier or later, causing the aggregate error across categories to approach zero.
    Consequently, these plots may give a misleading impression that concentration errors contain little useful information. In practice, however, our NN is trained on category-wise concentration errors, which evolve over time, even when the total concentration is saturated. Hence, the issue seen in the aggregated plots is expected to be far less pronounced in the actual learning setup. 
    Figure~\ref{fig:bins_error_maps} confirms that forecast errors at the category level remain non-negligible even when the total forecast ice concentration is equal to 1 and the total error is very small or almost zero. Conversely, when the total concentration approaches zero, the errors become negligible across categories. We nevertheless opted for training our NN in all regimes, including situations with zero forecast ice concentration (and volume); they may still exhibit non-negligible errors, as due for instance to delays in the onset of refreezing.
    \\
    To characterize
    the average behavior of each model configuration, we compute the RMSE
    and BIAS of ice volume and concentration errors from the respective datasets (Figure~\ref{fig:rmse-bias_lt60data}).
    An approximately linear relationship can be observed between ice volume RMSE
    and BIAS amplitude (Figure~\ref{fig:rmse-bias_lt60data}a), with the proportion of positively and negatively biased model configurations being nearly equal. On the other hand, ice concentration errors present a markedly
    different behavior (Figure~\ref{fig:rmse-bias_lt60data}b): biases are extremely small --- around three orders of magnitude
    lower than the RMSEs --- with no clear relationship between RMSE and BIAS, and
    the majority of configurations exhibit negative biases. As noted above, the aggregated ice concentration error is zero in most instances, even if the underlying ITD is inaccurate, so the overall bias remains relatively small.
    %

    %------------------------------------------------------------------------
    \subsection{Neural networks training and offline performance} \label{ssec:offline}
\begin{figure}
        \centering
        \begin{subfigure}
            {0.5\linewidth}
            \captionsetup{position=above,justification=raggedright, skip=0pt, margin=15pt}
            \caption{Ice volume}
            \centering
            \includegraphics[scale=0.65, trim={0.2cm 0.3cm 0.1cm 0cm}, clip]{
                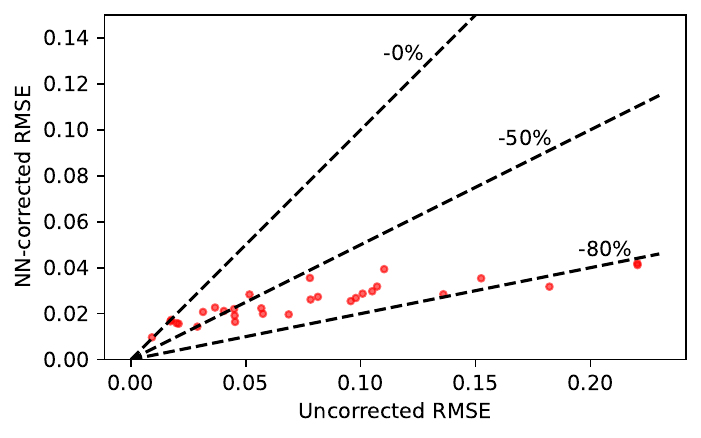
            }
            \label{fig:offline_rmse_red_vice}
        \end{subfigure} \hfill
        \begin{subfigure}
            {0.49\linewidth}
            \captionsetup{position=above,justification=raggedright, skip=0pt, margin=15pt}
            \caption{Ice volume}
            \centering
            \includegraphics[scale=0.65, trim={0 0.3cm 0 0}, clip]{
                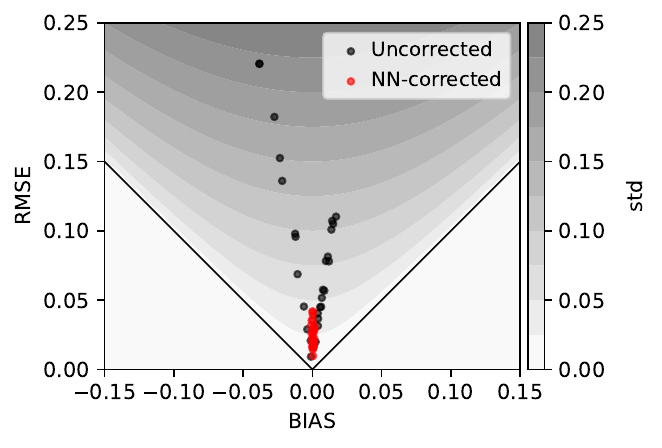
            }
            \label{fig:offline_rmse-bias_vice}
        \end{subfigure}
        \caption{Left panels (a) shows scatter plot of the residual RMSE after
        NN correction versus the uncorrected Icepack RMSE for ice volume on the test dataset. Dashed lines indicate three levels of RMSE
        reduction, as labeled in the figure. Right panel (b) displays scatter
        plots of BIAS and RMSE before and after correction for ice volume, with shaded contours representing the corresponding standard
        deviation.}
    \end{figure}

    %------------------------------------------------------------------------
    %
    Motivated by the exploratory analysis in the previous section, separate NNs are trained for each perturbed model configuration using an 8-year dataset spanning 1994 to 2001, totaling 2508 instances. Data from 2002
    (312 instances) serve as the validation set, while those from 2003 (312 instances)
    are held out for testing. The maximum number of training epochs is set to 2000
    and validation loss is monitored to apply early stopping and prevent overfitting.
    \begin{figure}
    \centering
        \begin{subfigure}{0.49\textwidth}
        \captionsetup{position=above,justification=raggedright, skip=-1pt, margin=18pt}
        \caption{}
        \includegraphics[scale=0.57,  trim={0cm 0.3cm 0.2cm 0.0cm}, clip]{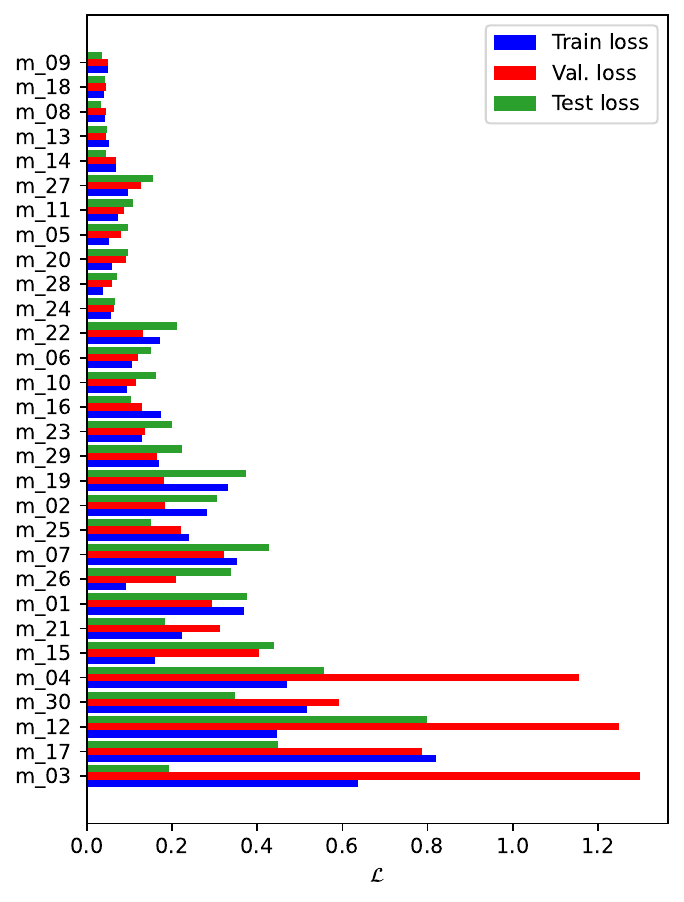}
        \end{subfigure}
        \begin{subfigure}{0.49\textwidth}
        \captionsetup{position=above,justification=raggedright, skip=-1pt, margin=15pt}
        \caption{}
        \includegraphics[scale=0.57,  trim={0.2cm 0.3cm 0 0.0cm}, clip]{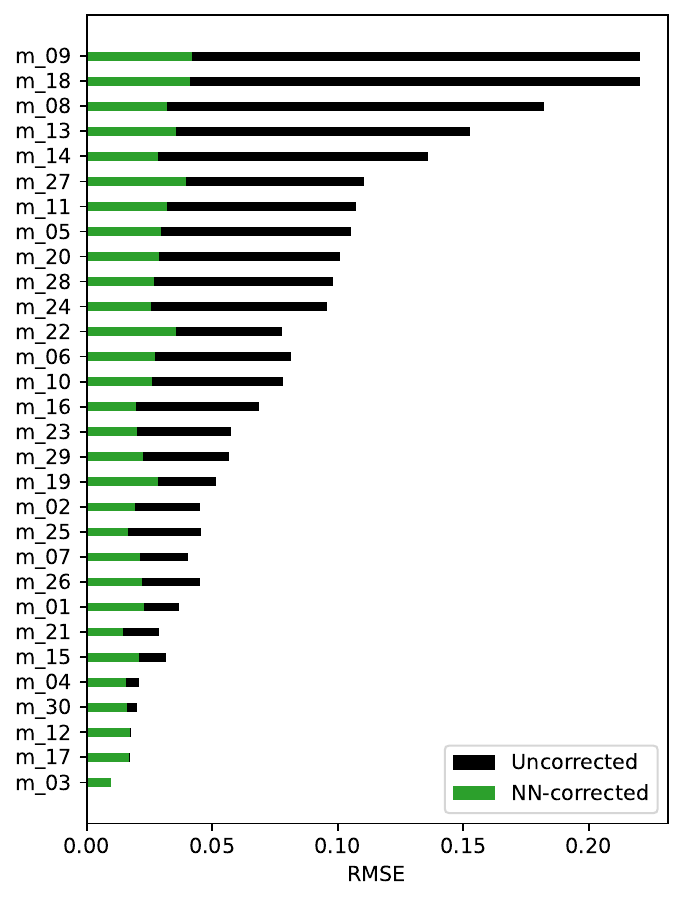}
        \end{subfigure}
        \caption{(a) Loss function $\mathcal{L}$ relative to ice
        volume error prediction, computed over training, validation and test
        datasets using the best-performing NN for each model configuration. (b) Ice volume RMSE of uncorrected forecasts
        and its residual value after correction by the NN computed on the test dataset.
        Model configurations are sorted as in Figure~\ref{fig:rmse-bias_lt60data}(a).}
        \label{fig:offline_vice}
    \end{figure}
    Figure~\ref{fig:offline_vice} and \ref{fig:offline_aice} show the results of NNs training performed separately on each dataset corresponding to a given model configuration, for ice volume (Figure~\ref{fig:offline_vice})
    and ice concentration (Figure~\ref{fig:offline_aice}) respectively. The values of the
    loss function $\mathcal{L}$, computed on training, validation and test
    datasets using the best-performing NNs (i.e., those achieving the lowest validation
    loss during training), indicate that the networks perform well for configurations
    with large forecast errors: where there is room for improvements. When the magnitude of the target forecast
    errors decreases, their performance deteriorates, even showing overfitting in a few cases.
    This trend is evident in the right panels of Figures~\ref{fig:offline_vice} and \ref{fig:offline_aice},
    where both the uncorrected baseline forecast RMSE and the residual RMSE after NN-correction are shown. As forecast RMSE decreases, the relative improvement
    from NN correction diminishes, and in some cases, the correction even worsens
    the prediction. \\
    A further quantification of RMSE reduction
    through the NN correction is presented in Figures~\ref{fig:offline_rmse_red_vice}
    and \ref{fig:offline_rmse_red_aice}, which show that both the NNs are able to
    reduce the ice volume and concentration RMSEs by over 80\% when the RMSE of the uncorrected models is large. In contrast, for near-perfect models with minimal forecast errors, the NN corrections
    have a negligible impact.
    Additionally, Figure~\ref{fig:offline_rmse-bias_vice} shows that, for ice volume, the NN eliminates the bias almost completely and the RMSE is reduced irrespective of the initial bias sign. For ice concentration (Figure~\ref{fig:offline_rmse-bias_aice}), the figure only displays a reduction in RMSE, as the uncorrected forecast bias is already minimal, consistent with the earlier commentary on Figure~\ref{fig:rmse-bias_lt60data}b.
    %

    %------------------------------------------------------------------------
    \subsection{Hybrid model online testing} \label{sec:online_testing}
    %------------------------------------------------------------------------

    %------------------------------------------------------------------------
    \subsubsection{Test protocol} \label{sec:test_protocol}
    %------------------------------------------------------------------------
    %
    Hybrid \mbox{Icepack-NN} models are tested over a four-year
    period following the data used for training, validation, and offline testing.
    The primary objective is to evaluate the performance of the hybrid models on long lead time forecasts, assessing, concurrently, their numerical stability, physical consistency and the capabilities to correct forecast error originated by the simultaneous presence of initial condition, atmospheric forcings and model error. 
    Initial condition errors, absent during training, emerge after
    the first hybrid iteration because NN predictions have inherent errors and only
    two state variables are updated.
    Atmospheric-forcing perturbations are introduced to mimic realistic forecasting conditions, in which the forcings provided to the operational hybrid system intrinsically differ from those used during training. This may occur when NNs are trained on analysis increments from reanalyses: the atmospheric forcings used in training come from the reanalysis, whereas during forecasting, the hybrid model is driven by forcing forecasts, which inevitably contain additional errors.
    \\
    For each of the six locations, 
    we produce 52 forecasts, initialized weekly throughout the year 2004.
    Each forecast spans 1080 days, corresponding to 18 hybrid model cycles. Hybrid \mbox{Icepack–NN} forecasts are evaluated against both the uncorrected Icepack model and the benchmark \mbox{Icepack–wclim} configuration.
    Time-correlated perturbations are applied to wind speed and air temperature from ERA5, using a decorrelation time scale of $2$~days and standard deviations of $3$\textrm{K} and $\sqrt{3}$\textrm{m/s}, respectively~\cite{Sakov2012, Cheng2020}.
    These perturbations are identical across all locations and forecasting systems (Icepack, \mbox{Icepack–NN}, and \mbox{Icepack–wclim}), but differ across the 52 start dates.
    Metrics defined in Equations~\eqref{eq:rmse_met}--\eqref{eqn:nrmse_leadtime} are, then, computed over datasets comprising $C=5$ ice categories, $L=6$ locations
    and $S=52$ simulations with distinct start date, resulting in a total of
    $N=C\times L \times S = 1560$ data points, for each of the $M=30$ model configurations.
    %
    
    %------------------------------------------------------------------------
    \subsubsection{Performance evaluation}
\begin{figure}
        \centering
        \includegraphics[scale=0.7]{
            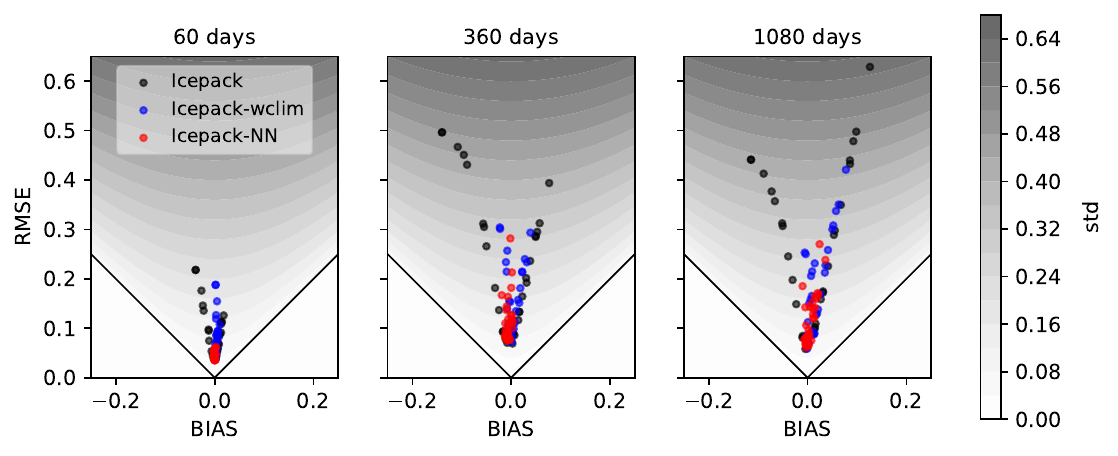
        }
        \caption{Scatter plots showing ice volume BIAS and RMSE for the
        uncorrected Icepack models (black dots), and the corrected hybrid models
        \mbox{Icepack-NN} (red dots) and \mbox{Icepack-wclim} (blue dots), at three different
        lead times (60, 360, 1080 days).}
        \label{fig:vice_bias-rmse_online}
    \end{figure}

\begin{figure}
        \centering
        \includegraphics[scale=0.7]{
            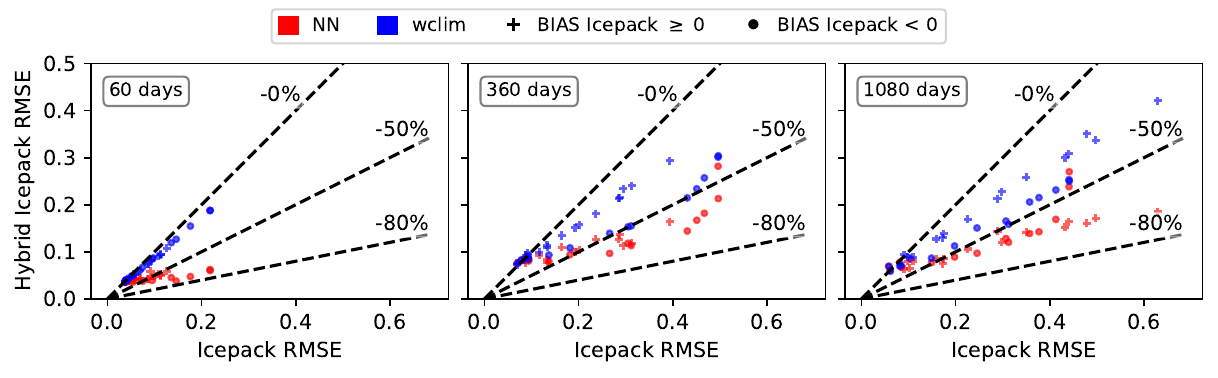
        }
        \caption{Scatter plots showing the ice volume RMSE for the hybrid
        Icepack models against the uncorrected Icepack RMSE at three different
        lead times (60, 360, 1080 days). Crosses/circles mark positive/negative BIAS in the uncorrected Icepack models. The dashed lines mark three percentages
        of RMSE reduction, which are indicated on the figure.}
        \label{fig:vice_rmse_red_online}
    \end{figure}

\begin{figure}
        \centering
        \includegraphics[scale=0.65, trim={0.2cm 0cm 0 0}, clip]{
            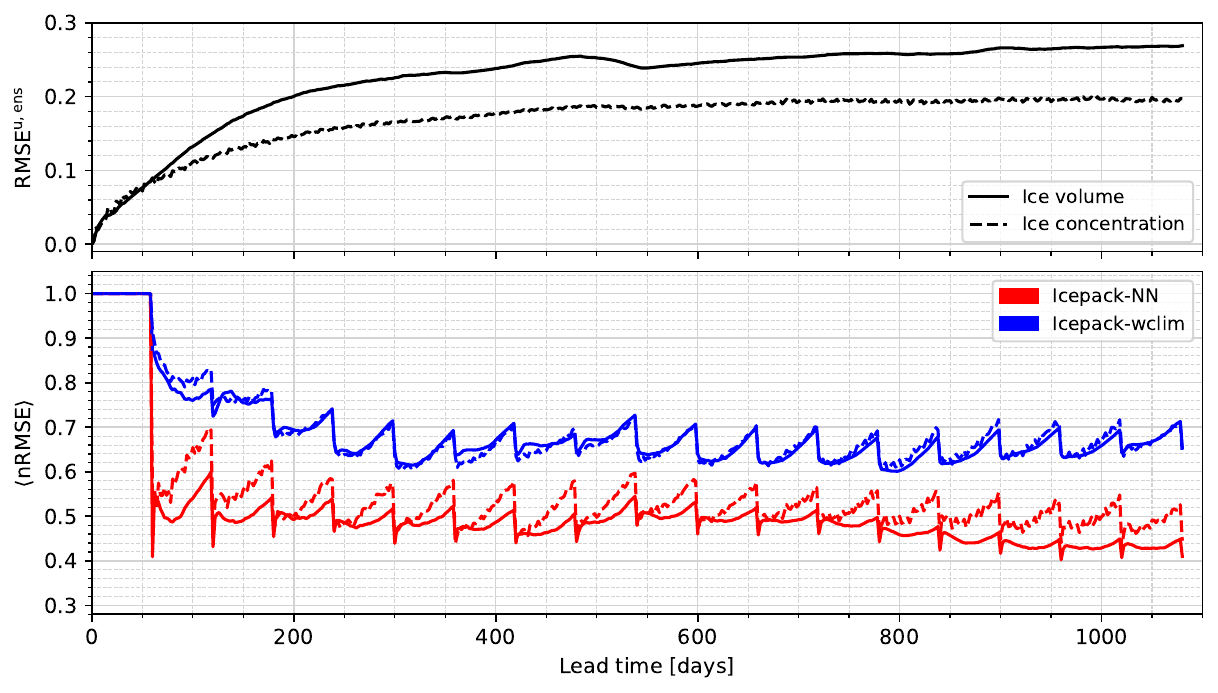
        }
        \caption{The top panel shows $\mathrm{RMSE}^{\mathrm{u,ens}}$ for the uncorrected Icepack ensemble (Eq.~\ref{eqn:ens_rmse}) as a function of lead time. 
        The bottom panel displays corresponding $\langle \mathrm{nRMSE} \rangle$ for the \mbox{Icepack-NN} and \mbox{Icepack-wclim} forecasts (Eq.~\ref{eqn:nrmse_leadtime}). 
        In both panels, solid and dashed lines denote the ice-volume and ice-concentration metrics, respectively.}
        \label{fig:nrmse_leadtime_fp}
    \end{figure}

\begin{figure}
        \centering
        \includegraphics[scale=0.65]{
            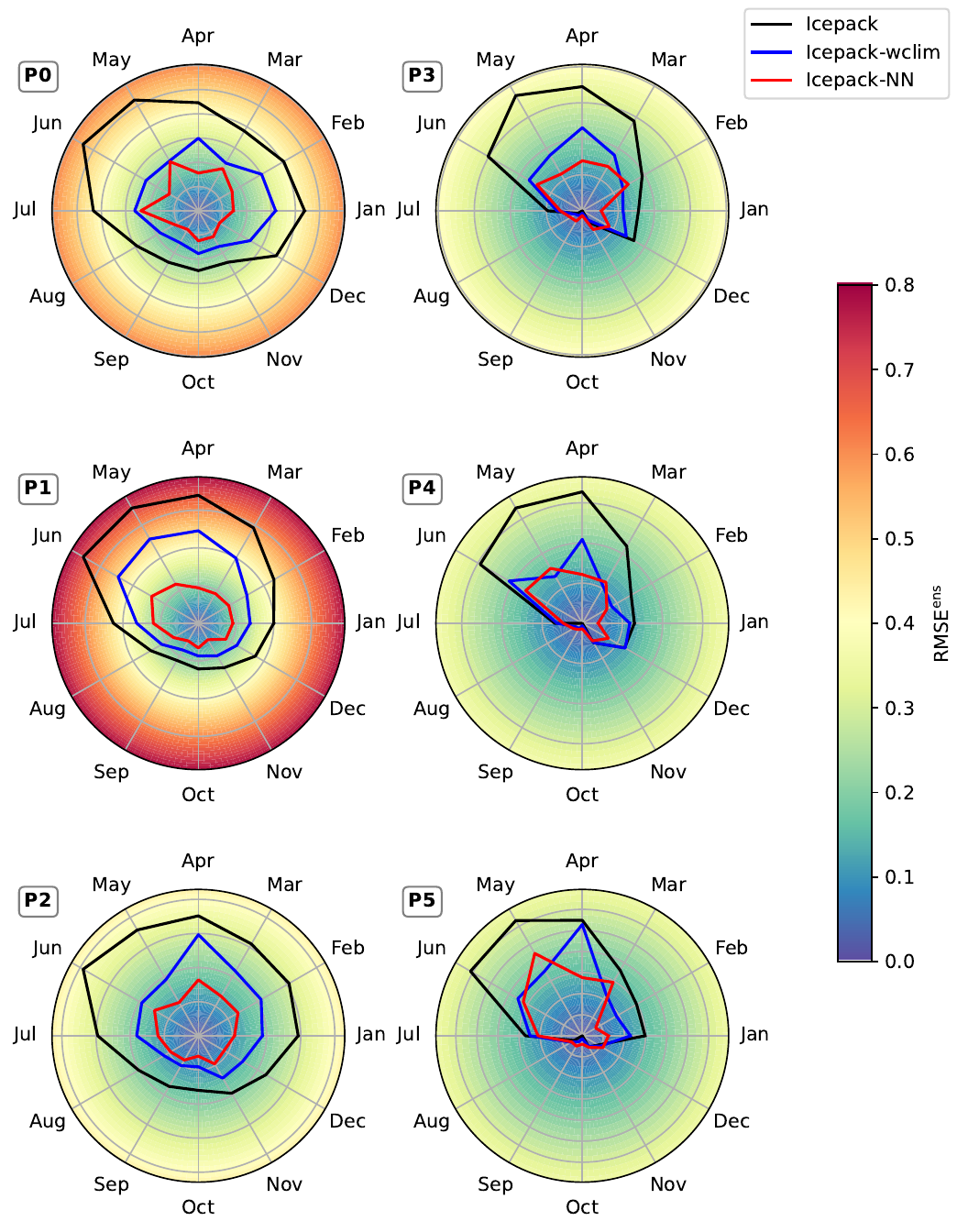
        }
        \caption{Ice volume $\text{RMSE}^{\textrm{ens}}$ as function of forecast month at lead
        time $\tau = 1080$ days for Icepack, \mbox{Icepack-NN} and \mbox{Icepack-wclim} predictions,
        at the six locations simulated.}
        \label{fig:online_RMSE_locs}
    \end{figure}

    %------------------------------------------------------------------------
    %
    Figure~\ref{fig:vice_bias-rmse_online} presents ice volume RMSE and BIAS for
    all model configurations at lead times of 60, 360 and 1080 days,
    corresponding to 1, 6, and 18 hybrid loops, while Figure~\ref{fig:vice_rmse_red_online} highlights the corresponding RMSE reduction achieved by the two hybrid systems.
    At 60 days, after the first
    correction, \mbox{Icepack-NN} models display RMSE and BIAS values consistent with those
    seen in offline evaluations (Figures~\ref{fig:offline_rmse_red_vice},~\ref{fig:offline_rmse-bias_vice}),
    and effectively reduce the forecast errors of the uncorrected Icepack model. In contrast, at the
    same lead time, \mbox{Icepack-wclim} models, although effective in reducing BIAS,
    show limited ability to reduce RMSE, as evident in the first panel of Figure~\ref{fig:vice_rmse_red_online},
    where the blue dots are almost aligned with the $-0\%$ reference line.
    At longer lead times, error magnitudes increase for all model configurations. Yet the \mbox{Icepack–NN} models continue to outperform the \mbox{Icepack–wclim} models, with the performance gap between the two hybrid approaches being more pronounced for configurations with positive BIAS (i.e. those that overestimate ice volume), which are generally associated with high snow conductivity (low insulation).
    In these cases, the \mbox{Icepack–wclim} models perform noticeably worse, as shown by the blue cross markers in the 360- and 1080-day panels of Figure~\ref{fig:vice_rmse_red_online}, which indicate a systematically smaller RMSE reduction. Such behaviour is reflected in the corresponding panels of Figure~\ref{fig:vice_bias-rmse_online}, where the blue dots display an asymmetric distribution, with larger RMSE values predominantly associated with positive BIAS on the right side of the plot.
    While our analyses do not provide a definitive explanation, we conjecture that this performance gap is likely associated with an increased sensitivity to initial conditions at the beginning of the melting season, which leads to degraded \mbox{Icepack–wclim} predictions.
    In contrast, the Icepack-NN models retain high skill, indicating a greater ability to cope with the resulting internal variability.
    \\
    To provide an ensemble-level perspective on forecast errors and their reduction by the hybrid systems, Figure~\ref{fig:nrmse_leadtime_fp} shows the time series of the $\text{RMSE}^{\textrm{ens}}$ for the uncorrected Icepack models (top panel) and the $\langle \text{nRMSE} \rangle$ (bottom panel) for the \mbox{Icepack-NN} and \mbox{Icepack-wclim} models. For both ice concentration and ice volume, the $\text{RMSE}^{\textrm{ens}}$ of the uncorrected models increases rapidly during the first $200$ days before stabilizing, approaching an almost stationary value after about one year as the models drift toward their own stable annual sea-ice cycles.
    The absence of seasonal variability in the top panel arises from the definition of the metric in Eqs.~\eqref{eq:rmse_met}-\eqref{eqn:ens_rmse}, which averages over forecasts initialized across the annual cycle and thus removes the seasonal signal.
    In the bottom panel, the $\langle \text{nRMSE} \rangle$ values --- corresponding to the slope of a least-squares
    regression line fitted to the scatter plot data of Figure~\ref{fig:vice_rmse_red_online} --- highlight the discrete-time nature of the hybrid systems, characterized by periods of error growth between successive impulsive corrections. The superior performance of the Icepack–NN models is evident, with $\langle \text{nRMSE} \rangle$ for ice volume reaching values close to $0.4$ at long lead times, whereas the corresponding \mbox{Icepack–wclim} value stays above $0.6$. Moreover, although the error reduction provided by \mbox{Icepack-wclim} remains relatively stable, a slight upward trend in its $\langle \text{nRMSE} \rangle$ values is visible. In contrast, the \mbox{Icepack-NN} curves show a slow decline after approximately 600 days, indicating a gradual improvement over long lead times.
    The performance assessment is completed by examining how the ice-volume RMSE varies with geographic location. Figure~\ref{fig:online_RMSE_locs} displays the seasonal evolution of RMSE at a fixed lead time of 1080 days. For the uncorrected Icepack model, all locations exhibit a pronounced annual cycle, with generally lower RMSE values from July to November and higher values from December to June. A similar seasonal pattern is also visible in the hybrid systems, although it is less pronounced, particularly at the higher-latitude sites P0, P1, and P2.
    Overall, while both hybrid approaches reduce errors relative to the uncorrected forecast, the \mbox{Icepack–NN} configuration delivers the best performance. At the lower-latitude locations P3, P4, and P5, the errors to be corrected are smaller, and the performance of the two hybrid systems becomes more similar. This is partly because the MSE loss used during training directs the NN to prioritise larger target values.
    In these locations, during the late-summer to early-autumn period, when the true ice volume is zero, \mbox{Icepack–NN} is systematically worse than both the uncorrected model and \mbox{Icepack–wclim}, although the resulting error remains very small.
    Here, the uncorrected forecasts are already highly accurate, and the climatological correction does not introduce additional error, whereas the NN retains a small amount of residual variability despite the true target being zero.

    %

    %%%%%%%%%%%%%%%%%%%%%%%%%%%%%%%%%%%%%%%%%%%%%%%%%%%%%%%%%%%%%%%%%%%%%%%%%
    \section{Adapting to physical model updates: transfer learning} \label{sec:transfer_learning}
    %%%%%%%%%%%%%%%%%%%%%%%%%%%%%%%%%%%%%%%%%%%%%%%%%%%%%%%%%%%%%%%%%%%%%%%%%
    %
    %------------------------------------------------------------------------
    \subsection{Motivation}
    %------------------------------------------------------------------------
    We have so far worked under the common assumption that the process underlying our dataset is autonomous (i.e. it does not explicitly depend on time) or stationary. Under this condition, offline training is, in principle, effective and its accuracy typically improves asymptotically as the size of the training dataset increases.
    The stationary/autonomous assumption does however not hold in many realistic scenarios: notably in the case of climate change.
    A number of solutions have been explored in the literature, in the context of model error correction.
    When the process statistics do not change too quickly, online learning~\cite{Farchi2021comparison} and running training strategies~\cite{He2025} have shown promising results.
    
    Here we envisage a qualitatively different scenario: the model or the DA system on which the training has been performed undergoes an update (e.g. a new parameter calibration), producing a sudden shift in the process statistics. In such cases, while avoiding a new offline training may be desirable for computational or practical reasons, methods designed for smooth, incremental adaptation (such as online learning) provide little advantage. 
    In this work we focus on transfer learning as a means to adapt a pretrained NN to a forecast model with a new set of (yet incorrect) calibrated parameters, and we assess under which conditions this adaptation provides better performance than retraining a model from scratch. The ML model’s ability to adapt and generalize across such changes is loosely related to how strongly the underlying dynamical system responds to parameter modifications~\cite{carrassi_model_2008}. In general, one might expect some degree of generalization as long as the system does not cross a bifurcation (also referred to as a tipping point in non-autonomous systems~\cite{ghil_sciamarella-2023}), which would induce qualitatively different behavior from the pretrained model and DA version. For this reason, parameter distance is not considered as a reliable practical criterion for transferability, since even small parameter changes may produce substantial changes in the dynamics. Moreover, its practical implementation is limited given that it would require prior knowledge of the parameter distributions. Nevertheless, in the absence of bifurcations, transfer learning should in general be more effective when the source and target configurations are more similar. \\
    Transfer learning has recently been studied as a tool to update the weights of a variational autoencoder in the context of a non-autonomous system~\cite{pasmans2025ensemble}.
    %

    %------------------------------------------------------------------------
    \subsection{Methodology and results}
\begin{figure}
        \centering
        \begin{subfigure}
            {0.49\textwidth}    
            \captionsetup{belowskip=-3pt}
            \caption{$\mathrm{m\_09}\to \mathrm{m\_24}$}
            \includegraphics[scale=0.62, trim={0 0 0 5pt}, clip]{   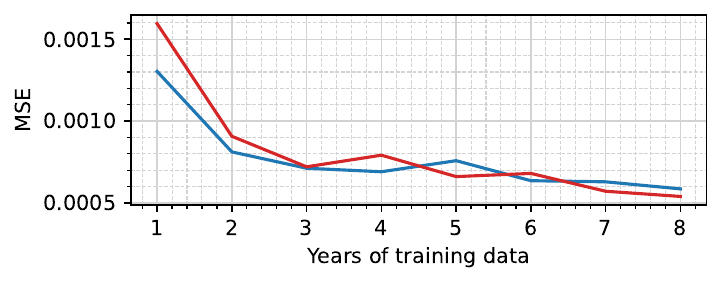
            }
            \label{fig:transf_lines:a}
        \end{subfigure}
        \hfill
        \begin{subfigure}
            {0.49\textwidth}
            \captionsetup{belowskip=-3pt}
            \caption{$\mathrm{m\_09}\to \mathrm{m\_22}$}
            \includegraphics[scale=0.62, trim={0 0 0 5pt}, clip]{   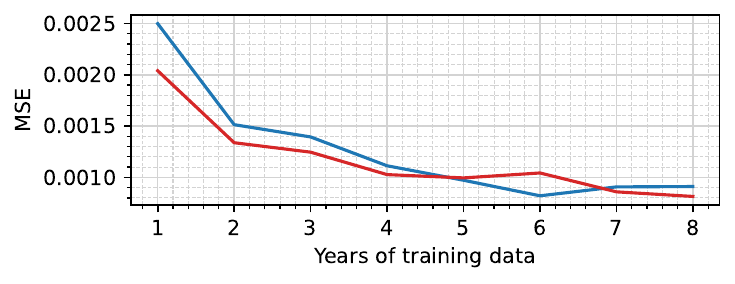
            }
            \label{fig:transf_lines:b}
        \end{subfigure}
        \caption{ Validation MSE obtained with fine tuning (\textcolor{tab:blue}{\sol})
        and retraining (\textcolor{tab:red}{\sol}) as a function of the number of
        training years. The pretrained NN is trained on configuration
        $\mathrm{m\_09}$ and fine-tuned for configurations $\mathrm{m\_24}$ (a) and
        $\mathrm{m\_22}$ (b). }
        \label{fig:transf_lines}
    \end{figure}

\begin{figure}
        \centering
        \includegraphics[width=0.62\linewidth]{
            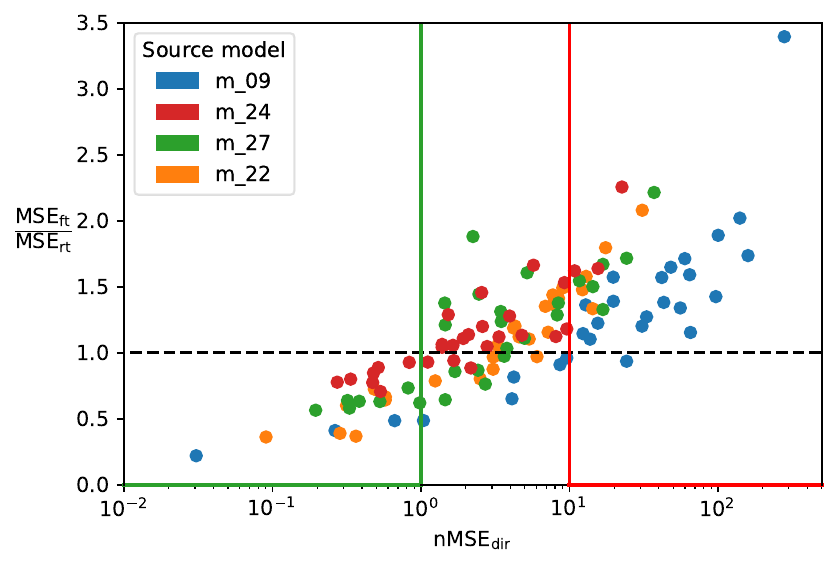
        }
        \caption{Scatter plot of the ratio between fine-tuning and retraining
        MSE, $\nicefrac{\text{MSE}_{\textrm{ft}}}{\text{MSE}_{\textrm{rt}}}$, on the validation set, plotted
        against the metric $\text{nMSE}_{\textrm{dir}}$~(Eq.\ref{eq:ftune_crit}).
        Both fine-tuning and retraining are performed using one year of training data.
        Each point refers
        to a single source-target configuration pair.
        Vertical green and red lines indicate mark $\text{nMSE}$ thresholds below (above) which fine-tuning is expected to perform better (worse) than retraining from scratch.
        }
        \label{fig:ftuning_criterion}
    \end{figure}

    %------------------------------------------------------------------------
    %
    We adopt a fine-tuning transfer learning
    strategy~\cite{iman2023review}, whereby NNs pretrained on a given \textit{source} model are further trained on as little as possible data from a different \textit{target} model, while retaining the original network architecture and using a reduced learning rate to preserve previously learned features.
    Their performance is then compared to NNs trained from scratch with progressively larger training sets.
    The same datasets previously described~(Section~\ref{sec:data_analysis}) are used,
    focusing on ice volume error correction. All models are retrained over progressively
    longer training sets: the last two years are reserved for validation and
    testing, while earlier years are incrementally added to the training set (up to eight years in
    total), as illustrated in Table~\ref{tab:ftune_exp}. Fine-tuning follows the same procedure, except training is initialized from pretrained models. 
    \begin{table}[ht]
        \small
        \centering
        \caption{Time-based dataset split used in the transfer learning experiments, illustrating the progressive enlargement of the training set. \legendsquare{blue!30}~Training set, \legendsquare{red!30}~Validation set, \legendsquare{green!30}~Test set}
        \label{tab:ftune_exp}
        \arrayrulecolor{gray!80}
        \rowcolors{1}{}{white}
        \renewcommand{\arraystretch}{0.75} % reduces row height globally
        \begin{tabular}{M{1.5cm}|c|c|c|c|c|c|c|c|c|c|}
            \textbf{Number of training years} & \textbf{1994} & \textbf{1995} &
            \textbf{1996} & \textbf{1997} & \textbf{1998} &
            \textbf{1999} & \textbf{2000} & \textbf{2001} & \textbf{2002} & \textbf{2003}\\
            \midrule
            1 & \multicolumn{7}{c|}{\cellcolor{gray!20}} & \cellcolor{blue!30} & \cellcolor{red!30} & \cellcolor{green!30} \\
            2 & \multicolumn{6}{c|}{\cellcolor{gray!20}} & \multicolumn{2}{c|}{\cellcolor{blue!30}} & \cellcolor{red!30} & \cellcolor{green!30}\\
            & \multicolumn{5}{c|}{\cellcolor{gray!20}\rule{0pt}{1pt}} & \multicolumn{3}{c|}{\cellcolor{blue!30}\rule{0pt}{1pt}} & \cellcolor{red!30}\rule{0pt}{1pt} & \cellcolor{green!30}\rule{0pt}{1pt} \\ 
            [-5pt] % reduces vertical space
             & \multicolumn{4}{c|}{\cellcolor{gray!20}\rule{0pt}{0pt}} & \multicolumn{4}{c|}{\cellcolor{blue!30}} & \cellcolor{red!30} & \cellcolor{green!30} \\
            [-5pt]
             & \multicolumn{3}{c|}{\cellcolor{gray!20}\rule{0pt}{0pt}} & \multicolumn{5}{c|}{\cellcolor{blue!30}} & \cellcolor{red!30} & \cellcolor{green!30} \\
            [-5pt]
             & \multicolumn{2}{c|}{\cellcolor{gray!20}\rule{0pt}{0pt}} & \multicolumn{6}{c|}{\cellcolor{blue!30}} & \cellcolor{red!30} & \cellcolor{green!30} \\
            [-5pt]
            \smash{\raisebox{7.pt}{$\vdots$}}
             & \cellcolor{gray!20} & \multicolumn{7}{c|}{\cellcolor{blue!30}} & \cellcolor{red!30} & \cellcolor{green!30} \\
            [-5pt]
            8 & \multicolumn{8}{c|}{\cellcolor{blue!30}} & \cellcolor{red!30} & \cellcolor{green!30}\\
            \bottomrule
        \end{tabular}
    \end{table}
    Importantly, when transferring a pretrained NN to a different task (i.e., the task of correcting model error of an incorrect physical model the NN has not seen in its training), the input and output normalization parameters (means and
    standard deviations) computed for pretraining, must be preserved to ensure consistency.
    Pretrained source models are those
    previously trained on the full eight-year training set~(Section~\ref{ssec:offline}).
    For each training dataset size, both retraining and fine-tuning are evaluated using different learning rates, as shown in Figure~\ref{fig:transfer_opt}. The
    comparison is then based on the best-performing learning rate (minimum validation
    loss) for each case. \\
    Results for two source-target configuration pairs are shown in Figure~\ref{fig:transf_lines}.
    The two selected pairs ($\mathrm{m\_09}\to \mathrm{m\_24}$ and $\mathrm{m\_09}\to \mathrm{m\_22}$) illustrate a scenario where a poorly calibrated model with large errors ($\mathrm{m\_09}$) is refined to achieve configurations that are either more ($\mathrm{m\_24}$) or less ($\mathrm{m\_22}$) similar to the initial configuration within the parameter space (cf.~Section~\ref{sec:param_err}).
    In both cases, the performance of fine-tuning and retraining tends to converge as the number of training years increases; however, fine-tuning delivers a clear advantage when a limited amount of training data is available~(Figure~\ref{fig:transf_lines:a}), which may translate into significant computational savings by reducing the need for extensive retraining.
    This advantage is not guaranteed (Figure~\ref{fig:transf_lines:b}), as the effectiveness of fine-tuning appears to depend on the similarity between the source and target tasks.\\
    Although the advantage of fine-tuning cannot be quantified strictly \emph{a priori}, in this section, we propose a criterion to estimate its potential before performing additional training.
    To this end, a normalized MSE is introduced, $\text{nMSE}_{\textrm{dir}}$,
    defined as follows,
    \begin{equation}
        \label{eq:ftune_crit} \text{nMSE}_{\textrm{dir}}= \frac{\overline{\left( g \left(\hat{\V{\theta}}_{\textrm{src}},  \V{z}_{\textrm{trg}} \right) - \V{e}_{\textrm{trg}} \right)^2}}{\overline{\left( \V{e}_{\textrm{trg}} \right)^2}}
        ,
    \end{equation}
    where $g(\hat{\V{\theta}}_{\textrm{src}}, \V{z}_{\textrm{trg}})$ denotes the prediction
    by the NN pretrained on the source task with weights $\hat{\V{\theta}}_{\textrm{src}}$, evaluated on inputs $\V{z}_{\textrm{trg}}$ from the target task. The term $\V{e}_{\textrm{trg}}$ represents the corresponding forecast error of the target model configuration.
    Unlike the NN loss function, this metric is computed in physical space, enabling a fair comparison between training strategies. The metric $\text{nMSE}_{\textrm{dir}}$, where ``\textrm{dir}" stands for \emph{direct transfer}, measures the MSE obtained by directly applying the pretrained NN 
    normalized by the uncorrected forecast MSE.
    Although a distance in parameter space may seem a natural way to quantify similarity between tasks in this context, we do not regard it as reliable, as prior parameter distributions are often unavailable and the mapping from parameter variations to dynamical response is non-linear.
    To assess the relevance of the ratio defined in Eq.~\eqref{eq:ftune_crit} as a criterion for estimating fine-tuning potential, we
    analyze multiple source-target configuration pairs: four pretrained source
    configurations, each targeting all remaining configurations. Fine-tuning and retraining are both performed using only one year of new training data. The results are summarized in Figure~\ref{fig:ftuning_criterion}, which presents
    scatter plots of the ratio between fine-tuning and retraining MSE, $\nicefrac
    {\text{MSE}_{\textrm{ft}}}{\text{MSE}_{\textrm{rt}}}$, computed on validation sets, plotted against $\text{nMSE}_{\textrm{dir}}$.
    As expected, for $\text{nMSE}_{\textrm{dir}}$ smaller than one, fine-tuning outperforms retraining ($\nicefrac
    {\text{MSE}_{\textrm{ft}}}{\text{MSE}_{\textrm{rt}}} < 1$). This behavior is natural, since $\text{nMSE}{\textrm{dir}}<1$ indicates that the pretrained model already provides a better-than-baseline prediction, implying a meaningful alignment between source and target dynamics. In such cases, fine-tuning benefits from a favorable initialization that already captures relevant structure, whereas retraining must learn this structure from scratch with limited data.
    On the other hand, when $\text{nMSE}_{\textrm{dir}}$ exceeds ten, fine-tuning performs worse than retraining. For intermediate values between one and ten, outcomes are mixed, and no reliably predictive rule emerges, although a slight dependence on the pretrained source model is noticeable.
    To quantify how consistently the validation and test outcomes align within each $\text{nMSE}_{\textrm{dir}}$ region ([0,1), [1,10], (10, $\infty$)), we computed an agreement score, defined as the fraction of instances in which both sets indicate the same preference (fine-tuning or retraining). The resulting agreements for the three regions were 1.00, 0.70, and 0.97, respectively, with the reduced agreement in the intermediate range reflecting the increased ambiguity when performance ratios lie close to one.
    Overall, $\text{nMSE}_{\textrm{dir}}$ offers a simple, though possibly conservative, criterion to assess the potential benefit of fine-tuning before additional training, since the point at which fine-tuning ceases to be advantageous can shift to higher values depending on the pretrained source model.

    %%%%%%%%%%%%%%%%%%%%%%%%%%%%%%%%%%%%%%%%%%%%%%%%%%%%%%%%%%%%%%%%%%%%%%%%%
    \section{Feature importance and physical interpretation} \label{sec:feat_imp}
\begin{figure}
        \centering
        \includegraphics[width=0.67\linewidth]{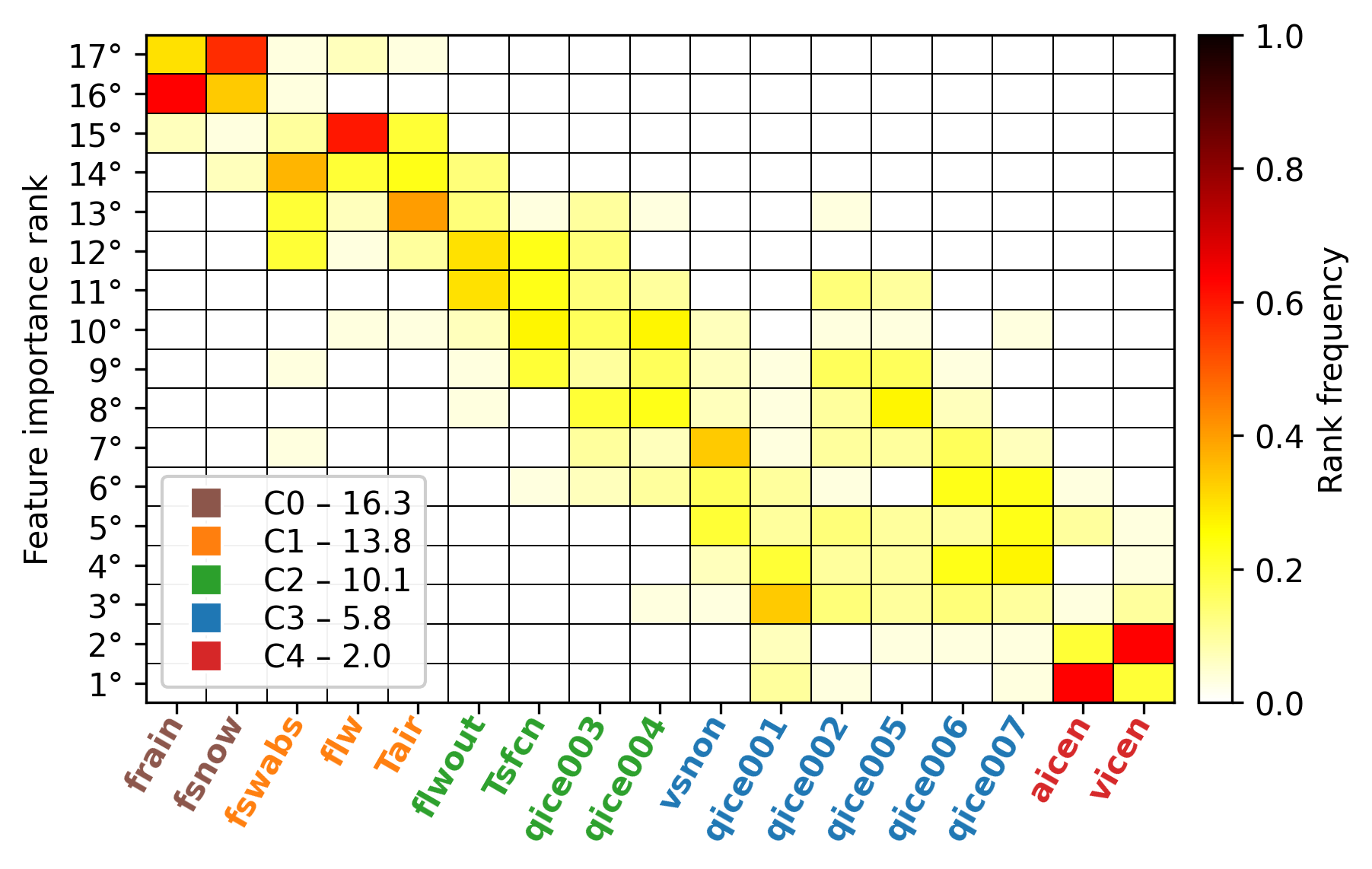}
        \caption{Feature-importance rank frequency map. Each column shows the frequency distribution of importance ranks over all the model configurations for the feature indicated on the x-axis (cf.~Eq.~\eqref{eqn:rank_freq}). Feature labels are colored according to their cluster of similarly important features, and the corresponding mean ranks are reported in the legend.}
        \label{fig:featrank}
    \end{figure}

\begin{figure}
        \centering
        \includegraphics[width=0.65\linewidth]{
            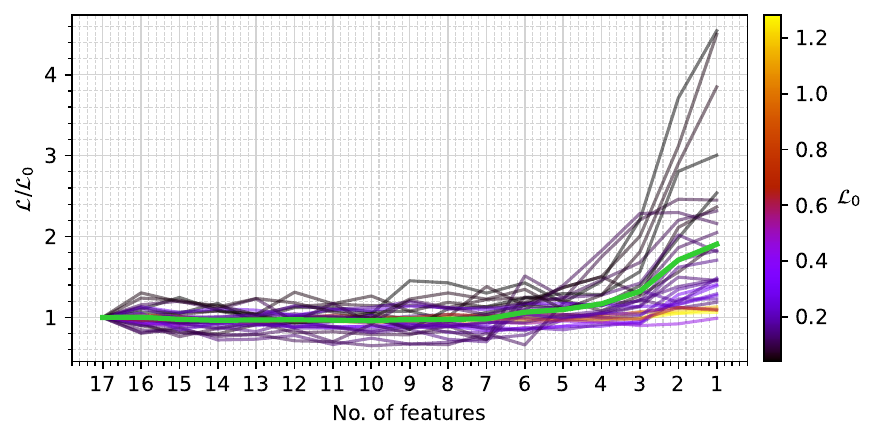
        }
        \caption{Normalized validation loss
        $\nicefrac{\mathcal{L}}{\mathcal{L_0}}$ against number of features. Colormap
        lines refer to individual models, with color indicating loss obtained
        using the full feature set. The green line denotes the ensemble average.}
        \label{fig:valloss-nfeat}
    \end{figure}

    %%%%%%%%%%%%%%%%%%%%%%%%%%%%%%%%%%%%%%%%%%%%%%%%%%%%%%%%%%%%%%%%%%%%%%%%%
    %
    We complete the study by performing a feature importance analysis to identify the most effective predictor features. This may guide and lead to the identification of a minimal model depending on the degree of desired accuracy. Minimal models are not only computationally lighter but generally more interpretable.
    The analysis employs a Permutation-Importance-based Recursive Feature Elimination (PI-RFE) algorithm, which combines Recursive Feature Elimination (RFE) with Permutation feature Importance (PI,~\citet{breiman2001random}).
    At each
    iteration, the network is trained and the importance of each feature is
    estimated by randomly permuting its values in the validation set and
    measuring the corresponding drop in predictive performance. The least important
    feature is eliminated and the process is repeated until only one feature is left.
    Figure~\ref{fig:featelim} illustrates the recursive feature elimination process
    for data relative to a specific  model configuration.
    Recursive feature elimination establishes a feature-importance ranking, with
    features eliminated last occupying higher positions.
    Monitoring the validation loss during feature elimination can also help identify a reduced subset of relevant predictors. However, the order in which features are eliminated and the corresponding increase in validation loss varies across networks trained on different model configurations. By applying \mbox{PI-RFE} to each configuration and then aggregating the results, we can identify consistent patterns in feature importance and validation-loss behavior across all configurations.
    Specifically, we compute a feature-importance-rank frequency map (Figure~\ref{fig:featrank}), which shows, for each
    feature, the frequencies of its rankings across all models. The normalized
    frequency of feature $X$ having rank $k$ across models is defined as
    \begin{equation}
        \label{eqn:rank_freq}\hat{f}_{k}(X) = \frac{1}{M}\sum_{m=1}^{M} \mathbf{1}
        _{\{ R_m(X) = k \}},
    \end{equation}
    where $M$ is the number of perturbed model configurations, $m$ indexes the models, and $R_{m}(X)$
    denotes the rank assigned to feature $X$ by PI-RFE for model $m$.
    Features are grouped into five clusters of similar importance using agglomerative clustering based on the Wasserstein distance, implemented using utilities from SciPy~\cite{2020SciPy-NMeth} and scikit-learn~\cite{scikit-learn}. These clusters are highlighted in Figure~\ref{fig:featrank} through the color of the feature labels, with each cluster’s mean rank reported in the legend.
    The least important cluster (C0) includes only precipitation rates (rain and snow), while the next cluster (C1) groups variables related to atmospheric forcing: air temperature, longwave radiation, and absorbed shortwave radiation.
    Since atmospheric forcing features are instantaneous snapshots, their high variability, and the additional intermittency of precipitation in particular, can plausibly explain their low rank. 
    However, they might  have achieved higher rankings if they had been represented by quantities integrated over the forecast window.
    At the opposite end of the spectrum, the most relevant features for ice-volume error prediction (cluster C4) are ice concentration and ice volume forecasts, in agreement with with previous findings on online error correction~\cite{Finn2023, Gregory2023}.
    Interestingly, snow volume does not belong to this top-importance cluster, even though the model error in our experiments is rooted in the snow-physics parameterizations. Instead, together with a subset of ice-layer enthalpies, it forms cluster C3, within which the surface and bottom-layer enthalpies emerge as the most influential features.
    By integrating rapid surface-process variability, ice enthalpies retain long-term information that remains relevant at the 60-day forecast horizon.
    The remaining variables form a medium-importance cluster (C2), which includes the ice-layer enthalpies not assigned to C3, the outgoing longwave radiation, and the ice surface temperature.\\
    The impact of feature elimination on the predictive performance of the NNs is assessed by examining how validation loss varies with a decreasing number of features during PI-RFE across all models.
    As shown in Figure~\ref{fig:offline_vice},
    validation loss values obtained with the full feature set differ across models.
    To extract an averaged trend of validation loss as a function of feature count, each curve is normalized by the corresponding validation loss
    $\mathcal{L}_{0}$ computed with the full feature set: Figure~\ref{fig:valloss-nfeat}
    reports the individual model curves (colored by colormap) and the ensemble average (green).
    Validation loss begins to increase consistently after the removal
    of 6–7 features, which in most cases contain the atmospheric forcing variables (Figure~\ref{fig:featrank}), thereby confirming their negligible impact on predictive performance.
    The sharpest increase occurs when
    reducing from three to two features, indicating that acceptable performance requires,
    in addition to ice volume and concentration, at least one ice-layer enthalpy
    (most likely the surface layer).
    Finally, ML models with inherently poor performance
    (high $\mathcal{L}_{0}$) appear largely insensitive to feature elimination.
    In these cases, as discussed in Section~\ref{ssec:offline}, our NN corrections have trouble estimating the very small errors of the original uncorrected models, regardless of the number of features.
    Overall, these results suggest that a minimal feature set capable of matching the performance obtained with all variables would include the features in clusters C3 and C4 and possibly the two remaining enthalpies from C2 for physical consistency, yielding a total of ten feature variables.
    While alternative experimental setups employing longer or shorter time windows may alter the relative importance of intermittent versus long-memory variables, and thus which ice layers are most important in terms of their enthalpy, the aggregating effect of ice enthalpy is expected to persist across setups, granting it greater predictive power than atmospheric variables. 

    %%%%%%%%%%%%%%%%%%%%%%%%%%%%%%%%%%%%%%%%%%%%%%%%%%%%%%%%%%%%%%%%%%%%%%%%%
    \section{Conclusions} \label{sec:conclusions}
    %%%%%%%%%%%%%%%%%%%%%%%%%%%%%%%%%%%%%%%%%%%%%%%%%%%%%%%%%%%%%%%%%%%%%%%%%
    We investigated the use of ML to estimate and correct model error arising from parameter misspecification in the state-of-the-art sea-ice column physical model Icepack. Experiments were conducted in an idealized setup, where true trajectories were generated using a reference Icepack configuration, while forecasts came from perturbed configurations with modified snow thermodynamics and radiative properties.
    For each perturbed configuration, NNs were trained to predict \mbox{$60$-day} forecast errors in ice volume and concentration across different spatial locations.
    Our approach followed the hybrid-modeling strategy introduced in \citet{brajard2021combining} and \citet{Farchi2021}. A key distinction from these earlier works is that, while they focused on chaotic autonomous systems, the model considered here (Icepack) is a forced, non-autonomous system, similar to those examined in \citet{Finn2023} and \citet{Gregory2023}. Moreover, to isolate the impact of parametric error from DA and initial condition errors and to enable the characterization of error statistics within the perturbed parameters subspace, we intentionally assumed perfect and complete observations. This allowed us to bypass DA and train NNs directly on exact forecast errors rather than analysis increments.
    The physical model and NN corrections were integrated into hybrid frameworks (\mbox{Icepack-NN}), applying NN-based error corrections iteratively in a manner analogous to sequential DA.
    Ad-hoc post-processing ensured physical consistency, model restartability and numerical stability.
    Hybrid models were tested for forecasts up to 1080 days and compared with a climatology-based benchmark (\mbox{Icepack-wclim}). These experiments showed that NN corrections are most effective for models exhibiting substantial deviations from the truth, without degrading the performance of nearly perfect models. Hybrid simulations remained stable and robust to atmospheric-forcing and initial-condition errors, despite their absence during training.
    Across all configurations, NNs consistently outperformed the climatology-based approach, with \mbox{Icepack-NN} achieving nearly a 60\% reduction in ice-volume RMSE, compared with less than 40\% for \mbox{Icepack-wclim}.\\
    Although the setup is idealized, our focus on errors connected to snow thermodynamics highlights that sufficiently long lead times are required for these errors to emerge and become detectable by ML methods. In operational systems, however, the training lead time is not a free parameter but is dictated by the DA cycle, which typically ranges from 5 to 7 days in sea-ice applications~\cite{Sakov2012, zhang2021assimilation, cipollone_bivariate_2023}. As a result, DA-based ML bias-correction approaches applied to realistic sea-ice models are expected to primarily capture fast-growing errors associated with horizontal dynamics, whereas slower thermodynamic error components may require additional calibration or ad-hoc strategies to be effectively addressed. \\
    Transfer learning experiments were conducted to assess how effectively a NN trained to correct one physical model configuration can be adapted to another. In the present study, configuration-specific training was adopted because the statistics of analysis increments depend on the underlying forecast model.
    Generally, a single NN trained across different forecast models represents an unavoidable compromise and should not be expected to be optimal for any individual model; likewise, a NN trained to correct a given model or configuration is not expected to generalize reliably to other contexts.
    This model dependence suggests that NNs typically need to be updated when the physical model is modified and, more broadly, whenever changes introduced in the DA system may alter the statistics of analysis increments.
    Updating NNs becomes particularly difficult when new observational products are introduced, since past reanalysis data, required for training a new NN and producing an updated hybrid model, cannot be reconstructed.
    We relied on fine-tuning of pretrained networks, whose performance is compared with networks retrained from scratch.
    For small dataset sizes, fine-tuning can be advantageous, though its effectiveness depends on the similarity between source and target tasks.
    Task similarity is not uniquely defined and cannot be inferred \emph{a priori} from parameter differences, because even small parameter changes may trigger bifurcations and substantially alter error statistics.
    For this reason, a criterion was sought to estimate fine-tuning potential benefit, before actual training, using only one year of new training data. 
    In particular, we considered the ratio between the pretrained network’s MSE on the target task and the forecast MSE of the target model.
    Results showed that if the ratio is below one, fine-tuning is preferable, whereas if it is larger than 10, fine-tuning performs worse than retraining. For intermediate values, results are mixed; therefore, testing the two approaches is necessary.
    While this transferability metric is case-specific, its first threshold is expected to hold more generally: if the direct application of a pretrained model retains any meaningful skill, even marginal, this suggests some degree of alignment between source and target tasks and makes fine-tuning a reasonable choice. Notably, when applicable, fine-tuning can help mitigate overfitting on limited datasets and improve generalization to unseen data.
    Overall, these findings represent a first step toward developing practical guidelines and a systematic approach for the efficient and rapid adaptation of pretrained NNs through transfer learning, particularly in data-limited operational settings. \\
    Finally, we carried out a feature importance analysis to assess the relative impact of input features on NN error prediction. Using permutation feature importance, a feature-importance-rank frequency map is derived.
    Atmospheric-forcing variables showed negligible importance, which is likely due to the high variability of the instantaneous snapshots used as inputs, particularly for precipitation, rather than to a lack of influence of the forcing on error growth. Their impact might increase if integrated over the forecast window or in settings where forcing biases play a more prominent role. Conversely, ice concentration and volume emerged as the most influential features, as also reported in earlier studies~\cite{Finn2023}.
    Moreover, the importance of enthalpy variables highlighted the relevance of thermodynamic state information for ML-based bias correction, particularly in regimes where forecast errors are dominated by thermodynamic processes. Their relevance in more realistic configurations, where errors are dominated by horizontal dynamics, remains to be assessed.
    
    Future work will extend this framework to more realistic applications, incorporating horizontal sea-ice dynamics and relaxing the perfect-DA assumption. We will address challenges associated with the assimilation of satellite-derived sea-ice thickness, and further explore the generalization of atmospheric-forcing inputs from reanalyses to forecasts, which can exhibit systematic and lead-time-dependent differences. Another aspect worth exploring concerns the behavior of hybrid models in fully coupled ocean–sea-ice systems; as noted by \citet{gregory2026advancing}, correcting only one component can expose the NN to out-of-sample inputs during free forecasts, since reanalysis constrains both components during training. Understanding and mitigating such mismatches will be essential for operational deployment.
    
    \FloatBarrier
    
    %------------------------------------------------------------------------
    \subsection*{ACKNOWLEDGEMENTS}
    %------------------------------------------------------------------------
    GD and AC were supported by the MELTED project ({\it Machine Learning for Arctic ice prediction project}), funded by the Italian Ministry of University and Research (MUR) through the 2022 PRIN (Progetti di Rilevante Interesse Nazionale) call, including contributions from the National Recovery and Resilience Plan (NRRP). 
    GD and AC also acknowledge support from the Scale-Aware Sea Ice Project (SASIP) funded by Schmidt Sciences, a philanthropic initiative that seeks to improve societal outcomes through the development of emerging science and technologies. CEREA is a member of Institut Pierre-Simon Laplace (IPSL).
    
    %------------------------------------------------------------------------
    \subsection*{DATA AVAILABILITY STATEMENT}
    %------------------------------------------------------------------------
    The data that support the findings of this study are available from the corresponding author upon reasonable request.
    
    %------------------------------------------------------------------------
    \subsection*{CONFLICT OF INTEREST STATEMENT}
    %------------------------------------------------------------------------
    The authors have no conflict of interest to declare.

    \bibliography{mybib.bib}

@article{driscoll2024parameter,
  title={Parameter sensitivity analysis of a sea ice melt pond parametrisation and its emulation using neural networks},
  author={Driscoll, Simon and Carrassi, Alberto and Brajard, Julien and Bertino, Laurent and Bocquet, Marc and {\'O}lason, Einar {\"O}rn},
  journal={Journal of Computational Science},
  volume={79},
  pages={102231},
  year={2024},
  publisher={Elsevier}
}

@article{hunke2013level,
  title={Level-ice melt ponds in the Los Alamos sea ice model, CICE},
  author={Hunke, Elizabeth C and Hebert, David A and Lecomte, Olivier},
  journal={Ocean Modelling},
  volume={71},
  pages={26--42},
  year={2013},
  publisher={Elsevier}
}

@article{urrego2016uncertainty,
  title={Uncertainty quantification and global sensitivity analysis of the Los Alamos sea ice model},
  author={Urrego-Blanco, Jorge R and Urban, Nathan M and Hunke, Elizabeth C and Turner, Adrian K and Jeffery, Nicole},
  journal={Journal of Geophysical Research: Oceans},
  volume={121},
  number={4},
  pages={2709--2732},
  year={2016},
  publisher={Wiley Online Library}
}

@misc{elizabeth_hunke_2024_11223920,
  author       = {Elizabeth Hunke and
                  Richard Allard and
                  David A. Bailey and
                  Philippe Blain and
                  Anthony Craig and
                  Frederic Dupont and
                  Alice DuVivier and
                  Robert Grumbine and
                  David Hebert and
                  Marika Holland and
                  Nicole Jeffery and
                  Jean-Francois Lemieux and
                  Robert Osinski and
                  Poulsen, Jacob and
                  Steketee, Anton and
                  Till Rasmussen and
                  Mads Ribergaard and
                  Roach, Lettie and
                  Andrew Roberts and
                  Matthew Turner and
                  Michael Winton and
                  Worthen, Denise},
  title        = {CICE-Consortium/CICE: CICE Version 6.5.1},
  month        = may,
  year         = 2024,
  publisher    = {Zenodo},
  version      = {6.5.1},
  doi          = {10.5281/zenodo.11223920},
  url          = {https://doi.org/10.5281/zenodo.11223920},
}

@misc{elizabeth_hunke_2025_16422921,
  author       = {Elizabeth Hunke and
                  Richard Allard and
                  David A. Bailey and
                  Philippe Blain and
                  Clemens-Sewall, David and
                  Anthony Craig and
                  Frederic Dupont and
                  Alice DuVivier and
                  Robert Grumbine and
                  David Hebert and
                  Marika Holland and
                  Nicole Jeffery and
                  Jean-Francois Lemieux and
                  Robert Osinski and
                  Till Rasmussen and
                  Mads Ribergaard and
                  Roach, Lettie and
                  Andrew Roberts and
                  Steketee, Anton and
                  Matthew Turner and
                  Michael Winton and
                  Zhao, Bin},
  title        = {CICE-Consortium/Icepack: Icepack 1.5.1},
  month        = jul,
  year         = 2025,
  publisher    = {Zenodo},
  version      = {1.5.1},
  doi          = {10.5281/zenodo.16422921},
  url          = {https://doi.org/10.5281/zenodo.16422921},
  swhid        = {swh:1:dir:9891d69e69c78e7d26eef8db839d77e73dd7f90b
                   ;origin=https://doi.org/10.5281/zenodo.1213462;vis
                   it=swh:1:snp:9b19e65d5bf4462f656be8f2a34d45d97ce31
                   bcf;anchor=swh:1:rel:cd8c9209f667b0dc7dbf38c6665d3
                   2529c71c715;path=Icepack-Icepack1.5.1
                  },
}

@article{Sakov2012,
  title={TOPAZ4: an ocean-sea ice data assimilation system for the North Atlantic and Arctic},
  author={Sakov, Pavel and Counillon, F and Bertino, L and Lis{\ae}ter, KA and Oke, PR and Korablev, A},
  journal={Ocean Science},
  volume={8},
  number={4},
  pages={633--656},
  year={2012},
  publisher={Copernicus Publications G{\"o}ttingen, Germany}
}

@article{Cheng2020,
   author = {Sukun Cheng and Ali Aydoğdu and Pierre Rampal and Alberto Carrassi and Laurent Bertino},
   doi = {10.3390/oceans1040022},
   issn = {26731924},
   issue = {4},
   journal = {Oceans},
   pages = {326-342},
   title = {Probabilistic Forecasts of Sea Ice Trajectories in the Arctic: Impact of Uncertainties in Surface Wind and Ice Cohesion},
   volume = {1},
   year = {2020},
}

@article{brajard2021combining,
  title={Combining data assimilation and machine learning to infer unresolved scale parametrization},
  author={Brajard, Julien and Carrassi, Alberto and Bocquet, Marc and Bertino, Laurent},
  journal={Philosophical Transactions of the Royal Society A},
  volume={379},
  number={2194},
  pages={20200086},
  year={2021},
  publisher={The Royal Society Publishing}
}

@article{Farchi2021comparison,
   author = {Alban Farchi and Marc Bocquet and Patrick Laloyaux and Massimo Bonavita and Quentin Malartic},
   doi = {10.1016/j.jocs.2021.101468},
   issn = {18777503},
   issue = {July},
   journal = {Journal of Computational Science},
   pages = {101468},
   publisher = {Elsevier B.V.},
   title = {A comparison of combined data assimilation and machine learning methods for offline and online model error correction},
   volume = {55},
   url = {https://doi.org/10.1016/j.jocs.2021.101468},
   year = {2021},
}

@article{Farchi2021,
   author = {Alban Farchi and Patrick Laloyaux and Massimo Bonavita and Marc Bocquet},
   doi = {10.1002/qj.4116},
   issn = {1477870X},
   issue = {739},
   journal = {Quarterly Journal of the Royal Meteorological Society},
   pages = {3067-3084},
   title = {Using machine learning to correct model error in data assimilation and forecast applications},
   volume = {147},
   year = {2021},
}

@article{Finn2023,
   author = {Tobias Sebastian Finn and Charlotte Durand and Alban Farchi and Marc Bocquet and Yumeng Chen and Alberto Carrassi and Véronique Dansereau},
   doi = {10.5194/tc-17-2965-2023},
   issn = {19940424},
   issue = {7},
   journal = {Cryosphere},
   pages = {2965-2991},
   title = {Deep learning subgrid-scale parametrisations for short-term forecasting of sea-ice dynamics with a Maxwell elasto-brittle rheology},
   volume = {17},
   year = {2023},
}

@article{Gregory2023,
   author = {William Gregory and Mitchell Bushuk and Alistair Adcroft and Yongfei Zhang and Laure Zanna},
   doi = {10.1029/2023MS003757},
   issn = {19422466},
   issue = {10},
   journal = {Journal of Advances in Modeling Earth Systems},
   title = {Deep Learning of Systematic Sea Ice Model Errors From Data Assimilation Increments},
   volume = {15},
   year = {2023},
}

@article{Gregory2024,
   author = {William Gregory and Mitchell Bushuk and Yongfei Zhang and Alistair Adcroft and Laure Zanna},
   doi = {10.1029/2023GL106776},
   issn = {19448007},
   issue = {3},
   journal = {Geophysical Research Letters},
   title = {Machine Learning for Online Sea Ice Bias Correction Within Global Ice-Ocean Simulations},
   volume = {51},
   year = {2024},
}

@article{Watt-Meyer2021,
   author = {Oliver Watt-Meyer and Noah D. Brenowitz and Spencer K. Clark and Brian Henn and Anna Kwa and Jeremy McGibbon and W. Andre Perkins and Christopher S. Bretherton},
   doi = {10.1029/2021GL092555},
   issn = {19448007},
   issue = {15},
   journal = {Geophysical Research Letters},
   month = {8},
   publisher = {John Wiley and Sons Inc},
   title = {Correcting Weather and Climate Models by Machine Learning Nudged Historical Simulations},
   volume = {48},
   year = {2021},
}

@article{Watson2019,
   author = {Peter A.G. Watson},
   doi = {10.1029/2018MS001597},
   issn = {19422466},
   issue = {5},
   journal = {Journal of Advances in Modeling Earth Systems},
   month = {5},
   pages = {1402-1417},
   publisher = {Blackwell Publishing Ltd},
   title = {Applying Machine Learning to Improve Simulations of a Chaotic Dynamical System Using Empirical Error Correction},
   volume = {11},
   year = {2019},
}

@article{Bolton2019,
   author = {Thomas Bolton and Laure Zanna},
   doi = {10.1029/2018MS001472},
   issn = {19422466},
   issue = {1},
   journal = {Journal of Advances in Modeling Earth Systems},
   month = {1},
   pages = {376-399},
   publisher = {Blackwell Publishing Ltd},
   title = {Applications of Deep Learning to Ocean Data Inference and Subgrid Parameterization},
   volume = {11},
   year = {2019},
}

@article{krizhevsky2012imagenet,
  title={Imagenet classification with deep convolutional neural networks},
  author={Krizhevsky, Alex and Sutskever, Ilya and Hinton, Geoffrey E},
  journal={Advances in neural information processing systems},
  volume={25},
  year={2012}
}

@article{pathak2022fourcastnet,
  title={Fourcastnet: A global data-driven high-resolution weather model using adaptive fourier neural operators},
  author={Pathak, Jaideep and Subramanian, Shashank and Harrington, Peter and Raja, Sanjeev and Chattopadhyay, Ashesh and Mardani, Morteza and Kurth, Thorsten and Hall, David and Li, Zongyi and Azizzadenesheli, Kamyar and others},
  journal={arXiv preprint arXiv:2202.11214},
  year={2022}
}

@article{lam2023learning-graphcast,
  title={Learning skillful medium-range global weather forecasting},
  author={Lam, Remi and Sanchez-Gonzalez, Alvaro and Willson, Matthew and Wirnsberger, Peter and Fortunato, Meire and Alet, Ferran and Ravuri, Suman and Ewalds, Timo and Eaton-Rosen, Zach and Hu, Weihua and others},
  journal={Science},
  volume={382},
  number={6677},
  pages={1416--1421},
  year={2023},
  publisher={American Association for the Advancement of Science}
}

@article{bi2023accurate-pangu,
  title={Accurate medium-range global weather forecasting with 3D neural networks},
  author={Bi, Kaifeng and Xie, Lingxi and Zhang, Hengheng and Chen, Xin and Gu, Xiaotao and Tian, Qi},
  journal={Nature},
  volume={619},
  number={7970},
  pages={533--538},
  year={2023},
  publisher={Nature Publishing Group UK London}
}

@article{chen2025operational-fengwu,
  title={The operational medium-range deterministic weather forecasting can be extended beyond a 10-day lead time},
  author={Chen, Kang and Han, Tao and Ling, Fenghua and Gong, Junchao and Bai, Lei and Wang, Xinyu and Luo, Jing-Jia and Fei, Ben and Zhang, Wenlong and Chen, Xi and others},
  journal={Communications Earth \& Environment},
  volume={6},
  number={1},
  pages={518},
  year={2025},
  publisher={Nature Publishing Group UK London}
}

@article{Bonavita2024,
   author = {Massimo Bonavita},
   doi = {10.1029/2023GL107377},
   issn = {19448007},
   issue = {12},
   journal = {Geophysical Research Letters},
   month = {6},
   publisher = {John Wiley and Sons Inc},
   title = {On Some Limitations of Current Machine Learning Weather Prediction Models},
   volume = {51},
   year = {2024},
}

@article{hersbach2020era5,
  title={The ERA5 global reanalysis},
  author={Hersbach, Hans and Bell, Bill and Berrisford, Paul and Hirahara, Shoji and Hor{\'a}nyi, Andr{\'a}s and Mu{\~n}oz-Sabater, Joaqu{\'\i}n and Nicolas, Julien and Peubey, Carole and Radu, Raluca and Schepers, Dinand and others},
  journal={Quarterly journal of the royal meteorological society},
  volume={146},
  number={730},
  pages={1999--2049},
  year={2020},
  publisher={Wiley Online Library}
}

@article{chevallier1998neural,
  title={A neural network approach for a fast and accurate computation of a longwave radiative budget},
  author={Chevallier, F and Ch{\'e}ruy, F and Scott, NA and Ch{\'e}din, A},
  journal={Journal of applied meteorology},
  volume={37},
  number={11},
  pages={1385--1397},
  year={1998}
}

@article{Horvat2022,
   author = {Christopher Horvat and Lettie A. Roach},
   doi = {10.5194/gmd-15-803-2022},
   issn = {19919603},
   issue = {2},
   journal = {Geoscientific Model Development},
   month = {1},
   pages = {803-814},
   publisher = {Copernicus GmbH},
   title = {WIFF1.0: A hybrid machine-learning-based parameterization of wave-induced sea ice floe fracture},
   volume = {15},
   year = {2022},
}

@article{Carrassi2018,
   author = {Alberto Carrassi and Marc Bocquet and Laurent Bertino and Geir Evensen},
   doi = {10.1002/wcc.535},
   issn = {17577799},
   issue = {5},
   journal = {Wiley Interdisciplinary Reviews: Climate Change},
   pages = {1-79},
   title = {Data assimilation in the geosciences: An overview of methods, issues, and perspectives},
   volume = {9},
   year = {2018},
}

@article{Lipscomb2001,
   author = {William H Lipscomb},
   doi = {https://doi.org/10.1029/2000JC000518},
   issue = {C7},
   journal = {Journal of Geophysical research},
   pages = {13989-14000},
   title = {Remapping the thickness distribution in sea ice models},
   volume = {106},
   url = {https://agupubs.onlinelibrary.wiley.com/doi/abs/10.1029/2000JC000518},
   year = {2001},
}

@article{thorndike1975thickness,
  title={The thickness distribution of sea ice},
  author={Thorndike, Alan S and Rothrock, Drew A and Maykut, Gary A and Colony, Roger},
  journal={Journal of Geophysical Research},
  volume={80},
  number={33},
  pages={4501--4513},
  year={1975},
  publisher={Wiley Online Library}
}

@article{turner2013two,
  title={Two modes of sea-ice gravity drainage: A parameterization for large-scale modeling},
  author={Turner, Adrian K and Hunke, Elizabeth C and Bitz, Cecilia M},
  journal={Journal of Geophysical Research: Oceans},
  volume={118},
  number={5},
  pages={2279--2294},
  year={2013},
  publisher={Wiley Online Library}
}

@article{He2025,
   author = {Zikang He and Yiguo Wang and Julien Brajard and Xidong Wang and Zheqi Shen},
   doi = {10.5194/tc-19-3279-2025},
   issn = {1994-0424},
   issue = {8},
   journal = {The Cryosphere},
   month = {8},
   pages = {3279-3293},
   title = {Correcting errors in seasonal Arctic sea ice prediction of Earth system models with machine learning},
   volume = {19},
   url = {https://tc.copernicus.org/articles/19/3279/2025/},
   year = {2025},
}

@misc{kingma2017adammethodstochasticoptimization,
      title={Adam: A Method for Stochastic Optimization}, 
      author={Diederik P. Kingma and Jimmy Ba},
      year={2017},
      eprint={1412.6980},
      archivePrefix={arXiv},
      primaryClass={cs.LG},
      url={https://arxiv.org/abs/1412.6980}, 
}

@article{finn2024generative,
  title={Generative diffusion for regional surrogate models from sea-ice simulations},
  author={Finn, Tobias Sebastian and Durand, Charlotte and Farchi, Alban and Bocquet, Marc and Rampal, Pierre and Carrassi, Alberto},
  journal={Journal of Advances in Modeling Earth Systems},
  volume={16},
  number={10},
  pages={e2024MS004395},
  year={2024},
  publisher={Wiley Online Library}
}

@article{breiman2001random,
  title={Random forests},
  author={Breiman, Leo},
  journal={Machine learning},
  volume={45},
  number={1},
  pages={5--32},
  year={2001},
  publisher={Springer}
}

@article{carrassi2011treatment,
  title={Treatment of the error due to unresolved scales in sequential data assimilation},
  author={Carrassi, Alberto and Vannitsem, St{\'e}phane},
  journal={International Journal of Bifurcation and Chaos},
  volume={21},
  number={12},
  pages={3619--3626},
  year={2011},
  publisher={World Scientific}
}

@article{bodnar2025foundation,
  title={A foundation model for the Earth system},
  author={Bodnar, Cristian and Bruinsma, Wessel P and Lucic, Ana and Stanley, Megan and Allen, Anna and Brandstetter, Johannes and Garvan, Patrick and Riechert, Maik and Weyn, Jonathan A and Dong, Haiyu and others},
  journal={Nature},
  pages={1--8},
  year={2025},
  publisher={Nature Publishing Group UK London}
}

@article{rasp2018deep,
  title={Deep learning to represent subgrid processes in climate models},
  author={Rasp, Stephan and Pritchard, Michael S and Gentine, Pierre},
  journal={Proceedings of the national academy of sciences},
  volume={115},
  number={39},
  pages={9684--9689},
  year={2018},
  publisher={National Academy of Sciences}
}

@article{rasp2024weatherbench,
  title={WeatherBench 2: A benchmark for the next generation of data-driven global weather models},
  author={Rasp, Stephan and Hoyer, Stephan and Merose, Alexander and Langmore, Ian and Battaglia, Peter and Russell, Tyler and Sanchez-Gonzalez, Alvaro and Yang, Vivian and Carver, Rob and Agrawal, Shreya and others},
  journal={Journal of Advances in Modeling Earth Systems},
  volume={16},
  number={6},
  pages={e2023MS004019},
  year={2024},
  publisher={Wiley Online Library}
}

@article{pasmans2025ensemble,
  title={Ensemble Kalman filter in latent space using a variational autoencoder pair},
  author={Pasmans, Ivo and Chen, Yumeng and Finn, Tobias Sebastian and Bocquet, Marc and Carrassi, Alberto},
  journal={arXiv preprint arXiv:2502.12987},
  year={2025}
}

@article{iman2023review,
  title={A review of deep transfer learning and recent advancements},
  author={Iman, Mohammadreza and Arabnia, Hamid Reza and Rasheed, Khaled},
  journal={Technologies},
  volume={11},
  number={2},
  pages={40},
  year={2023},
  publisher={MDPI}
}

@article{ogorman2018,
  title={Using Machine Learning to Parameterize 1012 Moist Convection: Potential for Modeling of Climate, Climate Change, and 1013 Extreme Events},
  author={O’Gorman, PA and Dwyer, JG},
  journal={Journal of Advances in Modeling Earth Systems},
  volume={10},
  number={10},
  pages={1014},
  year={2018}
}

@article{du2025reducing,
  title={Reducing Model Biases with Machine Learning Corrections Derived from Ocean Data Assimilation Increments},
  author={Du, Danni and Lu, Feiyu and Adcroft, Alistair},
  journal={Authorea Preprints},
  year={2025},
  publisher={Authorea}
}

@article{ghil_sciamarella-2023,
AUTHOR = {Ghil, M. and Sciamarella, D.},
TITLE = {Review article: Dynamical systems, algebraic topology and the climate sciences},
JOURNAL = {Nonlinear Processes in Geophysics},
VOLUME = {30},
YEAR = {2023},
NUMBER = {4},
PAGES = {399--434},
URL = {https://npg.copernicus.org/articles/30/399/2023/},
DOI = {10.5194/npg-30-399-2023}
}

@article{scikit-learn,
  title={Scikit-learn: Machine Learning in {P}ython},
  author={Pedregosa, F. and Varoquaux, G. and Gramfort, A. and Michel, V.
          and Thirion, B. and Grisel, O. and Blondel, M. and Prettenhofer, P.
          and Weiss, R. and Dubourg, V. and Vanderplas, J. and Passos, A. and
          Cournapeau, D. and Brucher, M. and Perrot, M. and Duchesnay, E.},
  journal={Journal of Machine Learning Research},
  volume={12},
  pages={2825--2830},
  year={2011}
}

@ARTICLE{2020SciPy-NMeth,
  author  = {Virtanen, Pauli and Gommers, Ralf and Oliphant, Travis E. and
            Haberland, Matt and Reddy, Tyler and Cournapeau, David and
            Burovski, Evgeni and Peterson, Pearu and Weckesser, Warren and
            Bright, Jonathan and {van der Walt}, St{\'e}fan J. and
            Brett, Matthew and Wilson, Joshua and Millman, K. Jarrod and
            Mayorov, Nikolay and Nelson, Andrew R. J. and Jones, Eric and
            Kern, Robert and Larson, Eric and Carey, C J and
            Polat, {\.I}lhan and Feng, Yu and Moore, Eric W. and
            {VanderPlas}, Jake and Laxalde, Denis and Perktold, Josef and
            Cimrman, Robert and Henriksen, Ian and Quintero, E. A. and
            Harris, Charles R. and Archibald, Anne M. and
            Ribeiro, Ant{\^o}nio H. and Pedregosa, Fabian and
            {van Mulbregt}, Paul and {SciPy 1.0 Contributors}},
  title   = {{{SciPy} 1.0: Fundamental Algorithms for Scientific
            Computing in Python}},
  journal = {Nature Methods},
  year    = {2020},
  volume  = {17},
  pages   = {261--272},
  adsurl  = {https://rdcu.be/b08Wh},
  doi     = {10.1038/s41592-019-0686-2},
}

@article{gregory2026advancing,
    author = {William Gregory  and Mitchell Bushuk  and Yong-Fei Zhang  and Alistair Adcroft  and Laure Zanna  and Colleen McHugh  and Liwei Jia },
    title = {Advancing global sea ice prediction capabilities using a fully coupled climate model with integrated machine learning},
    journal = {Science Advances},
    volume = {12},
    number = {1},
    pages = {eady8957},
    year = {2026},
    doi = {10.1126/sciadv.ady8957},
    URL = {https://www.science.org/doi/abs/10.1126/sciadv.ady8957},
    eprint = {https://www.science.org/doi/pdf/10.1126/sciadv.ady8957},
}

@article{cipollone_bivariate_2023,
	title = {Bivariate sea-ice assimilation for global-ocean analysis-reanalysis},
	volume = {19},
	issn = {18120792},
	doi = {10.5194/os-19-1375-2023},
	number = {5},
	journal = {Ocean Science},
	author = {Cipollone, Andrea and Banerjee, Deep Sankar and Iovino, Doroteaciro and Aydogdu, Ali and Masina, Simona},
	month = sep,
	year = {2023},
	note = {Publisher: Copernicus Publications},
	pages = {1375--1392}
}

@article{zhang2021assimilation,
  title={Assimilation of satellite-retrieved sea ice concentration and prospects for September predictions of Arctic sea ice},
  author={Zhang, Yong-Fei and Bushuk, Mitchell and Winton, Michael and Hurlin, Bill and Yang, Xiaosong and Delworth, Tom and Jia, Liwei},
  journal={Journal of Climate},
  volume={34},
  number={6},
  pages={2107--2126},
  year={2021}
}

@article{carrassi_model_2008,
	title = {Model error and sequential data assimilation: {A} deterministic formulation},
	volume = {134},
	issn = {0035-9009, 1477-870X},
	shorttitle = {Model error and sequential data assimilation},
	url = {https://rmets.onlinelibrary.wiley.com/doi/10.1002/qj.284},
	doi = {10.1002/qj.284},
	language = {en},
	number = {634},
	urldate = {2025-12-12},
	journal = {Quarterly Journal of the Royal Meteorological Society},
	author = {Carrassi, A. and Vannitsem, S. and Nicolis, C.},
	month = jul,
	year = {2008},
	pages = {1297--1313},
}

@article{farchi2025,
title = "Development of an offline and online hybrid model for the Integrated Forecasting System",
author = "Farchi, A. and Chrust, M. and Bocquet, M. and Bonavita, M.",
volume = "151",
journal = "Q. J. R. Meteorol. Soc.",
pages = "e4934",
year = "2025",
doi = "10.1002/qj.4934"
}

@article{durand2025,
author = "Durand, C. and Finn, T. S. and Farchi, A. and Bocquet, M. and Brajard, J. and Bertino, L.",
title = "Four-dimensional variational data assimilation with a sea-ice thickness emulator",
journal = "The Cryosphere",
voume = "19",
year = "2025",
pages = "5613--5637",
doi = "10.5194/tc-19-5613-2025"
}

@article{chapman2025improving,
  title={Improving climate bias and variability via CNN-based state-dependent model-error corrections},
  author={Chapman, William E and Berner, Judith},
  journal={Geophysical Research Letters},
  volume={52},
  number={6},
  pages={e2024GL114106},
  year={2025},
  publisher={Wiley Online Library}
}

@article{bocquet2020_fds,
title = {Bayesian inference of chaotic dynamics by merging data assimilation, machine learning and expectation-maximization},
journal = {Foundations of Data Science},
volume = {2},
number = {1},
pages = {55-80},
year = {2020},
issn = {},
doi = {10.3934/fods.2020004},
author = {Marc Bocquet and Julien Brajard and Alberto Carrassi and Laurent Bertino},
}

@article{bocquet2023surrogate,
  title={Surrogate modeling for the climate sciences dynamics with machine learning and data assimilation},
  author={Bocquet, Marc},
  journal={Frontiers in Applied Mathematics and Statistics},
  volume={9},
  pages={1133226},
  year={2023},
  publisher={Frontiers Media SA}
}

@misc{li2020asha,
      title={A System for Massively Parallel Hyperparameter Tuning}, 
      author={Liam Li and Kevin Jamieson and Afshin Rostamizadeh and Ekaterina Gonina and Moritz Hardt and Benjamin Recht and Ameet Talwalkar},
      year={2020},
      eprint={1810.05934},
      archivePrefix={arXiv},
      primaryClass={cs.LG},
      url={https://arxiv.org/abs/1810.05934}, 
}

@article{kochkov2024neural,
  title={Neural general circulation models for weather and climate},
  author={Kochkov, Dmitrii and Yuval, Janni and Langmore, Ian and Norgaard, Peter and Smith, Jamie and Mooers, Griffin and Kl{\"o}wer, Milan and Lottes, James and Rasp, Stephan and D{\"u}ben, Peter and others},
  journal={Nature},
  volume={632},
  number={8027},
  pages={1060--1066},
  year={2024},
  publisher={Nature Publishing Group UK London}
}

@article{peng2024hybrid,
  title={A hybrid deep learning and data assimilation method for model error estimation},
  author={Peng, Ziyi and Lei, Lili and Tan, Zhe-Min},
  journal={Science China Earth Sciences},
  volume={67},
  number={12},
  pages={3655--3670},
  year={2024},
  publisher={Springer}
}

%%%% SUPPORTING INFORMATION
    \clearpage
    \appendix
    
    % Reset counters
    \setcounter{section}{0}
    \setcounter{figure}{0}
    \setcounter{table}{0}
    \setcounter{equation}{0}
    
    % Prefix counters with S
    \renewcommand{\thesection}{S\arabic{section}}
    \renewcommand{\thefigure}{S\arabic{figure}}
    \renewcommand{\thetable}{S\arabic{table}}
    \renewcommand{\theequation}{S\arabic{equation}}
    
    \begin{center}
      {\LARGE\bfseries Supporting Information}
    \end{center}
    \vspace{1em}

    \begin{figure}[H]
        \centering
        \begin{subfigure}{\textwidth}
        \centering
        \includegraphics[scale=0.6]{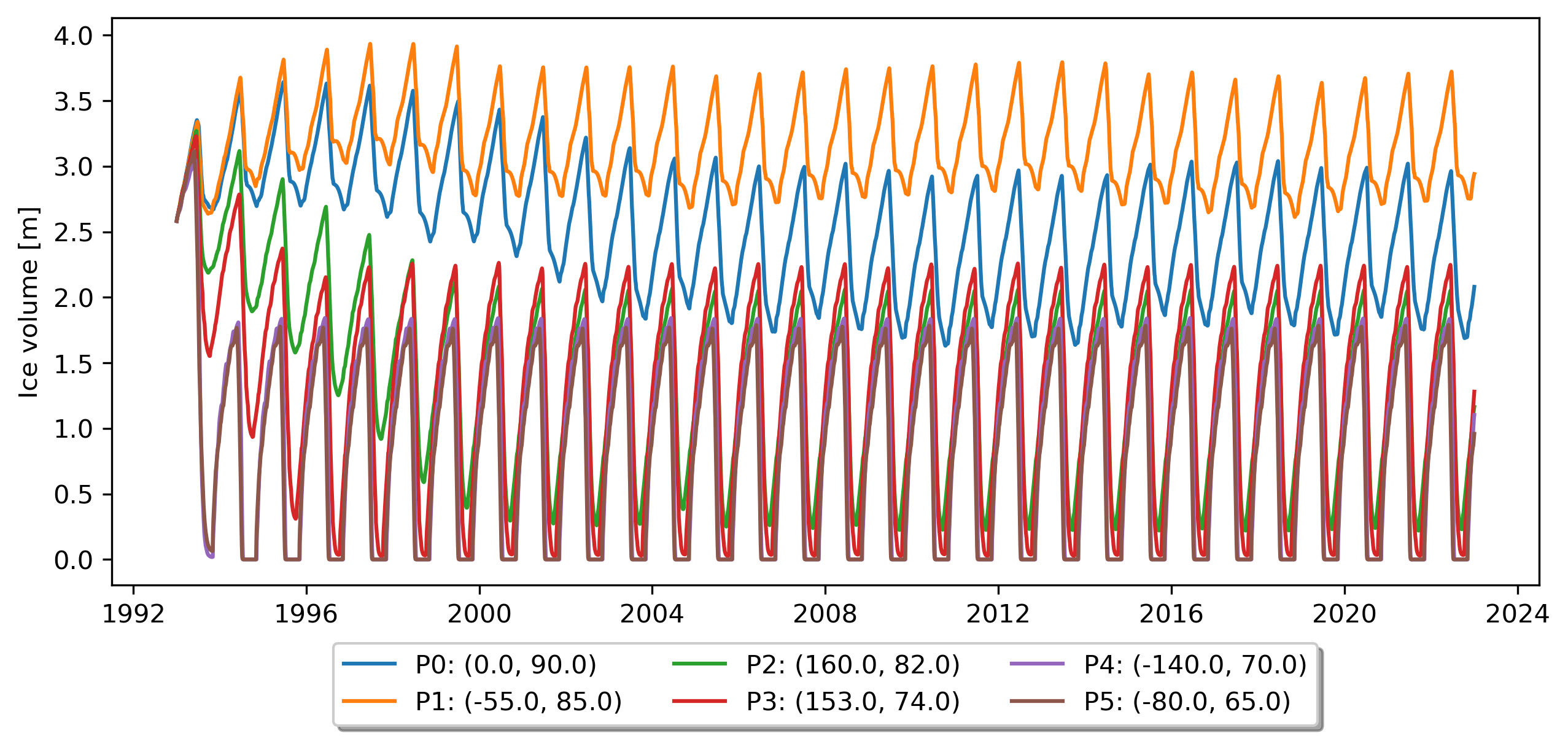}
        \caption{}  
        \label{fig:long_runs}
        \end{subfigure}
        \begin{subfigure}{\textwidth}
        \centering
        \includegraphics[scale=0.6]{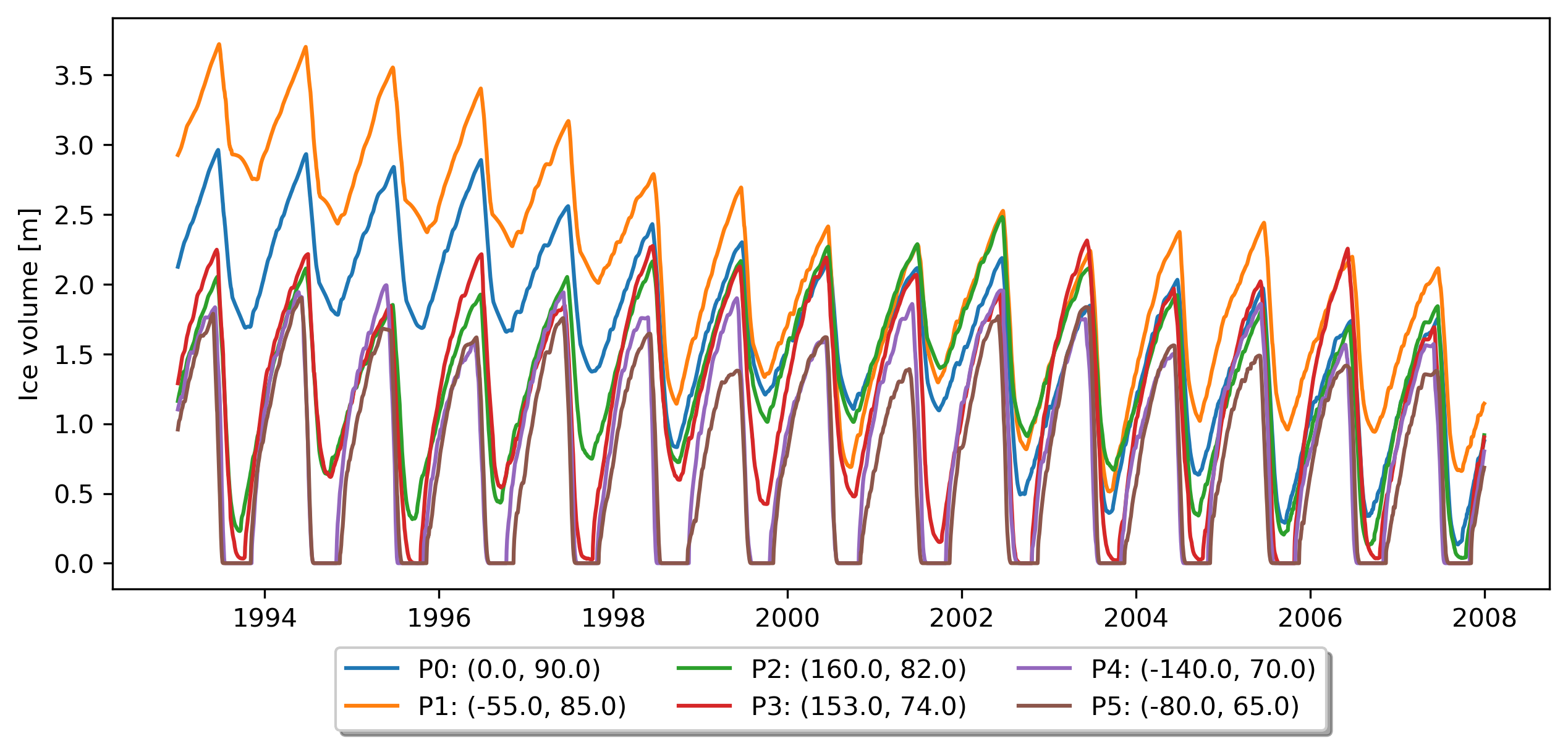}
        \caption{}    
        \label{fig:long_forecasts}
        \end{subfigure}
        \caption{(a) Sea ice volume time series at the six selected locations, obtained from the ‘truth’ reference model forced cyclically with atmospheric conditions from 1993. (b) Same as (a), but with simulations initialized from the final yearly cycle of (a) and driven by time-varying atmospheric forcings.}
    \end{figure}

    \section{Parametric error metrics}
        \label{sec:param_err}
\begin{figure}
            \centering
            \begin{subfigure}
                {0.31\textwidth}
                \includegraphics[
                    height=7.2cm,
                    trim={0.25cm 0cm 0.25cm 0cm},
                    clip
                ]{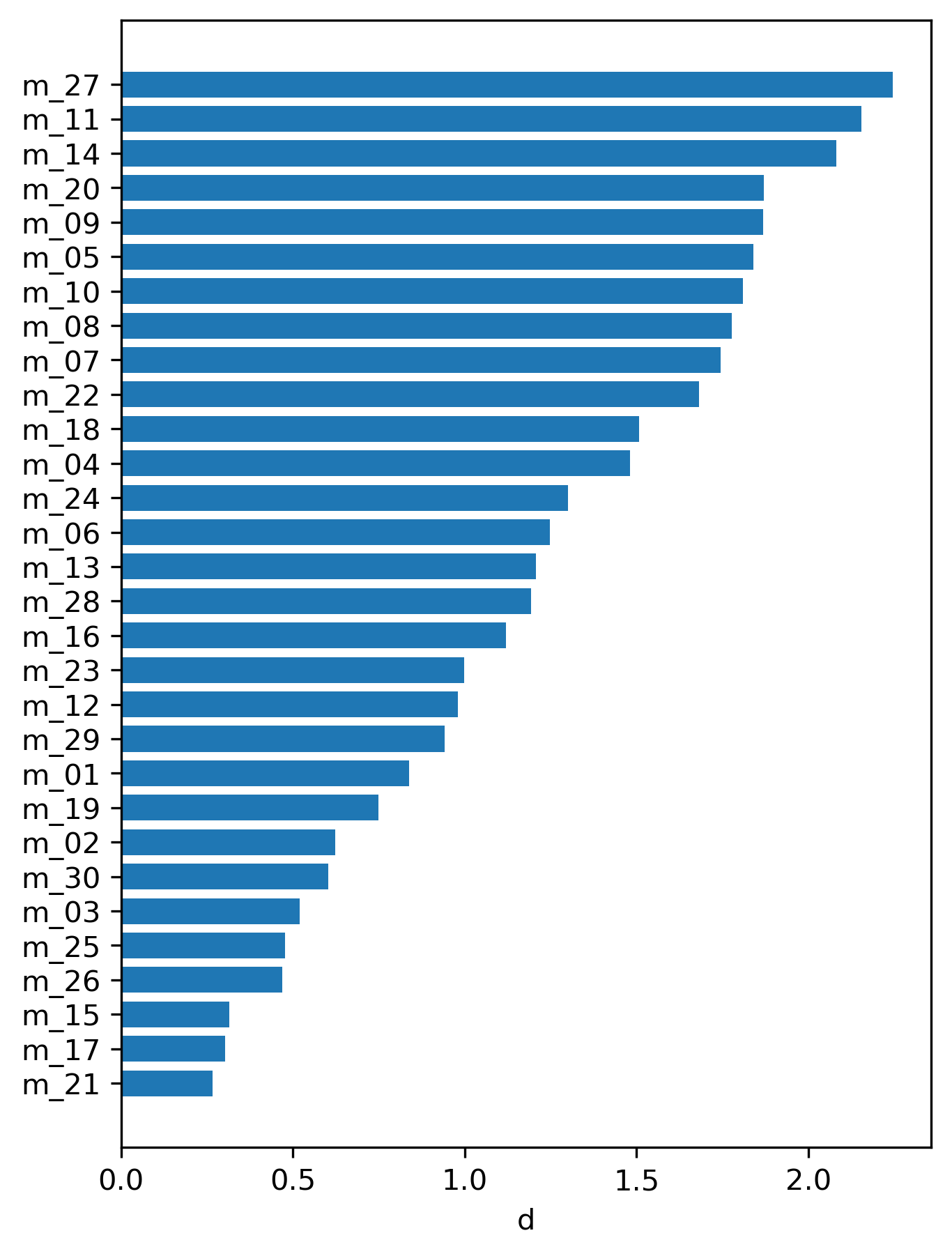}
                \caption{}
            \end{subfigure}
            \begin{subfigure}
                {0.68\textwidth} \raggedleft
                \includegraphics[height=7.2cm, trim={0cm 0cm 0.3cm 0cm}, clip]{
                    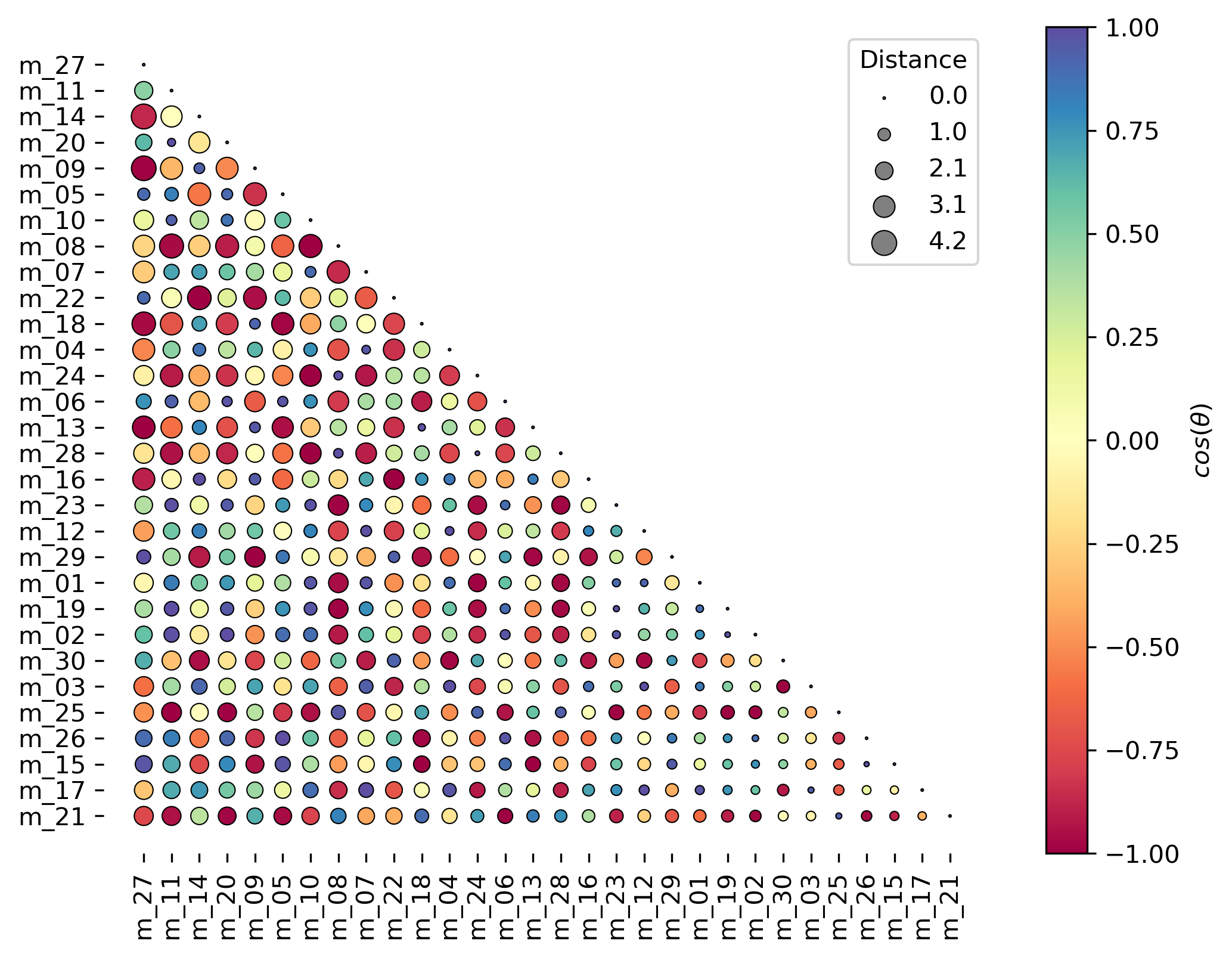
                }
                \caption{}
            \end{subfigure}
            \caption{PC2: (a) Distances computed according to \eqref{eqn:confdist}
            between each member in PC2 and reference configuration. (b) Marker size
            represent the distance between configuration pairs, whereas coloring
            represents the cosine of the angle between configuration vector
            pairs according to \eqref{eqn:confangle} }
            \label{fig:summ_PC2}
        \end{figure}

        Perturbed configurations are characterized by the Mahalanobis
        distance from the reference configuration and between each other. Such
        distance $d$ between two generic configurations $\bm{c}_{i}, \bm{c}_{j}$,
        is defined as follows:
        \begin{subequations}
            \begin{eqnarray}
                d(\bm{c}_i, \bm{c}_j) = \|\left( \bm{c}_i - \bm{c}_j\right)\|_{\bm{P}}
                \label{eqn:confdist}
            \end{eqnarray}
        \end{subequations}
        where $\bm{c}_{i} = (p_{1i}, p_{2i}, \dots, p_{Ni})$ corresponds to a
        point in an N-dimensional parameter subspace, representing the i-th parameters'
        configuration and $\bm{P}= \bm{S}^{-1}$ the precision matrix. Moreover,
        to evaluate the similarity between configurations, it is useful to evaluate
        the angle cosine $\cos{\!(\theta)}$ between configuration position vectors
        in the metric space defined by the Mahalanobis distance, whose origin is
        represented by the reference configuration, as shown in Equation~\eqref{eqn:confangle}.
        \begin{subequations}
            \begin{eqnarray}
                \bm{e}_k = \bm{c}_k - \bm{c}_0 \\
                cos(\theta)_{i,j} =
                \frac{\langle \bm{e}_{i}, \bm{e}_{j} \rangle_{\bm{P}}}{\| \bm{e}_{i}
                \|_{\bm{P}}\| \bm{e}_{j} \|_{\bm{P}}}.\label{eqn:confangle}
            \end{eqnarray}
        \end{subequations}
        The metrics defined in Equations~\eqref{eqn:confdist} and \eqref{eqn:confangle}
        are computed for each member relative to the reference configuration, as
        well as to any other member. The results, depicted in Figure~\ref{fig:summ_PC2}
        offer an overview of parameters' perturbations magnitude and similarities
        for the configurations generated.
    
    \section{Dependence of Icepack forecast errors on lead time and start date} \label{sec:lead-sdata_errors}
\begin{figure}
        \centering
        \begin{subfigure}
            {0.49\textwidth}            \captionsetup{position=above,justification=raggedright, skip=2pt, margin=20pt}
            \caption{P0}
            \includegraphics[width=\textwidth, trim={0cm 0.2cm 0 0.24cm}, clip]{
                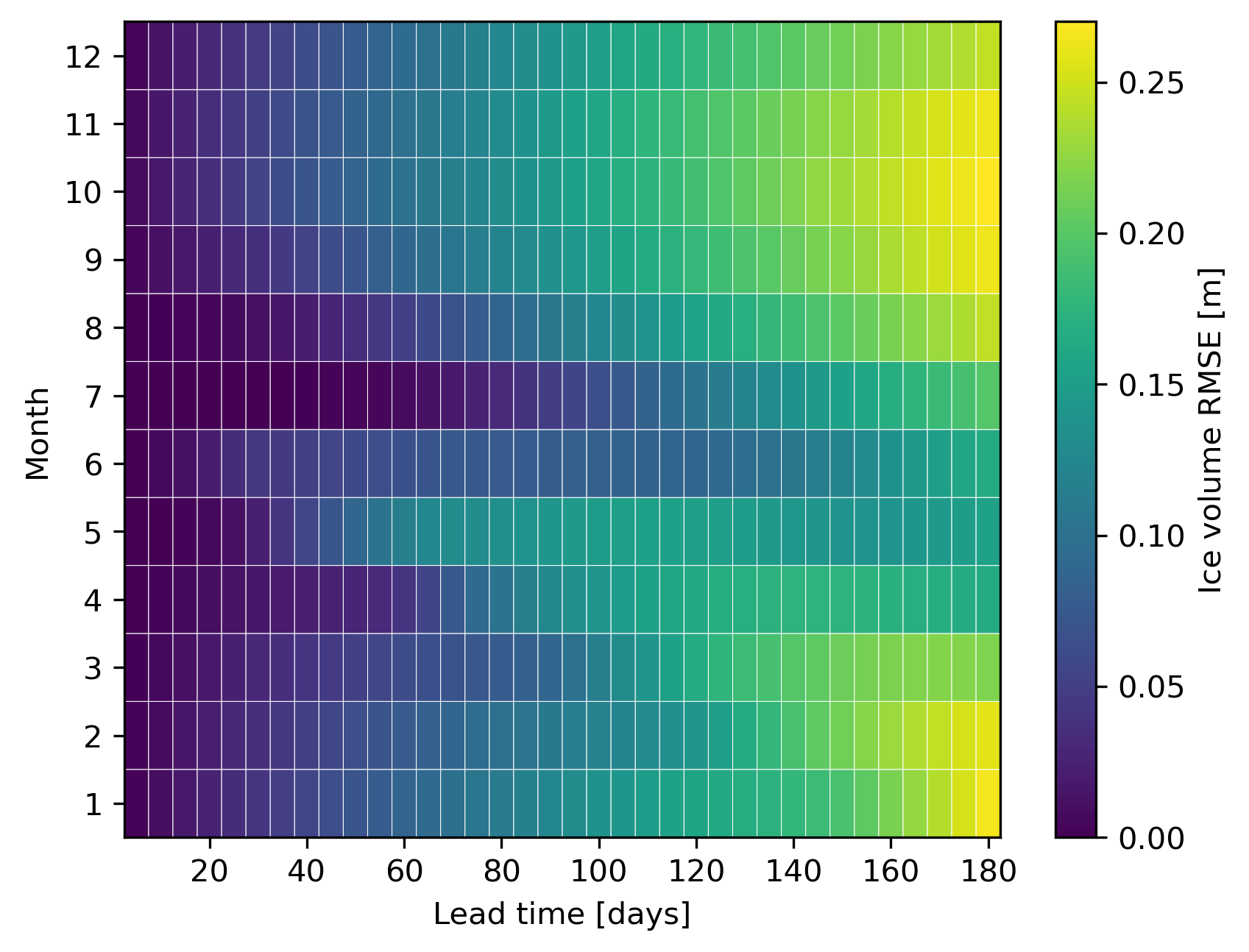
            }
        \end{subfigure}
        \begin{subfigure}
            {0.49\textwidth}            \captionsetup{position=above,justification=raggedright, skip=2pt, margin=20pt}
            \caption{P3}
            \includegraphics[width=\textwidth, trim={0cm 0.2cm 0 0.24cm}, clip]{
                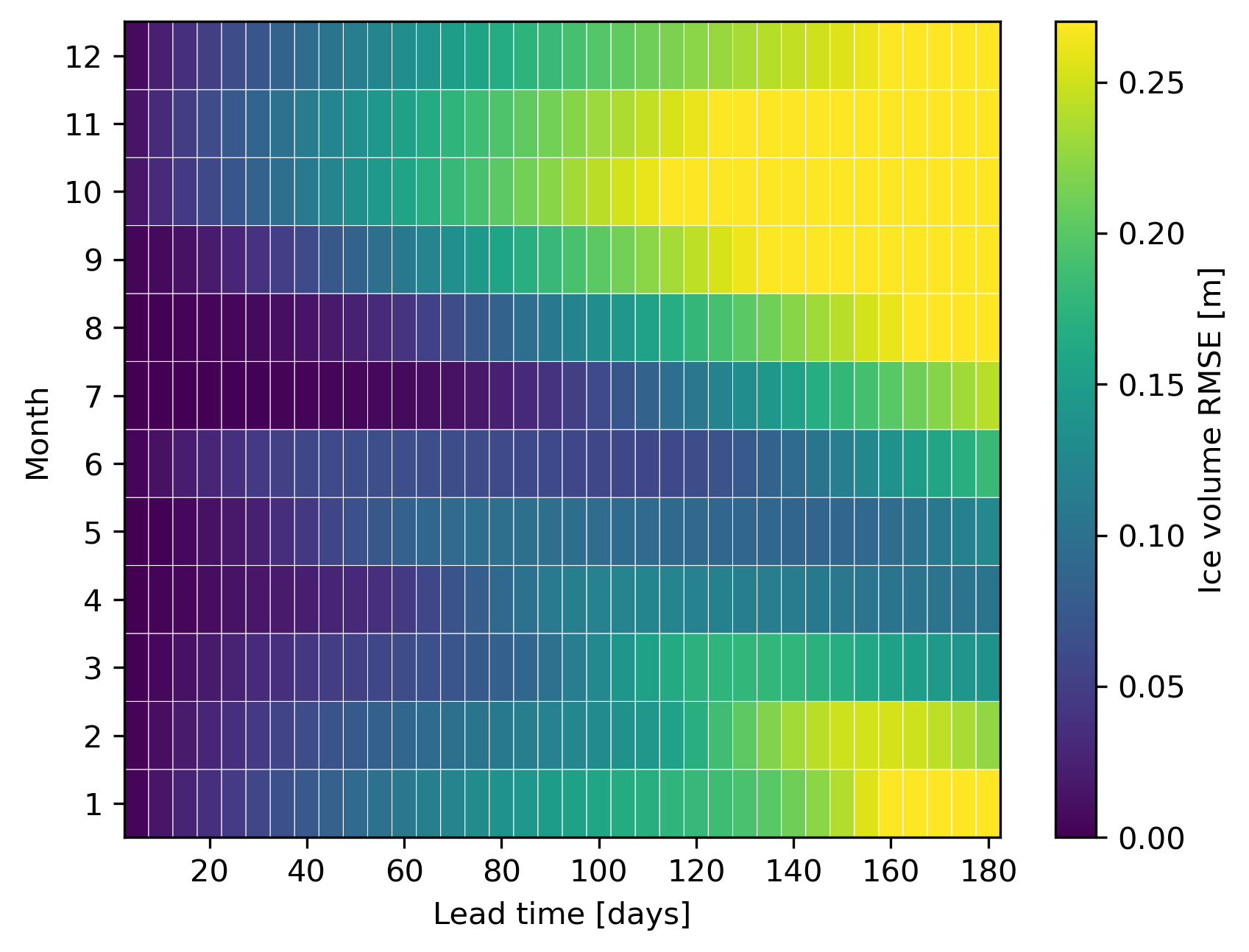
            }
        \end{subfigure}\\
        \vspace{0.1cm}
        \begin{subfigure}
            {0.49\textwidth}
            \captionsetup{position=above,justification=raggedright, skip=2pt, margin=20pt}
            \caption{P1}
            \includegraphics[width=\textwidth, trim={0cm 0.2cm 0 0.24cm}, clip]{
                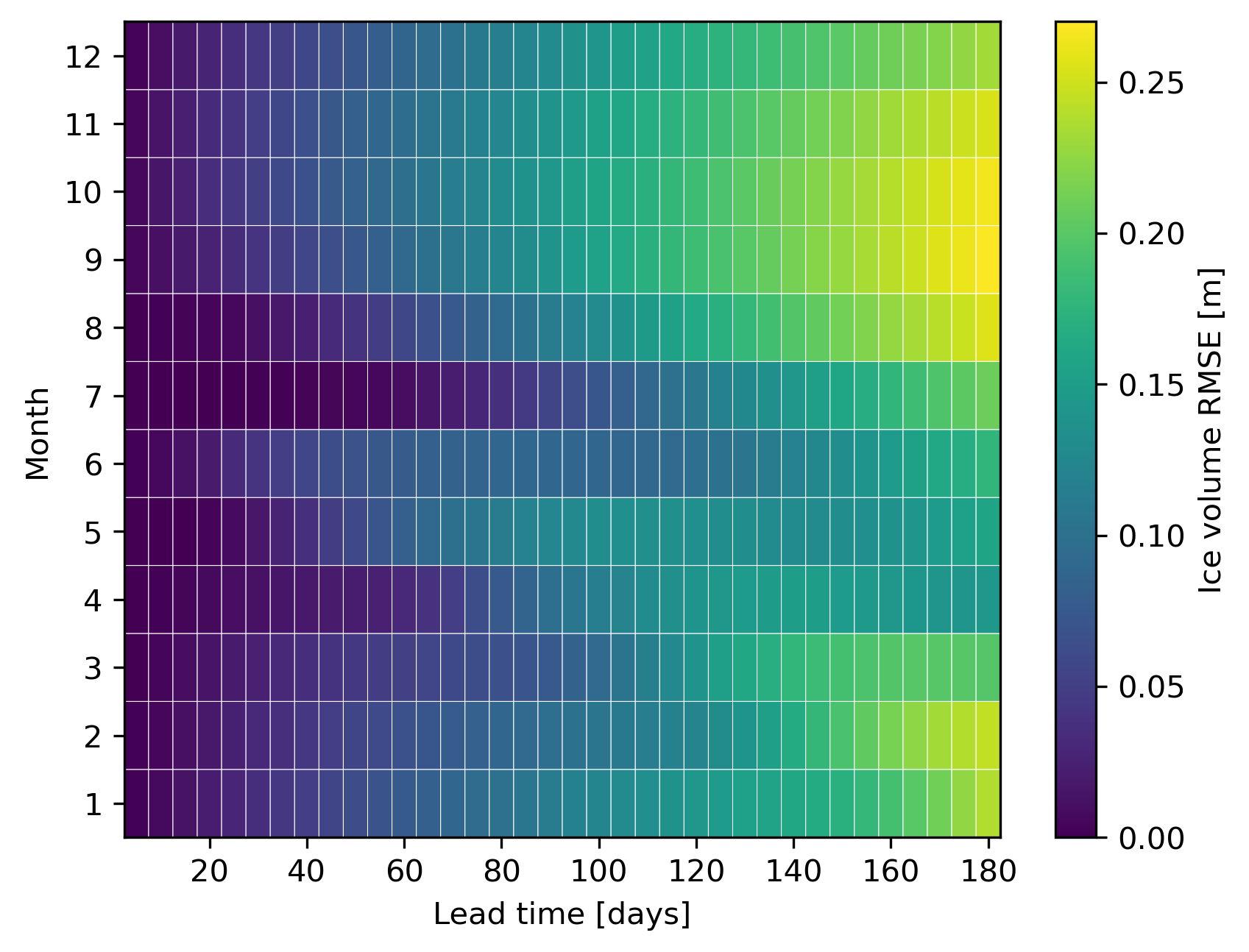
            }
        \end{subfigure}
        \begin{subfigure}
        {0.49\textwidth}
        \captionsetup{position=above,justification=raggedright, skip=2pt, margin=20pt}
        \caption{P4}
        \includegraphics[width=\textwidth, trim={0cm 0.2cm 0 0.24cm}, clip]{
            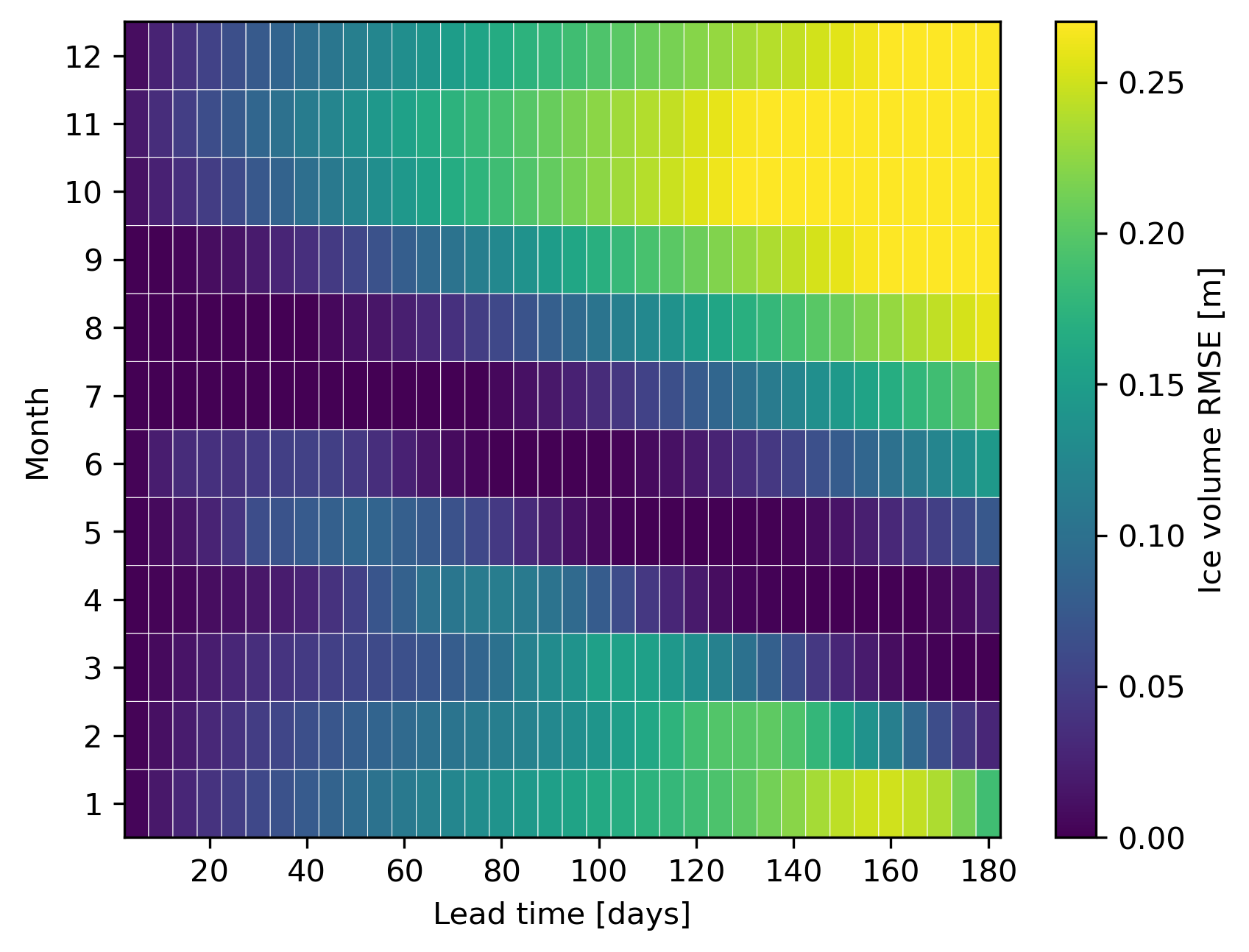
        }
    \end{subfigure}\\
    \vspace{0.1cm}
    \begin{subfigure}
        {0.49\textwidth}
        \captionsetup{position=above,justification=raggedright, skip=2pt, margin=20pt}
        \caption{P2}
        \includegraphics[width=\textwidth, trim={0cm 0.2cm 0 0.24cm}, clip]{
            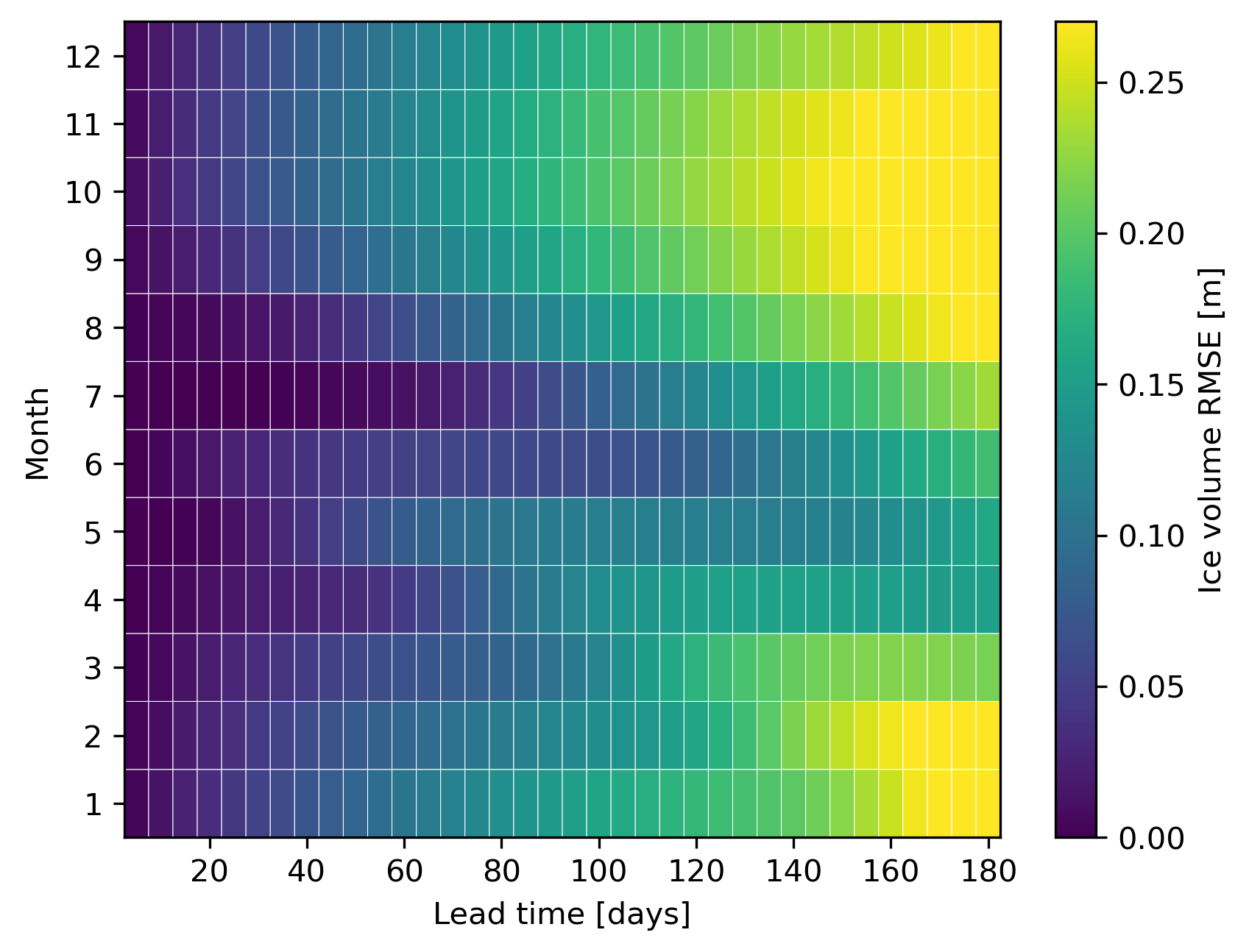
        }
    \end{subfigure}
    \begin{subfigure}
        {0.49\textwidth}
        \captionsetup{position=above,justification=raggedright, skip=2pt, margin=20pt}
        \caption{P5}
        \includegraphics[width=\textwidth, trim={0cm 0.2cm 0 0.24cm}, clip]{
            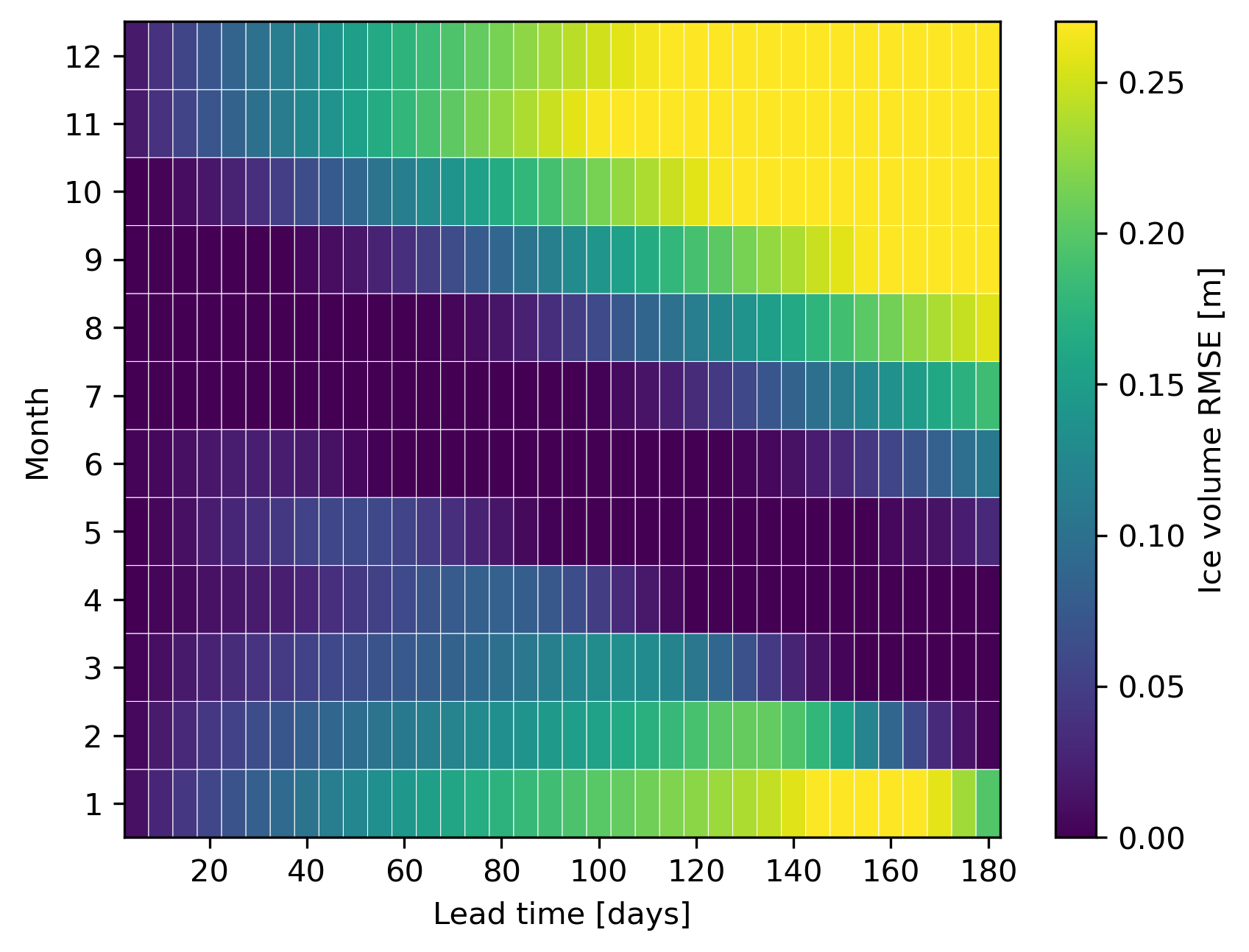
        }
    \end{subfigure}\\
    \caption{$\text{RMSE}^{\textrm{ens}}$ of the uncorrected Icepack forecasts, as function of lead time and starting month, at the six
    locations under study.}
    \label{fig:RMSE_maps_PC2}
    \end{figure}

    As detailed in Section~3.1, the uncorrected Icepack model is initialized from the true reference states at $522$ start dates across six locations, and subsequently integrated for 180 days.
    The dependence of the prediction errors on the lead time and start date is investigated in Fig.~\ref{fig:RMSE_maps_PC2}. It displays the $\text{RMSE}^{\textrm{ens}}$, averaged over start dates
    within each calendar month, separately for each of the six locations. It is evident that the error shows little dependence on 
    the start date. Nevertheless, a pronounced diagonal band of near-zero
    error appears, particularly in the panels along the right column, corresponding to lower-latitude locations P3, P4 and P5.
    This behavior of the error reflects the complete melt of sea ice during summer, when the error approaches zero, followed by the
    subsequent refreezing during which the misspecification of the snow physics parametrization progressively manifest and lead to the error growth. \\
    Based on the result in Fig.~\ref{fig:RMSE_maps_PC2}, we choose to design our ML-based bias correction to act at the lead time of 60~days (cf. Section~3.1). This choice ensures a sufficient
    signal-to-noise ratio and allows systematic errors to emerge, while avoiding excessive skill degradation between successive corrections.

    \begin{figure}
        \centering
        \begin{subfigure}{0.49\textwidth}
            \includegraphics[scale=0.45]{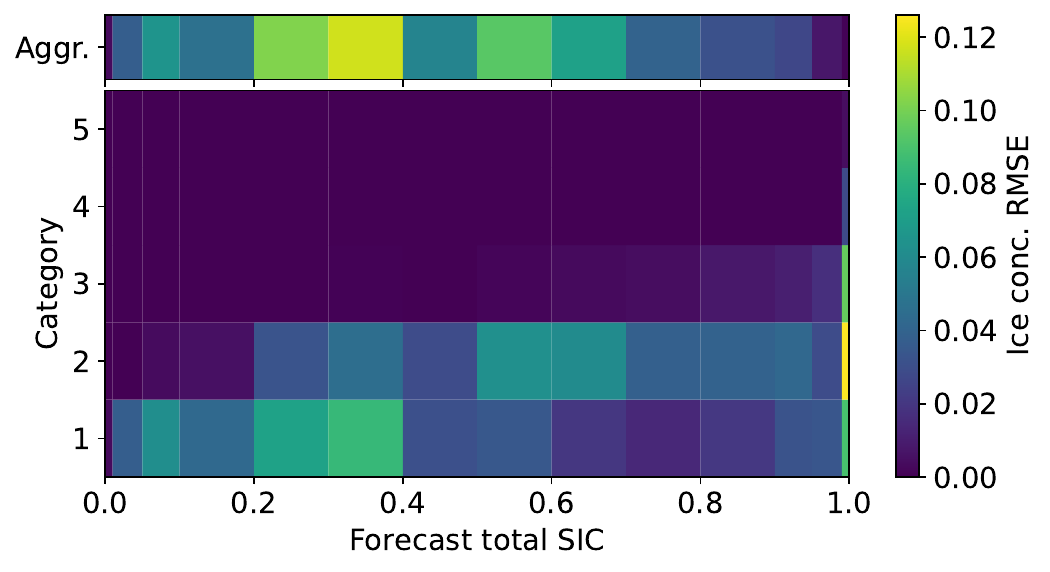}
            \caption{}
        \end{subfigure}
        \begin{subfigure}{0.49\textwidth}
            \includegraphics[scale=0.45]{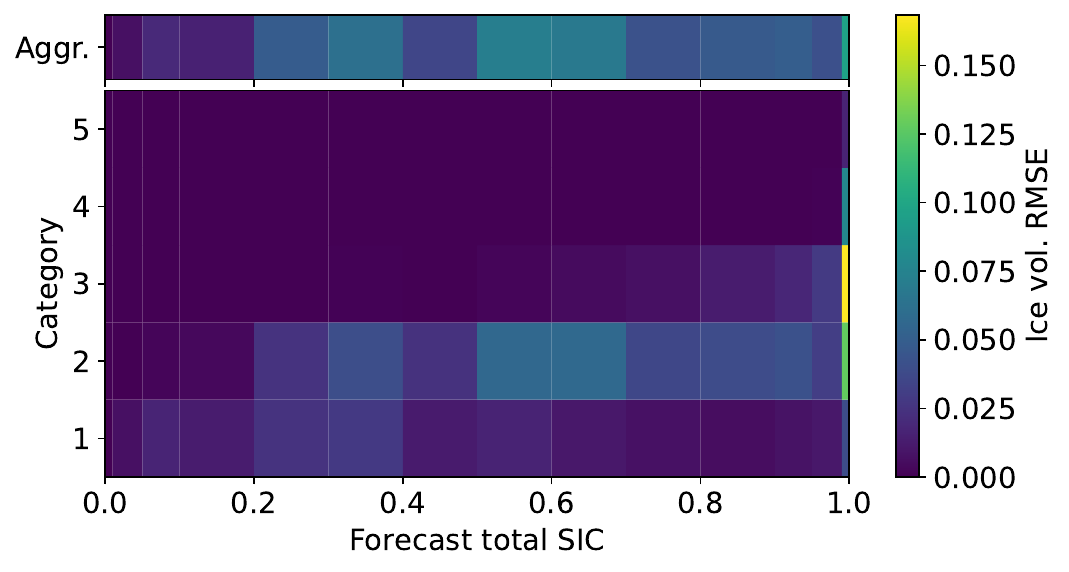}
            \caption{}
        \end{subfigure}
        \caption{Forecast $\text{RMSE}^{\textrm{ens}}$ at 60-day lead time across ice thickness categories as a function of intervals of total forecast sea ice concentration (non-uniform bins, finer near 0 and 1). Panel~(a) shows errors for ice concentration, and panel~(b) for ice volume. The top row (Aggr.) corresponds to the aggregated variables.}
        \label{fig:bins_error_maps}
    \end{figure}

    \begin{figure}
        \centering
        \begin{subfigure}
            {0.5\linewidth}
            \captionsetup{position=above,justification=raggedright, skip=0pt, margin=15pt}
            \caption{Ice concentration}
            \centering
            \includegraphics[scale=0.65, trim={0.2cm 0.3cm 0.1cm 0cm}, clip]{
                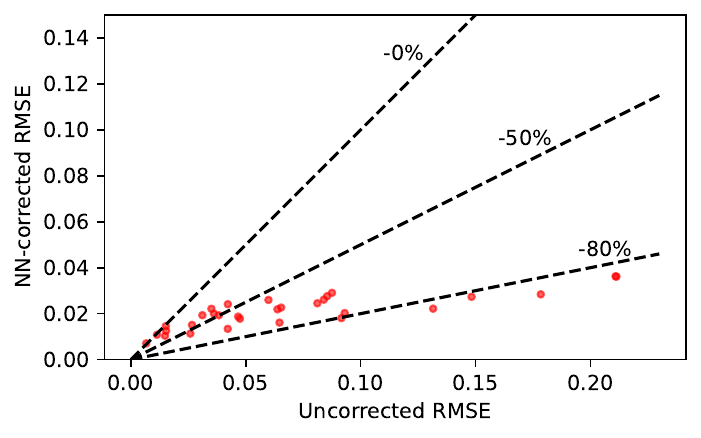
            }
            \label{fig:offline_rmse_red_aice}
        \end{subfigure}\hfill
        \begin{subfigure}
            {0.49\linewidth}
            \captionsetup{position=above,justification=raggedright, skip=0pt, margin=15pt}
            \caption{Ice concentration}
            \centering
            \includegraphics[scale=0.65, trim={0 0.3cm 0 0}, clip]{
                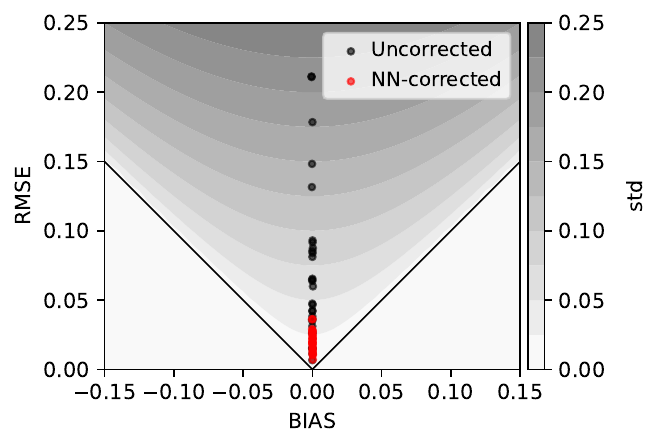
            }
            \label{fig:offline_rmse-bias_aice}
        \end{subfigure}
        \caption{Left panels (a) shows scatter plot of the residual RMSE after
        NN correction versus the uncorrected Icepack RMSE for ice concentration on the test dataset. Dashed lines indicate three levels of RMSE
        reduction, as labeled in the figure. Right panel (b) displays scatter
        plots of BIAS and RMSE before and after correction for ice concentration, with shaded contours representing the corresponding standard
        deviation.}
    \end{figure}
    \begin{figure}
        \centering
        \begin{subfigure}{0.49\textwidth}
        \captionsetup{position=above,justification=raggedright, skip=-1pt, margin=18pt}
        \caption{}
        \includegraphics[scale=0.57,  trim={0cm 0.3cm 0.2cm 0.0cm}, clip]{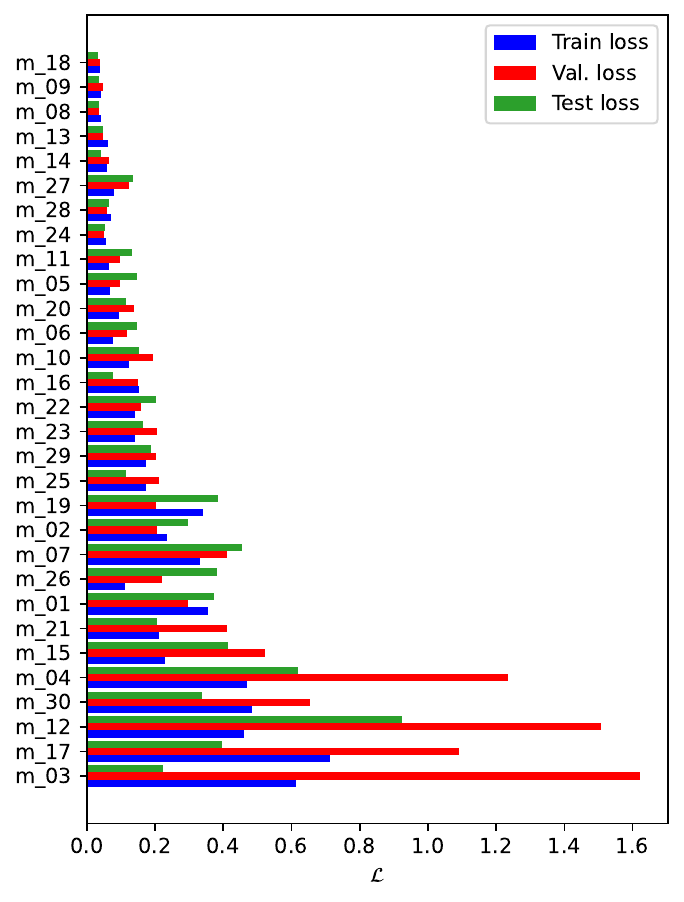}
        \end{subfigure}
        \begin{subfigure}{0.49\textwidth}
        \captionsetup{position=above,justification=raggedright, skip=-1pt, margin=15pt}
        \caption{}
        \includegraphics[scale=0.57,  trim={0.2cm 0.3cm 0 0.0cm}, clip]{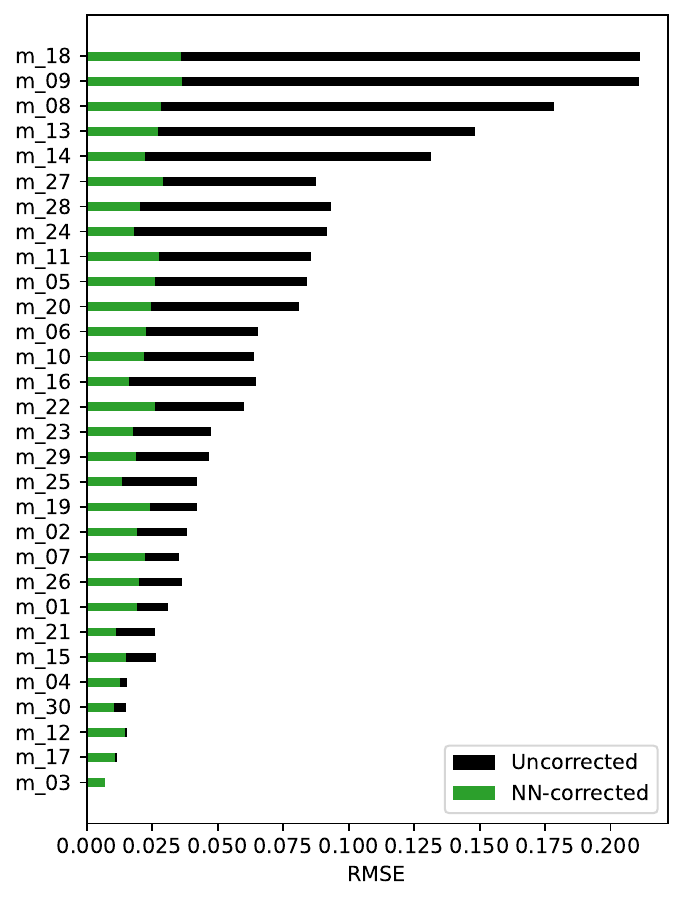}
        \end{subfigure}
        \caption{(a) Loss function $\mathcal{L}$ relative to ice concentration errors prediction, computed over training, validation and test datasets using the best-performing neural network for each ensemble member. (b) RMSE of ice concentration forecasts and the residual value after correction by the NN. Ensemble members are sorted as in Figure~5(b).}
        \label{fig:offline_aice}
    \end{figure}
    \begin{figure}
        \centering
        \includegraphics[width=0.9\linewidth]{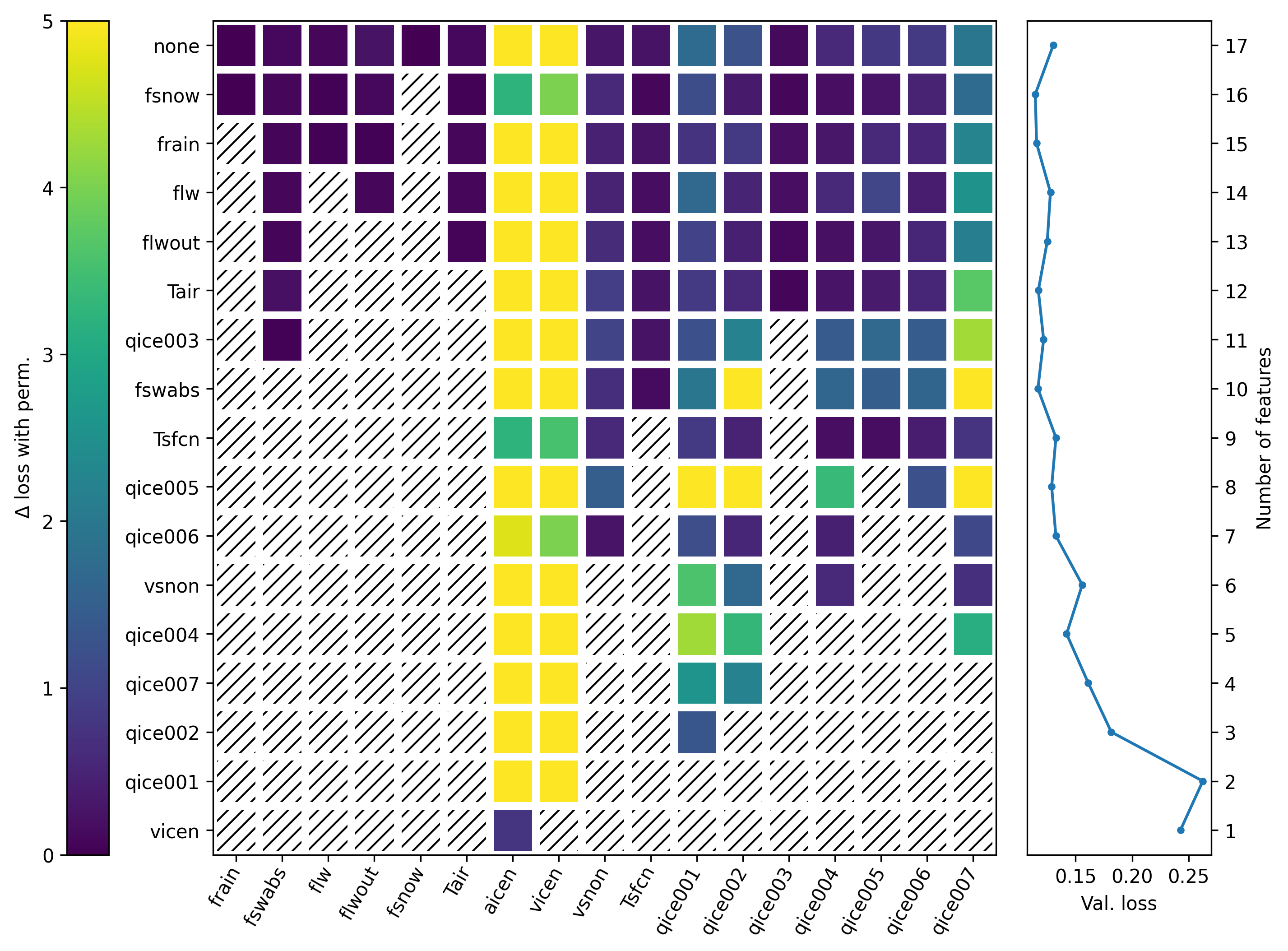}
        \caption{Recursive feature elimination for a single ensemble member. Rows represent successive iterations (top to bottom). Square colors denote the increase in validation loss caused by permuting each feature, relative to the reference obtained without permutations. Reference validation loss values are shown in the right panel. At each iteration, the feature whose permutation produces the smallest increase is removed, indicated by hatching in the subsequent row and labeled on the left.}
        \label{fig:featelim}
    \end{figure}
    \begin{figure}
        \centering
        \includegraphics[width=0.7\linewidth]{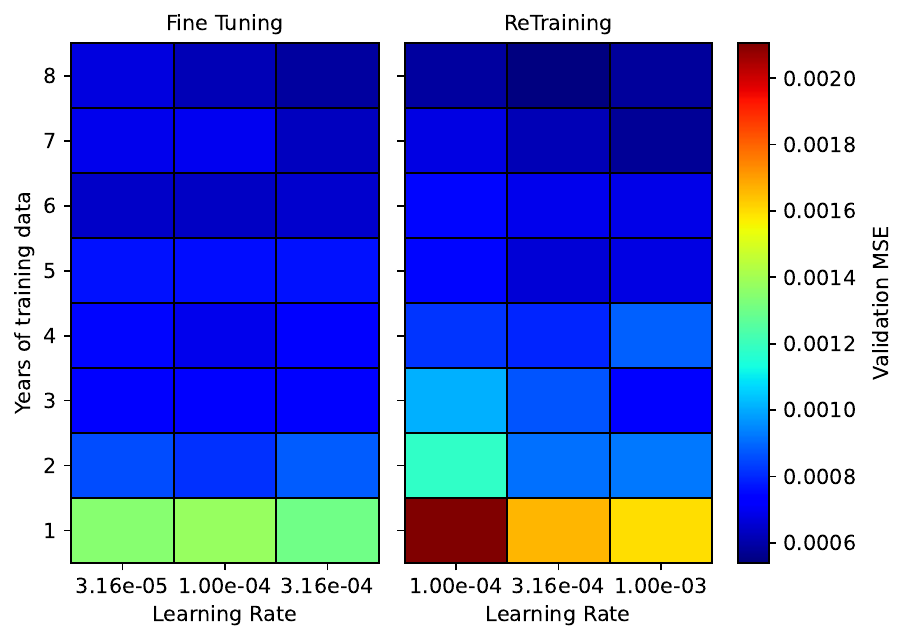}
        \caption{Validation MSE comparison between Fine-Tuning and Re-Training strategies for predicting ice volume forecast error of model $\mathrm{m\_24}$, across different learning rates and numbers of training years. In the fine-tuning experiments, the neural network was pretrained on data from model configuration $\mathrm{m\_09}$.}
        \label{fig:transfer_opt}
    \end{figure}
\end{document}